\documentclass[aps,prd,final,twocolumn,letterpaper,superscriptaddress,nofootinbib,notitlepage]{revtex4-1}

\usepackage{graphicx}                       
\usepackage[outdir=./]{epstopdf}
\usepackage{amsmath}
\usepackage{amssymb}
\usepackage{amsthm}
\usepackage{dsfont}       
\usepackage[usenames,dvipsnames,table]{xcolor}
\usepackage[format=plain, font=small, labelfont=bf, justification = raggedright]{caption}
\usepackage{array}  
\usepackage{overpic}
\usepackage{placeins} 
\usepackage{fancyhdr}
\usepackage{colortbl}
\usepackage{multirow} 
\usepackage{hyperref} 
\hypersetup{colorlinks = true,linkcolor = blue, breaklinks = true}

\newcommand{\eq}[1]{(\ref{#1})}
\newcommand{\Eq}[1]{Eq.~\eq{#1}}
\newcommand{\Eqs}[1]{Eqs.~\eq{#1}}
\newcommand{\Fig}[1]{Fig.~\ref{#1}}
\newcommand{\Sec}[1]{Sec.~\ref{#1}}
\newcommand{\Ref}[1]{Ref.~\cite{#1}}
\newcommand{\Refs}[1]{Refs.~\cite{#1}}
\newcommand{\App}[1]{Appendix~\ref{#1}}

\newcommand{\ie}{{i.e., }}

\newcommand{\mc}[1]{\mathcal{#1}}

\newcommand{\msf}[1]{\mathsf{#1}}

\newcommand{\mbb}[1]{\mathbb{#1}}

\newcommand{\oper}[1]{\smash{\hat{#1}}}
\newcommand{\unit}[1]{\smash{\check{#1}}}
\newcommand{\pd}{\partial}

\newcommand{\dd}{\mathrm{d}}

\newcommand{\bra}[1]{\langle#1 |}
\newcommand{\ket}[1]{|#1 \rangle}
\newcommand{\braket}[2]{\langle#1 |  #2 \rangle}

\newcommand{\dubdot}{\text{\large :}}
\newcommand{\tripdot}{ \text{\large $\therefore$}}
\newcommand{\Tr}{\text{tr}}

\newcommand{\NIMT}[1]{\mc{N}^{}_{#1}}
\newcommand{\IdentOp}{\oper{\mathds{1}}}
\newcommand{\deltaQO}{\Delta_\text{QO}}
\newcommand{\filter}{\mc{F}}
\newcommand{\cont}[1]{\mc{C}_{#1}}
\newcommand{\curv}{\mc{K}}

\newcommand{\Vect}[1]{{\boldsymbol{\rm #1}}}
\newcommand{\VectOp}[1]{\oper{\Vect{#1}}}

\newcommand{\Mat}[1]{\msf{#1}}
\newcommand{\IMat}[1]{\Mat{I}_{#1}}
\newcommand{\OMat}[1]{\Mat{0}_{#1}}

\newcommand{\Symb}[1]{\mc{#1}}
\newcommand{\Weyl}{\mbb{W}}
\newcommand{\WeylInv}{\mbb{W}^{-1}}
\newcommand{\QoSymb}{\Symb{D}_\text{c}}

\newcommand{\nullFrac}{\vphantom{\frac{}{}}}

\newcommand{\Stroke}[1]{\text{\ooalign{ $#1$\cr \hidewidth\raise.225ex \hbox{$-\mkern.5mu$}\cr}}}

\begin{document}
\setlength{\parskip}{0pt}
\setlength{\belowcaptionskip}{0pt}


\title{Restoring geometrical optics near caustics using sequenced metaplectic transforms}
\author{N. A. Lopez}
\affiliation{Department of Astrophysical Sciences, Princeton University, Princeton, New Jersey 08544, USA}
\author{I. Y. Dodin}
\affiliation{Department of Astrophysical Sciences, Princeton University, Princeton, New Jersey 08544, USA}
\affiliation{Princeton Plasma Physics Laboratory, Princeton, New Jersey 08540, USA}

\begin{abstract}

Geometrical optics (GO) is often used to model wave propagation in weakly inhomogeneous media and quantum-particle motion in the semiclassical limit. However, GO predicts spurious singularities of the wavefield near reflection points and, more generally, at caustics. We present a new formulation of GO, called metaplectic geometrical optics (MGO), that is free from these singularities and can be applied to any linear wave equation. MGO uses sequenced metaplectic transforms of the wavefield, corresponding to symplectic transformations of the ray phase space, such that caustics disappear in the new variables, and GO is reinstated. The Airy problem and the quantum harmonic oscillator are described analytically using MGO for illustration. In both cases, the MGO solutions are remarkably close to the exact solutions and remain finite at cutoffs, unlike the usual GO solutions.

\end{abstract}

\maketitle

\pagestyle{fancy}
\lhead{Lopez \& Dodin}
\rhead{Geometrical optics near caustics}
\thispagestyle{empty}


\section{Introduction}
\label{sec:intro}

The geometrical-optics (GO) approximation, sometimes called the Wentzel--Kramers--Brillouin (WKB) approximation, is commonly used to model wave propagation in weakly inhomogeneous media, a special case being the semiclassical motion of quantum particles~\cite{Landau81,Kravtsov90,Shankar94,Tracy14}. It is applicable when, loosely speaking, $kL \gg 1$, where $k$ is the local wavenumber and $L$ is the smallest scale among those that characterize the local properties of the medium, the wave envelope, and $k$ itself. However, even for initially smooth fields, GO often predicts the appearance of caustics, where $kL \to 0$. Examples of simple caustics include cutoffs (where $k \to 0$) and focal points (where $L \to 0$). The GO wavefield has spurious singularities at caustics, which is an issue for many applications of GO, such as calculating scattering cross sections~\cite{Ford59,Adam02} and modeling thermonuclear fusion experiments~\cite{Jaun07,Richardson10,Myatt17,Lopez18a}. Hence, developing reduced models for handling caustics is an important practical problem.

In some cases, this problem is solved by locally reducing a given wave equation to a simpler one with a known solution, such as Airy's equation~\cite{Kravtsov93,Ludwig66,Berry72}. In other cases, cutoffs are modeled as discrete interfaces, such as in specular-reflection or perfect-conductor approximations~\cite{Jackson75,Born99,Lopez18b}. However, such approaches assume that the spatial structure of the caustic is known \textit{a~priori}, which is not often the case. A more fundamental description of caustics was developed by Maslov~\cite{Maslov81} based on geometrical properties of GO solutions in the ray phase space. By occasionally rotating the phase space by $\pi/2$ using the Fourier transform in one or more spatial variables, one can remove a caustic and locally reinstate GO. Still, this approach is inconvenient for simulations because the rotation points have to be introduced \textit{ad~hoc}, requiring the simulations to be supervised. (Maslov's approach is discussed in more detail in \Sec{sec:maslov}.) Improvements have been made in the specific context of the Helmholtz equation~\cite{Alonso97b} and in wavepacket propagation~\cite{Richardson10,Littlejohn86a,Huber88a}, but not in the general case.

Here, we propose an alternate formulation of GO that can handle caustics without encountering these issues. We use the same general idea as in Maslov's method; however, instead of rotating the phase space by $\pi/2$ at select locations, we rotate the phase space \textit{continually} along a GO ray. This is accomplished with the metaplectic transform (MT)~\cite{Littlejohn86a,deGosson06}, or more precisely, the \textit{near-identity} metaplectic transform (NIMT)~\cite{Lopez19a}. Importantly, we assume neither the caustic structure nor a specific wave equation. Using the Weyl symbol calculus~\cite{Tracy14}, we formulate this procedure for any linear wave, including those governed by integro-differential equations. Hence, our approach can be useful for a wide variety of applications, such as in optics and in plasma physics.

This paper is organized as follows. In \Sec{sec:GO}, we review the basic equations of GO and introduce caustics. We also discuss Maslov's method in more detail. In \Sec{sec:MGO}, our new approach, called metaplectic GO or MGO, is derived. Subsidiary results include an algorithm to explicitly construct the rotation matrices that our method employs and also the GO equations in an arbitrarily rotated phase space. In \Sec{sec:examples}, we discuss two examples of one-dimensional ($1$-D) waves governed by Airy's equation and by Weber's equation. We study these systems analytically and show that MGO yields an accurate approximation at all locations, including near the cutoffs. In \Sec{sec:conclusion}, we summarize our main conclusions. Auxiliary calculations are presented in appendices.


\section{Geometrical optics and its breakdown near caustics}
\label{sec:GO}

\subsection{The geometrical-optics equations}

Consider an undriven linear scalar wave equation on an $N$-D configuration space with coordinates $\Vect{q}$, or $\Vect{q}$-space, which is assumed to be Euclidean%
\footnote{A generalization to a non-Euclidean space can be done using the machinery described in \Ref{Dodin19}.}. %
In general, such an equation can be written as
\begin{equation}
    \int \dd \Vect{q}' \, D(\Vect{q}, \Vect{q}') \psi(\Vect{q}') = 0 \, ,
    \label{eq:intWAVE}
\end{equation}

\noindent where $\psi(\Vect{q})$ is the wavefield on $\Vect{q}$, and $D(\Vect{q},\Vect{q}')$ is some dispersion kernel. Through use of Dirac $\delta$-functions, \Eq{eq:intWAVE} can describe local differential wave equations (but it can also describe integro-differential wave equations such as those common in plasma physics~\cite{Tracy14}). For example, choosing $D(\Vect{q},\Vect{q}') = \nabla'^2 \delta(\Vect{q}' - \Vect{q}) + n^2(\Vect{q}') \delta(\Vect{q}' - \Vect{q})$ leads to the Helmholtz equation with a spatially varying index of refraction $n(\Vect{q})$. Here, $\nabla$ and $\nabla'$ are the gradient with respect to $\Vect{q}$ and $\Vect{q}'$ respectively.

Let us introduce%
\footnote{Here, we use the bra-ket notation that is standard in quantum mechanics and optics~\cite{Stoler81}.} %
a Hilbert space of state vectors $\ket{\psi}$ such that $\psi(\Vect{q})$ is the projection of a given $\ket{\psi}$ onto the eigenbasis $\{ \ket{\Vect{q}} \}$ of the coordinate operator $\VectOp{q}$. We adopt the usual normalization such that
\begin{equation}
    \int \dd \Vect{q} \, \ket{\Vect{q}} \bra{\Vect{q}} = \IdentOp \, ,
\end{equation}

\noindent where $\IdentOp$ is the identity operator. Then,
\begin{equation}
    \psi(\Vect{q}) \doteq \braket{\Vect{q}}{\psi} \, ,
\end{equation}

\noindent where the symbol $\doteq$ denotes definitions. We define $\VectOp{q}$ through its action on the Hilbert space as $\VectOp{q}\ket{\Vect{q}'} = \Vect{q}' \ket{\Vect{q}'}$, or equivalently, through its matrix elements 
\begin{equation}
    \bra{\Vect{q}}\VectOp{q} \ket{\Vect{q}'} = \Vect{q}' \delta(\Vect{q}' - \Vect{q}) \, .
\end{equation}

\noindent The canonically conjugate momentum operator $\VectOp{p}$ is similarly defined through its matrix elements as 
\begin{equation}
    \bra{\Vect{q}}\VectOp{p} \ket{\Vect{q}'} = i\nabla' \delta(\Vect{q}' - \Vect{q}) \, .
\end{equation}

Let us further define the dispersion operator $\oper{D}$ through its matrix elements $\bra{\Vect{q}}\oper{D} \ket{\Vect{q}'} = D(\Vect{q},\Vect{q'})$. Then, \Eq{eq:intWAVE} can be represented as
\begin{equation}
    \oper{D}(\VectOp{q}, \VectOp{p}) \ket{\psi} = \ket{0} \, ,
    \label{eq:hilbertWAVE}
\end{equation}

\noindent with \Eq{eq:intWAVE} being simply the projection of \Eq{eq:hilbertWAVE} onto the coordinate eigenbasis. (Here, $\ket{0}$ is the null vector.) Note that $\oper{D}$ is expressed as a function of $\VectOp{q}$ and $\VectOp{p}$. When the dispersion kernel $D(\Vect{q},\Vect{q}')$ describes a local differential wave equation, the construction of $\oper{D}(\VectOp{q},\VectOp{p})$ is trivial. For example, the aforementioned Helmholtz equation has $\oper{D}(\VectOp{q},\VectOp{p}) = -\VectOp{p}^2 + n^2(\VectOp{q})$. However, constructing $\oper{D}(\VectOp{q},\VectOp{p})$ for integro-differential wave equations requires a pseudo-differential representation of $D(\Vect{q},\Vect{q}')$. Such a representation can be obtained using the Weyl symbol calculus, which we shall introduce momentarily.

GO is the asymptotic model of \Eq{eq:hilbertWAVE} for the short-wavelength limit (\Sec{sec:intro}). In this limit, the wavefield can be partitioned into a rapidly varying phase and a slowly varying envelope. Following \Ref{Dodin19}, we define the envelope state vector $\ket{\phi}$ via the unitary transformation
\begin{equation}
    \ket{\phi} \doteq e^{-i \theta(\VectOp{q})} \ket{\psi} \, ,
    \label{eq:phiENV}
\end{equation}

\noindent where $\theta(\VectOp{q})$ is a hermitian operator representing the phase of $\psi(\Vect{q})$. Under this transformation, \Eq{eq:hilbertWAVE} becomes
\begin{equation}
    e^{-i \theta(\VectOp{q})}
    \oper{D}(\VectOp{q}, \VectOp{p}) \,
    e^{i \theta(\VectOp{q})}
    \ket{\phi} = \ket{0} \, .
    \label{eq:hilbertENV}
\end{equation}

We shall now approximate the envelope dispersion operator of \Eq{eq:hilbertENV} in the GO limit. This is readily accomplished using the Weyl symbol calculus, which provides a mapping between functions and operators~\cite{McDonald88a}. Let $\Symb{A}(\Vect{z})$ be a function on classical phase space with coordinates $\Vect{z} \doteq (\Vect{q}, \Vect{p})^\intercal$. The Weyl transform $\Weyl$ maps $\Symb{A}(\Vect{z})$ to an operator $\oper{A}(\VectOp{z})$ on the Hilbert space as
\begin{align}
    \oper{A}(\VectOp{z}) &= \Weyl \left[ \Symb{A}(\Vect{z}) \right]
    \nonumber\\
    &\doteq
    \int \frac{\dd \Vect{z}' \, \dd \Vect{\zeta}}{(2\pi)^{2N}} \, 
    \Symb{A}(\Vect{z}') 
    e^{-i\Vect{\zeta}^\intercal \Mat{J} \Vect{z}'} 
    e^{i\Vect{\zeta}^\intercal \Mat{J} \VectOp{z}} \, ,
    \label{eq:Weyl}
\end{align}

\noindent where $\VectOp{z} \doteq (\VectOp{q}, \VectOp{p})^\intercal$ and we have introduced the matrix
\begin{equation}
    \Mat{J} \doteq 
    \begin{pmatrix}
        \OMat{N} & \IMat{N} \\
        -\IMat{N} & \OMat{N}
    \end{pmatrix}\, ,
    \label{eq:jMAT}
\end{equation}

\noindent with $\OMat{N}$ and $\IMat{N}$ being respectively the $N$-D null and identity matrices. (Here, $^\intercal$ denotes the matrix transpose, which also denotes the scalar dot product for vectors, \ie $\Vect{a}^\intercal \Vect{b} \equiv \Vect{a} \cdot \Vect{b}$.) The function $\Symb{A}(\Vect{z})$ is called the Weyl symbol of $\oper{A}(\VectOp{z})$. The inverse Weyl transform, sometimes called the Wigner transform, maps operators back to phase-space functions as
\begin{align}
    \Symb{A}(\Vect{z}) &= \WeylInv \left[\oper{A}(\VectOp{z}) \right] 
    \nonumber\\
    &\doteq \int \frac{\dd \Vect{z}'}{(2\pi)^N} \, e^{i\left(\Vect{z}'\right)^\intercal \Mat{J} \Vect{z}} \, \Tr \left[
    e^{-i\left(\Vect{z}'\right)^\intercal \Mat{J} \VectOp{z}}
    \oper{A}(\VectOp{z})\right] \, .
    \label{eq:Wigner}
\end{align}

\noindent The Weyl symbol calculus is reviewed in \App{app:Weyl}.

With the Weyl symbol calculus, approximating operators becomes as easy as approximating functions: one performs a Wigner transform to obtain the operator's Weyl symbol, approximates the symbol in the desired limit using, say, familiar Taylor expansions, then performs a Weyl transform to obtain the correspondingly approximated operator. As shown in \App{app:GO}, applying this procedure to \Eq{eq:hilbertENV} ultimately yields
\begin{equation}
    \left\{ 
        \Symb{D}\left[\VectOp{q}, \Vect{p}(\VectOp{q}) \right]
        + \Vect{v}(\VectOp{q})^\intercal \VectOp{p}
        - \frac{i}{2} \pd_\Vect{q} \cdot \Vect{v}(\VectOp{q})
    \right\} \ket{\phi} = \ket{0} \, ,
    \label{eq:approxENV}
\end{equation}

\noindent where 
\begin{subequations}
    \label{eq:GOdefs}
    \begin{align}
        \Symb{D}(\Vect{z}) &\doteq \WeylInv \left[\oper{D}(\VectOp{z}) \right] \, ,
        \\
        \label{eq:pGRADtheta}
        \Vect{p}(\Vect{q}) &\doteq \pd_\Vect{q} \theta(\Vect{q}) \, ,
        \\
        \Vect{v}(\Vect{q}) &\doteq \left.\pd_{\Vect{p}} \Symb{D}(\Vect{q}, \Vect{p}) \right|_{\Vect{p}=\Vect{p}(\Vect{q})}
    \end{align}
\end{subequations}

\noindent are interpreted respectively as the local dispersion relation, the local wavevector, and the local group velocity. Importantly, note that
\begin{equation}
    \nabla \times \Vect{p}(\Vect{q}) = \Vect{0} \, .
    \label{eq:curlp}
\end{equation}

\noindent Projecting \Eq{eq:approxENV} onto $\Vect{q}$-space then yields the GO equations,
\begin{subequations}
    \label{eq:GO}
    \begin{gather}
        \label{eq:GOrays}
        \Symb{D}\left[\Vect{q}, \Vect{p}(\Vect{q}) \right] = 0 \, , 
        \\
        \label{eq:GOenv}
        \Vect{v}(\Vect{q})^\intercal \pd_{\Vect{q}} \phi(\Vect{q}) 
        + \frac{1}{2} \left[\pd_\Vect{q} \cdot \Vect{v}(\Vect{q}) \right] \phi(\Vect{q}) = 0 \, ,
    \end{gather}
\end{subequations}

\noindent where $\phi(\Vect{q}) \doteq \braket{\Vect{q}}{\phi}$. Specifically, \Eq{eq:GOrays} represents a local dispersion relation, and \Eq{eq:GOenv} represents an envelope equation. Note that for vector fields, additional polarization terms can emerge~\cite{Dodin19,Oancea20}.

\subsection{Ray trajectories and caustics}
\label{sec:raysCAUSTICS}

Equations \eq{eq:GO} are commonly solved along the $1$-D characteristic ray trajectories of \Eq{eq:GOrays}. These rays are described by Hamilton's equations
\begin{equation}
    \pd_{\tau_1} \Vect{z} = \Mat{J} \, \pd_\Vect{z} \Symb{D}(\Vect{z}) \, ,
    \label{eq:rayHAM}
\end{equation}

\noindent where $\tau_1$ is some variable used for ray parameterization. Since the value of $\Symb{D}(\Vect{z})$ [\Eq{eq:GOrays}] and $\nabla \times \Vect{p}(\Vect{q})$ [\Eq{eq:curlp}] are conserved along each ray, the solutions to \Eq{eq:rayHAM} are confined to an $N$-D surface in the $2N$-D phase space called the dispersion manifold. A parametric representation of the dispersion manifold is readily obtained by integrating \Eq{eq:rayHAM}. Indeed, the formal solution to \Eq{eq:rayHAM} can be written as $\Vect{z}(\Vect{\tau})$, where $\Vect{\tau} \doteq (\tau_1, \Vect{\tau}_\perp)$ and $\Vect{\tau}_\perp$ are coordinates on the $(N-1)$-D sub-surface of initial conditions $\Vect{z}_0(\Vect{\tau}_\perp) \doteq \Vect{z}(0,\Vect{\tau}_\perp)$.

Importantly, since \Eq{eq:rayHAM} is a first-order autonomous system, rays can never cross in phase space; however, their projections onto $\Vect{q}$-space, $\Vect{q}(\Vect{\tau})$, have no such restriction. This can be problematic for the GO model [\Eqs{eq:GO}], which is constructed in $\Vect{q}$-space using $\Vect{q}(\Vect{\tau})$. To see why, let us integrate \Eq{eq:GOenv} along a ray. The first term in \Eq{eq:GOenv} is clearly the directional derivative of $\phi$ along the ray trajectory. Following \Ref{Kravtsov90}, the second term is simplified upon noting that
\begin{equation}
    \pd_\Vect{q} \cdot \Vect{v}(\Vect{\tau}) = \Tr \left[ \pd_\Vect{q} \Vect{v}(\Vect{\tau}) \right] \, ,
\end{equation}

\noindent where $\Tr$ denotes the matrix trace and $\Vect{v}(\Vect{\tau}) \doteq \Vect{v}\left[ \Vect{q}(\Vect{\tau}) \right]$. Then, since
\begin{align}
    \pd_\Vect{q} \Vect{v}(\Vect{\tau}) &= \pd_\Vect{\tau} \Vect{v}(\Vect{\tau})
    \left[\pd_\Vect{\tau} \Vect{q}(\Vect{\tau}) \right]^{-1} 
    \nonumber\\
    &= \pd_\Vect{\tau} \left[\pd_{\tau_1}\Vect{q}(\Vect{\tau}) \right]
    \left[\pd_\Vect{\tau} \Vect{q}(\Vect{\tau}) \right]^{-1}
    \nonumber\\
    &= \pd_{\tau_1} \left[\pd_\Vect{\tau} \Vect{q}(\Vect{\tau}) \right]
    \left[\pd_\Vect{\tau} \Vect{q}(\Vect{\tau}) \right]^{-1}
\end{align}

\noindent by the chain rule, Jacobi's formula for the derivative of the matrix determinant implies that
\begin{equation}
    \Tr\left[\pd_\Vect{q} \Vect{v}(\Vect{\tau}) \right]
    = \pd_{\tau_1} \log J(\Vect{\tau}) \, ,
    \label{eq:rayJACOB}
\end{equation}

\noindent where $J(\Vect{\tau}) \doteq \det \pd_\Vect{\tau} \Vect{q}(\Vect{\tau})$ is the Jacobian determinant of the ray evolution in $\Vect{q}$-space, and $\log(x)$ is the natural logarithm. 

Hence, \Eq{eq:GOenv} is written along a ray as
\begin{equation}
    \pd_{\tau_1} \phi(\Vect{\tau}) 
    + \frac{1}{2}\left[ \pd_{\tau_1} \log J(\Vect{\tau}) \right]
    \phi(\Vect{\tau}) = 0 \, .
    \label{eq:rayENV}
\end{equation}

\noindent The solution of \Eq{eq:rayENV} is
\begin{equation}
    \phi(\Vect{\tau}) = \phi_0(\Vect{\tau}_\perp) 
    \sqrt{ \frac{J_0(\Vect{\tau}_\perp)}{J(\Vect{\tau})} } \, ,
\end{equation}

\noindent where $\phi_0(\Vect{\tau}_\perp) \doteq \phi(0, \Vect{\tau}_\perp)$ and $J_0(\Vect{\tau}_\perp) \doteq J(0, \Vect{\tau}_\perp)$ are set by initial conditions. Clearly, $\phi(\Vect{\tau})$ has singularities where $J(\Vect{\tau}) = 0$. These locations are called caustics.

\begin{figure}
    \centering
    \includegraphics[width=0.9\linewidth]{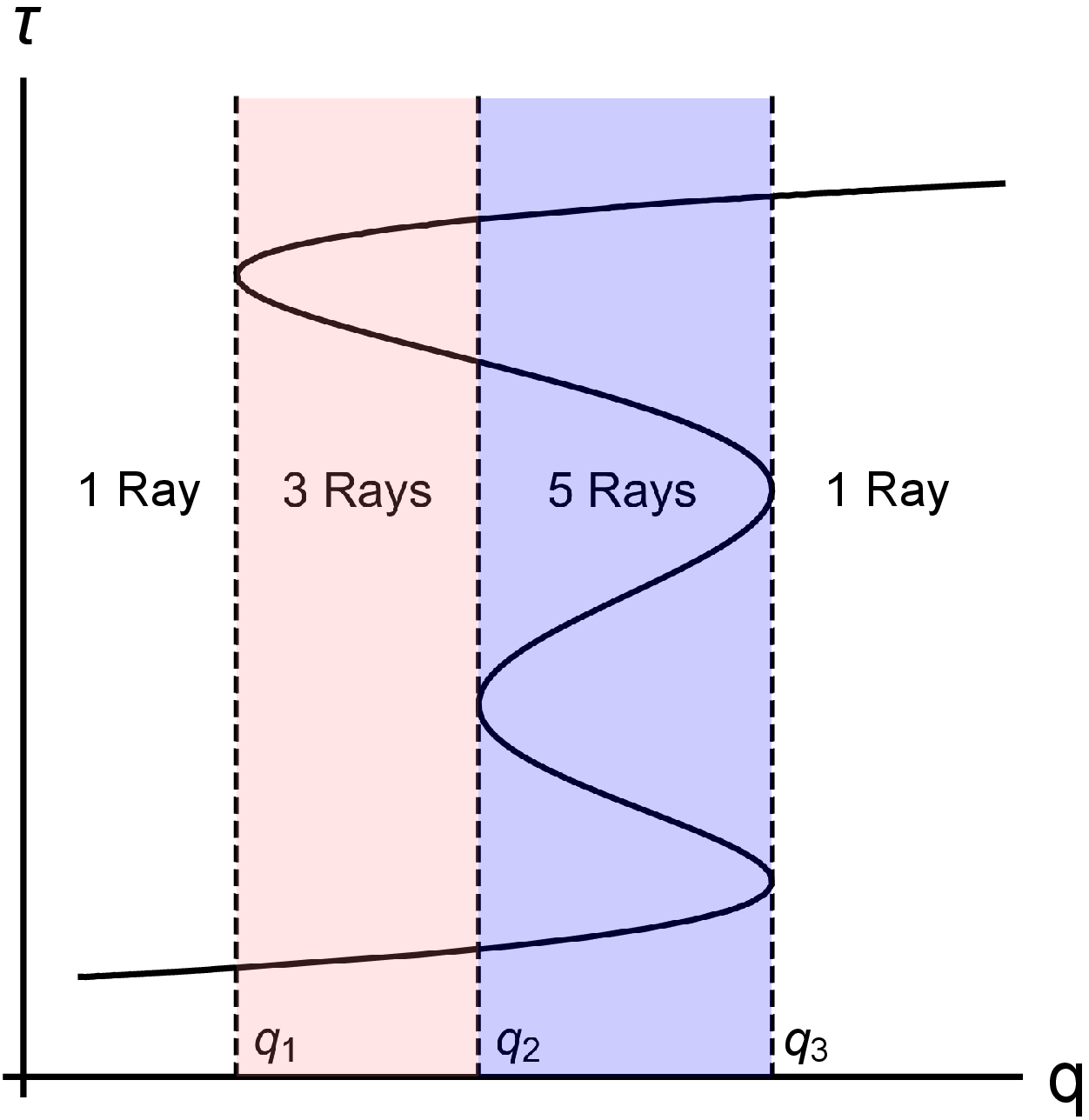}
    \caption{For a continuous function $\Vect{q}(\Vect{\tau})$, $\det \pd_\Vect{\tau} \Vect{q} = 0$ typically occurs at boundaries between regions where $\Vect{q}(\Vect{\tau}) = \Vect{q}_0$ has differing numbers of roots. This is illustrated above in $1$-D, where $q'(\tau) = 0$ at $q_1$, $q_2$, and $q_3$. An exception is when $\det \pd_\Vect{\tau} \Vect{q} = 0$ corresponds to a degenerate saddlepoint (an inflection point in $1$-D); however, these structures are not stable to small perturbations, so they are not often seen in physical systems.}
    \label{fig:caustPROJ}
\end{figure}

To better understand where and why caustics occur, let us consider the extended `ray parameter' space $(\Vect{q}, \Vect{\tau})$. In this space, the ray trajectories are represented by the graph $\Vect{\tau} = \Vect{\tau}(\Vect{q})$, which is obtained by a formal inversion of $\Vect{q}(\Vect{\tau})$. The condition $J(\Vect{\tau}) = 0$ has a geometric interpretation in this space: $J(\Vect{\tau}) = 0$ where the projection of $\Vect{\tau}(\Vect{q})$ onto $\Vect{q}$-space becomes singular. Importantly, caustics do not occur every time rays cross in $\Vect{q}$-space, but rather, when the number of rays crossing in $\Vect{q}$-space changes abruptly. In this sense, caustics appear as topological boundaries (see \Fig{fig:caustPROJ}).

Since $\Vect{\tau}$ are coordinates on the dispersion manifold, the same geometric interpretation of caustics must hold in phase space as well. Indeed, since
\begin{equation}
    \pd_\Vect{q}\Vect{p}(\Vect{\tau}) = 
    \pd_\Vect{\tau} \Vect{p}(\Vect{\tau}) 
    \left[\pd_\Vect{\tau} \Vect{q}(\Vect{\tau}) \right]^{-1} \, ,
\end{equation}

\noindent it follows that
\begin{equation}
    \det \pd_\Vect{q} \Vect{p}(\Vect{\tau})
    = \frac{\det \pd_\Vect{\tau} \Vect{p}(\Vect{\tau})}
    {J(\Vect{\tau})} \, .
\end{equation}

\noindent Hence, $J(\Vect{\tau}) = 0$ where the dispersion manifold has a singular projection onto $\Vect{q}$-space as well. Formulating caustics as projection singularities in phase space is advantageous because it highlights the arbitrariness in the initial choice to project \Eq{eq:hilbertWAVE} onto $\Vect{q}$-space. More generally, a caustic occurs wherever the dispersion manifold has a singular projection onto the chosen projection plane. Consequently, caustics can be removed by rotating the projection plane.

\subsection{Maslov's method for caustic removal}
\label{sec:maslov}

A popular paradigm for performing such phase-space rotations is Maslov's method~\cite{Maslov81,Ziolkowski84}. This method takes advantage of the fact that a well-behaved dispersion manifold cannot have a singular projection onto both $\Vect{q}$-space and $\Vect{p}$-space simultaneously. A caustic that appears in $\Vect{q}$-space is thus absent in $\Vect{p}$-space, and vice versa. By repeatedly switching between $\Vect{q}$-space and $\Vect{p}$-space as caustics are approached, one can construct a GO framework that does not produce singularities along a given ray.

\begin{figure}
    \centering
    \includegraphics[width=0.9\linewidth]{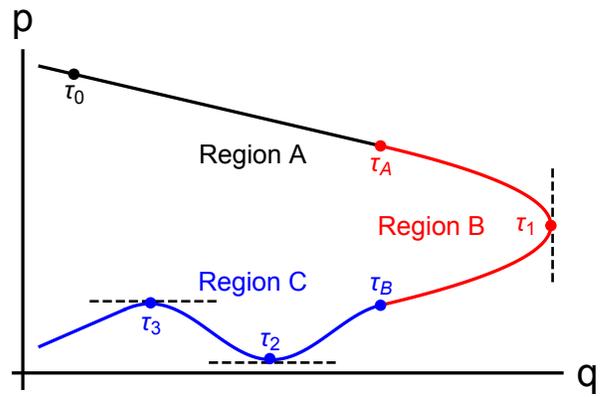}
    \caption{A $1$-D dispersion manifold with coordinate $\tau$ that exhibits a $\Vect{q}$-space caustic at $\tau = \tau_1$ and two $\Vect{p}$-space caustics at $\tau = \tau_2$ and $\tau = \tau_3$. Region A ($\tau \le \tau_A$) is far from the caustics, so both the $\Vect{q}$-space and $\Vect{p}$-space GO solutions are well-behaved. However, region B ($\tau_A < \tau < \tau_B$) is close to the $\Vect{q}$-space caustic, so the $\Vect{q}$-space GO solution is singular while the $\Vect{p}$-space GO solution is well-behaved. Similarly, region C ($\tau \ge \tau_B$) is close to the $\Vect{p}$-space caustics, so the $\Vect{q}$-space GO solution is well-behaved while the $\Vect{p}$-space GO solution is singular.}
    \label{fig:Maslov}
\end{figure}

To illustrate this method qualitatively, let us consider the dispersion manifold shown in \Fig{fig:Maslov}. A $\Vect{q}$-space caustic occurs at $q(\tau_1)$, while $\Vect{p}$-space caustics occur at $p(\tau_2)$ and $p(\tau_3)$. Consider a wave initially located at $q(\tau_0)$. In region A, that is, for $\tau$ between $\tau_0$ and some $\tau_a$ to be specified momentarily, $\Vect{q}$-space can be used as the projection plane. Consequently, the wave envelope is evolved using \Eq{eq:GOenv}. Near the $\Vect{q}$-space caustic at $q(\tau_1)$, however, \Eq{eq:GOenv} breaks down and cannot be used. Instead, the projection plane should be switched from $\Vect{q}$-space to $\Vect{p}$-space prior to encountering the caustic at $\tau_1$. The switching location, $\tau_a$, must be far enough from $\Vect{q}$-space and $\Vect{p}$-space caustics such that GO is accurate in both representations near $\tau_a$, but is otherwise arbitrary. 

Next, the wavefield $\psi(q)$ is transformed to its $\Vect{p}$-space representation $\Psi(p)$ at $\tau_a$. This is achieved using the Fourier transform (FT) subsequently evaluated via the stationary phase approximation (SPA)~\cite{Olver10}. Indeed, since
\begin{equation}
    \hspace{-1.2mm}
    \Psi(p) = \int \frac{\dd q}{\sqrt{2\pi}} \, \psi(q) e^{-i pq} = \int \frac{\dd q}{\sqrt{2\pi}} \, \phi(q) e^{i \theta(q) - i pq} \, ,
    \label{eq:fourier}
\end{equation}

\noindent the phase of the FT integrand is stationary where $\pd_q \theta(q) = p$, which is satisfied along the dispersion manifold by definition. When $\tau_a$ is chosen sufficiently far from both $\Vect{q}$-space caustics (such that $\phi$ is not singular) and $\Vect{p}$-space caustics (such that $\pd^2_q \theta \equiv \pd_q p$ is nonzero), the SPA of \Eq{eq:fourier} is
\begin{equation}
    \Psi\left[ p(\tau_a) \right] = \frac{\psi\left[ q(\tau_a) \right]}{\sqrt{\pd_q p(\tau_a) }}e^{i\frac{\pi}{4} - i p(\tau_a) q(\tau_a)} \, .
\end{equation}

\noindent Thus, the SPA has the important role in Maslov's method of localizing the FT to become a pointwise mapping from $\psi\left[q(\tau_a)\right]$ to $\Psi\left[p(\tau_a)\right]$. 

Being absent from $\Vect{p}$-space caustics, $\Psi(p)$ is evolved in region B from $\tau_a$ to $\tau_b$ using GO formulated in $\Vect{p}$-space. We shall derive the GO equations in various projection planes, including $\Vect{p}$-space, in the following section; for the moment, let us simply note that the $\Vect{p}$-space GO equations are not obtained by projecting \Eq{eq:approxENV} onto $\{\ket{\Vect{p}} \}$, but instead, more sophisticated machinery must be introduced. After propagating $\Psi(p)$ through region B, the projection plane must be switched back to $\Vect{q}$-space to avoid the $\Vect{p}$-space caustic at $p(\tau_2)$. This is accomplished by using the inverse FT evaluated via SPA. Since there are no remaining $\Vect{q}$-space caustics, $\psi(q)$ can be evolved using $\Vect{q}$-space GO for all subsequent $\tau > \tau_b$.

Maslov's method has been very successful for the theoretical analysis of caustics. However, the lack of rigorous criteria for choosing when to switch between $\Vect{q}$-space and $\Vect{p}$-space is unsatisfying and can be awkward for practical calculations. For a code, this selection must be performed using an external module that supervises the envelope evaluation, detects when a caustic is becoming `close' using some \textit{ad hoc} cost function, then triggers a switch in representation. A framework that could proceed unsupervised would be more desirable. In the remainder of the paper, we shall present such a framework.


\section{Restoring geometrical optics using phase-space rotations}
\label{sec:MGO}

\begin{figure}
    \centering
    \includegraphics[width=0.8\linewidth,trim={5mm 7mm 20mm 9mm},clip]{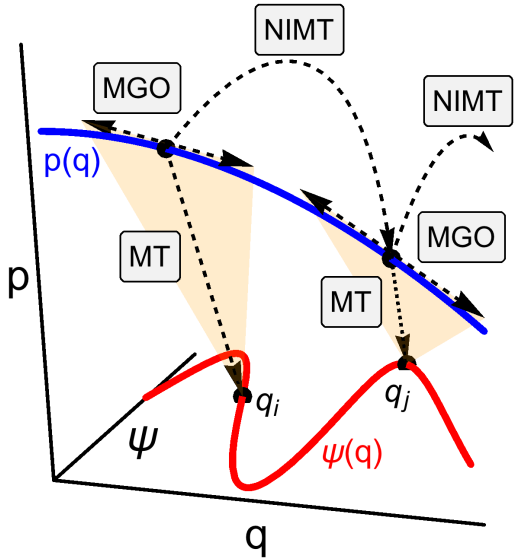}
    \caption{New framework for GO that is free from caustic singularities. This new approach consists of three steps: for a given point on the dispersion manifold, (i) the GO solution is calculated on the local tangent plane using metaplectic geometrical optics (MGO), (ii) the MGO solution is projected onto the tangent plane of the next point on the dispersion manifold using a near-identity metaplectic transform (NIMT) to initialize the next MGO calculation, (iii) the MGO solution is projected onto configuration space using a metaplectic transform (MT) subsequently evaluated using stationary phase or steepest-descent method. The process is repeated for all points on the dispersion manifold.}
    \label{fig:metMASLOV}
\end{figure}

Maslov's method for caustic removal uses only a small subset of all the possible projective planes; indeed, only $\Vect{q}$-space, $\Vect{p}$-space, and simple combinations such as $(q_x, p_y)$ are considered. This restriction ultimately results in an algorithm which must be supervised. However, allowing for a wider variety of possible projective planes can eliminate this need for supervision. This is the basis for our approach, which is outlined in \Fig{fig:metMASLOV}. 

In short, rather than \textit{sometimes} switching between $\Vect{q}$-space and $\Vect{p}$-space as in Maslov's method, we propose to \textit{always} switch between $\Vect{q}$-space and the local tangent plane of the dispersion manifold at a desired query point $\Vect{z}(\Vect{\tau})$. Each point on the dispersion manifold is thus treated equally, and there is no need to arbitrarily designate specific points as `switching' points. Also, there will never be a caustic near $\Vect{z}(\Vect{\tau})$ by definition. For these reasons, our approach should be easy to implement in a code. Let us now develop this idea in more detail.

\subsection{Phase-space rotations via metaplectic operators}

Let us first discuss how to transform between $\Vect{q}$-space and the local tangent plane of the dispersion manifold. Generally speaking, for a given plane in phase space to be a valid choice for a GO projective plane, it must be related to $\Vect{q}$-space by a linear symplectic transformation. This is because linear symplectic transformations preserve the Poisson bracket, and consequently define an equivalency class on phase space~\cite{Goldstein02}. 

A linear symplectic transformation is defined by a $2N\times2N$ matrix $\Mat{S}$ which satisfies
\begin{equation}
    \Mat{S} \Mat{J} \Mat{S}^\intercal = \Mat{J} \, .
    \label{eq:symplecS}
\end{equation} 

\noindent This matrix transforms the original phase space $\Vect{z}$ into a new phase space $\Stroke{\Vect{Z}}$ via
\begin{subequations}
    \begin{equation}
        \Stroke{\Vect{Z}} = \Mat{S} \Vect{z} \, ,
        \label{eq:symplecZ}
    \end{equation}

    \noindent or more explicitly, using $\Stroke{\Vect{Z}} \doteq (\Vect{Q}, \Vect{P})^\intercal$ and $\Vect{z} \doteq (\Vect{q}, \Vect{p})^\intercal$,
    \begin{equation}
        \Vect{Q} = \Mat{A} \Vect{q} + \Mat{B} \Vect{p} \, , 
        \quad 
        \Vect{P} = \Mat{C} \Vect{q} + \Mat{D} \Vect{p} \, ,
    \end{equation}
\end{subequations}

\noindent where $\Mat{A}$, $\Mat{B}$, $\Mat{C}$, and $\Mat{D}$ are $N \times N$ block matrices that comprise $\Mat{S}$ as follows:
\begin{equation}
    \Mat{S} \doteq
    \begin{pmatrix}
        \Mat{A} & \Mat{B} \\
        \Mat{C} & \Mat{D}
    \end{pmatrix} \, .
    \label{eq:sABCD}
\end{equation}

\noindent The inverse matrix, $\Mat{S}^{-1}$, has a similar block decomposition given as
\begin{equation}
    \Mat{S}^{-1} =
    \begin{pmatrix}
        \Mat{D}^\intercal & - \Mat{B}^\intercal \\
        - \Mat{C}^\intercal & \Mat{A}^\intercal 
    \end{pmatrix} \, ,
\end{equation}

\noindent which is readily derived using \Eqs{eq:symplecS} and \eq{eq:sABCD}.

Note that \Eq{eq:symplecZ} preserves the origin of phase space, \ie $\Vect{z} = \Vect{0}$ maps to $\Stroke{\Vect{Z}} = \Vect{0}$. Shifting the origin does not affect the projective properties of a plane; hence, for the purposes of developing GO, we can identify all projective planes which differ only by a shift in the origin as equivalent. As a result, when we speak of the `tangent plane' of the dispersion manifold at $\Vect{z}(\Vect{\tau})$, we really speak of the plane which is parallel to the tangent plane at $\Vect{z}(\Vect{\tau})$ and passes through the origin.

In the Hilbert space, linear symplectic transformations are performed using metaplectic operators. A metaplectic operator $\oper{M}$ is a unitary operator which, via conjugation, transforms the operator $\VectOp{z}$ to $\Stroke{\VectOp{Z}}$ as
\begin{subequations}
    \begin{equation}
        \label{eq:zTRANS}
        \Stroke{\VectOp{Z}} \doteq \oper{M}^\dagger \VectOp{z} \oper{M} = \Mat{S} \VectOp{z} \, ,
    \end{equation}

    \noindent or in terms of $\VectOp{q}$ and $\VectOp{p}$,
    \begin{align}
        \VectOp{Q} &\doteq \oper{M}^\dagger \VectOp{q} \oper{M} 
        = \Mat{A} \VectOp{q} + \Mat{B} \VectOp{p} \, , \\
        \VectOp{P} &\doteq \oper{M}^\dagger \VectOp{p} \oper{M} 
        = \Mat{C} \VectOp{q} + \Mat{D} \VectOp{p} \, .
    \end{align}
\end{subequations}

\noindent Configuration-space basis vectors are transformed as
\begin{equation}
    \ket{\Vect{Q}} = \oper{M}^\dagger \ket{\Vect{q}} \, .
\end{equation}

\noindent Correspondingly, wavefunctions of $\Vect{q}$-space are projected onto $\Vect{Q}$-space as~\cite{Lopez19a}
\begin{align}
    \Psi(\Vect{Q}) &= \int \dd \Vect{q} \, \braket{\Vect{Q}}{\Vect{q}} \braket{\Vect{q}}{\psi} \nonumber\\
    &= \int \frac{\sigma \, \dd \Vect{q}}{(2\pi i)^{N/2} \sqrt{\det{\Mat{B}}} } \, \psi(\Vect{q}) \, 
    e^{ i G_1(\Vect{q}, \Vect{Q}) } \, .
    \label{eq:metTRANS}
\end{align}

\noindent Here, $\sigma \doteq \pm 1$ is the overall sign of the MT, $G_1(\Vect{q}, \Vect{Q})$ is the generating function~\cite{Goldstein02}
\begin{equation}
    G_1(\Vect{q}, \Vect{Q}) \doteq 
    \frac{1}{2} \Vect{q}^\intercal \Mat{B}^{-1}\Mat{A} \Vect{q} 
    - \Vect{q}^\intercal \Mat{B}^{-1} \Vect{Q} + \frac{1}{2} \Vect{Q}^\intercal \Mat{D}\Mat{B}^{-1} \Vect{Q} \, ,
\end{equation}

\noindent and our branch-cut convention restricts all complex phases onto the interval $(-\pi, \pi]$. Thus, $\sigma$ determines the sign of $\sqrt{\det{\Mat{B}}}$. In writing \Eq{eq:metTRANS}, we have also assumed that $\Mat{B}$ is invertible for simplicity. The generalization for non-invertible $\Mat{B}$ is straightforward, and involves a $\delta$-function kernel within the nullspace of $\Mat{B}$~\cite{Littlejohn86a}.

Equation \eq{eq:metTRANS} defines $\Psi(\Vect{Q})$ as the metaplectic transform (MT) of $\psi(\Vect{q})$. The inverse MT is given as
\begin{equation}
    \psi(\Vect{q}) = \int \frac{ \sigma \, \dd \Vect{Q}}{(-2\pi i)^{N/2} \sqrt{\det{\Mat{B}}} } \, \Psi(\Vect{Q}) \,
    e^{ - i G_1(\Vect{q}, \Vect{Q}) } \, ,
    \label{eq:invMET}
\end{equation}

\noindent with complex phases restricted to the interval $[-\pi, \pi)$. Special cases of the MT include the FT and the fractional FT. When $\Mat{S}$ is near-identity, the integral transform of \Eq{eq:metTRANS} is well-approximated by a differential transform. For eikonal $\psi(\Vect{q})$, the near-identity MT (NIMT) which transforms $\psi(\Vect{q})$ to $\Psi(\Vect{Q})$ is given as~\cite{Lopez19a}
\begin{subequations}
    \label{eq:NIMT}
    \begin{align}
        \Psi(\Vect{Q}) &\approx 
            \frac{ e^{ i\Theta\left( \Mat{A}^{-1} \Vect{Q} \right) } }
            {\sqrt{\det\Mat{A}}} 
            \left\{ \phi\left( \Mat{A}^{-1} \Vect{Q} \right) \nullFrac \right. \nonumber\\
            &\left. \hspace{22mm} - \frac{1}{2} \Tr \left[ 
                \Mat{A}^{-1} \Mat{B} \, \Mat{F} \left( \Mat{A}^{-1} \Vect{Q} \right)
            \right] \right\} \, , \\
        \Mat{F}(\Vect{q}) &\doteq \pd_\Vect{q} \theta(\Vect{q}) \left[ \pd_\Vect{q} \phi(\Vect{q}) \right]^\intercal 
            + \pd_\Vect{q} \phi(\Vect{q}) \left[ \pd_\Vect{q} \theta(\Vect{q}) \right]^\intercal \nonumber\\
            &\hspace{30mm} + \phi(\Vect{q}) \pd^2_{\Vect{q} \Vect{q}} \theta(\Vect{q}) \, , \\
        \Theta(\Vect{q}) &\doteq \theta(\Vect{q}) 
            + \frac{1}{2} \Vect{q}^\intercal \Mat{A}^\intercal \Mat{C} \Vect{q} \nonumber\\
            &\hspace{11.5mm} - \frac{1}{2} \left[ \pd_\Vect{q} \theta(\Vect{q}) \right]^\intercal \Mat{A}^{-1} \Mat{B} \, \pd_\Vect{q} \theta(\Vect{q}) \, ,
    \end{align}
\end{subequations}

\noindent where we have dropped the term $\pd^2_{\Vect{q} \Vect{q}} \phi(\Vect{q})$ from $\Mat{F}(\Vect{q})$ because it is higher order in the GO parameter. 

Now, let us explicitly construct the symplectic matrix that maps $\Vect{q}$-space to the tangent plane of the dispersion manifold at some $\Vect{z}(\Vect{\tau})$. Recall that coordinates and coordinate axes transform oppositely (contravariantly versus covariantly). In other words, if the coordinates are transformed by $\Mat{S}$, then the coordinate axes are transformed by $\Mat{S}^{-1}$; hence, we desire $\Mat{S}^{-1}$ to map $\Vect{q}$-space to the local tangent space, rather than $\Mat{S}$. Let $\{ \unit{\Vect{T}}_j ( \Vect{t} ) \}$ and $\{ \unit{\Vect{N}}_j ( \Vect{t} ) \}$ be a symplectically dual set of $N$ orthonormal tangent vectors and normal vectors to the dispersion manifold at $\Vect{\tau} = \Vect{t}$, respectively. As can be readily verified, the matrix
\begin{equation}
    \Mat{R}_\Vect{t} = 
    \begin{pmatrix}
        \uparrow & & \uparrow 
	& \uparrow & & \uparrow\\
	\unit{\Vect{T}}_1(\Vect{t}) & \ldots & \unit{\Vect{T}}_N(\Vect{t}) 
	& \unit{\Vect{N}}_1(\Vect{t}) & \ldots & \unit{\Vect{N}}_N(\Vect{t})\\
	\downarrow & & \downarrow 
	& \downarrow & & \downarrow
    \end{pmatrix}
    \label{eq:uMAT}
\end{equation}

\noindent maps $\Vect{q}$-space to the local tangent space at $\Vect{\tau} = \Vect{t}$. (The arrows emphasize that $\{ \unit{\Vect{T}}_j ( \Vect{t} ) \}$ and $\{ \unit{\Vect{N}}_j ( \Vect{t} ) \}$ form the columns of $\Mat{R}_\Vect{t}$.) Since $\Mat{R}_\Vect{t}$ is unitary, we obtain
\begin{equation}
    \Mat{S}_\Vect{t} = \Mat{R}_\Vect{t}^\intercal \, .
    \label{eq:tangMAT}
\end{equation}

The vectors $\{ \unit{\Vect{T}}_j ( \Vect{t} ) \}$ and $\{ \unit{\Vect{N}}_j ( \Vect{t} ) \}$ can be constructed using a `symplectic Gram--Schmidt' method as follows. First, designate
\begin{equation}
    \unit{\Vect{T}}_1 ( \Vect{t} ) = \left.\left. \pd_{\tau_1} \Vect{z}(\Vect{\tau}) \right/
    \| \pd_{\tau_1} \Vect{z}(\Vect{\tau}) \| \right|_{\Vect{\tau} = \Vect{t}} \, .
    \label{eq:tangVEC1}
\end{equation}

\noindent Next, the complete set $\{ \unit{\Vect{T}}_j ( \Vect{t} ) \}$ is obtained as
\begin{equation}
    \unit{\Vect{T}}_j ( \Vect{t} ) = \mc{G}_s \left[
        \unit{\Vect{T}}_1(\Vect{t}), 
        \ldots, 
        \unit{\Vect{T}}_{j-1}(\Vect{t}), 
        \left. \pd_{\tau_j} \Vect{z}(\Vect{\tau}) \right|_{\Vect{\tau} = \Vect{t}}  
    \right] \, ,
    \label{eq:sympTANG}
\end{equation}

\noindent where $\mc{G}_s$ represents the `modified Gram--Schmidt operator', which returns the orthogonalized version of its final argument with respect to the previous $j-1$ arguments~\cite{Trefethen97}. Finally, the complete set $\{ \unit{\Vect{N}}_j ( \Vect{t} ) \}$ are obtained from $\{ \unit{\Vect{T}}_j ( \Vect{t} ) \}$ as
\begin{equation}
    \unit{\Vect{N}}_j(\Vect{t}) = - \Mat{J} \, \unit{\Vect{T}}_j(\Vect{t}) \, .
    \label{eq:sympNORM}
\end{equation}

Let us confirm that as constructed, $\Mat{S}_\Vect{t}$ is indeed both orthogonal and symplectic. By \Eq{eq:sympTANG},
\begin{subequations}
    \begin{equation}
        \left[ \unit{\Vect{T}}_j (\Vect{t}) \right]^\intercal \unit{\Vect{T}}_{j'} (\Vect{t}) 
        = \delta_{j j'} \, ,
        \label{eq:ttORTH}
    \end{equation}
    
    \noindent which immediately implies that
    \begin{align}
        \label{eq:nnORTH}
        \left[ \unit{\Vect{N}}_j (\Vect{t}) \right]^\intercal \unit{\Vect{N}}_{j'} (\Vect{t}) 
        &= \delta_{j j'} \, , \\
        \label{eq:ntSYMP}
        \left[ \unit{\Vect{N}}_j (\Vect{t}) \right]^\intercal \Mat{J} \, \unit{\Vect{T}}_{j'} (\Vect{t})
        &= - \delta_{j j'} \, .
    \end{align}
    
    \noindent Also, since the dispersion manifold is obtained by the gradient lift $\Vect{p} = \pd_\Vect{q} \theta(\Vect{q})$, the dispersion manifold is a Lagrangian manifold~\cite{Arnold89}. A Lagrangian manifold has the property that any set of tangent vectors satisfy 
    \begin{equation}
        \left[ \unit{\Vect{T}}_j (\Vect{t}) \right]^\intercal \Mat{J} \, \unit{\Vect{T}}_{j'} (\Vect{t}) = 0 \, .
        \label{eq:ttSYMP}
    \end{equation}
    
    \noindent Consequently, 
    \begin{align}
        \label{eq:nnSYMP}
        \left[ \unit{\Vect{N}}_j (\Vect{t}) \right]^\intercal \Mat{J} \, \unit{\Vect{N}}_{j'} (\Vect{t}) 
        &= 0 \, , \\
        \label{eq:ntORTH}
        \left[ \unit{\Vect{N}}_j (\Vect{t}) \right]^\intercal \unit{\Vect{T}}_{j'} (\Vect{t})
        &= 0 \, .
    \end{align}
\end{subequations}

\noindent Hence, $\Mat{S}_\Vect{t}$ is orthogonal per \Eqs{eq:ttORTH}, \eq{eq:nnORTH}, and \eq{eq:ntORTH}, and is symplectic per \Eqs{eq:ntSYMP}, \eq{eq:ttSYMP}, and \eq{eq:nnSYMP}.

\subsection{Geometrical optics in arbitrary projective plane}

Let us now develop the GO equations on a projective plane that is obtained from $\Vect{q}$-space via some symplectic matrix $\Mat{S}$. We again consider the general wave equation given in \Eq{eq:hilbertWAVE}, but, rather than introducing an eikonal ansatz on $\Vect{q}$-space as done in \Eq{eq:phiENV}, we now assume the wavefield is eikonal on the desired projective plane. Hence, we perform the unitary transformation
\begin{equation}
    \ket{\Phi} = e^{-i \Theta(\VectOp{Q}) } \ket{\psi}
    \label{eq:PhiENV}
\end{equation}

\noindent such that \Eq{eq:hilbertWAVE} becomes
\begin{equation}
    e^{-i \Theta(\VectOp{Q}) } 
    \oper{D}(\VectOp{z}) \, 
    e^{i \Theta(\VectOp{Q}) } 
    \ket{\Phi} = \ket{0} \, .
    \label{eq:wavePHI}
\end{equation}

In principle, \Eq{eq:wavePHI} is sufficient to develop GO; however, the simultaneous presence of $\VectOp{Q}$ and $\VectOp{z}$ is inconvenient. To rectify this, let us introduce into \Eq{eq:wavePHI} the metaplectic operator corresponding to $\Mat{S}$ as
\begin{equation}
    e^{-i \Theta(\VectOp{Q}) } 
    \oper{M}
    \oper{M}^\dagger
    \oper{D}(\VectOp{z}) \, 
    \oper{M}
    \oper{M}^\dagger
    e^{i \Theta(\VectOp{Q}) } 
    \ket{\Phi} = \ket{0} \, ,
    \label{eq:waveMET}
\end{equation}

\noindent where we have used the unitarity of $\oper{M}$. This allows us to rigorously transform $\VectOp{z}$ to $\Stroke{\VectOp{Z}}$. As shown in \App{app:metWEYL},
\begin{equation}
    \WeylInv\left[ 
        \oper{M}
        \oper{M}^\dagger
        \oper{D}(\VectOp{z}) \, 
        \oper{M}
        \oper{M}^\dagger 
    \right] 
    = \Symb{D}\left(
        \Mat{S}^{-1} \Stroke{\Vect{Z}} 
    \right) 
    = \Symb{D}(\Vect{z})
    \, ,
\end{equation}

\noindent which also demonstrates the well-known `symplectic covariance' property of the Weyl symbol~\cite{Littlejohn86a,deGosson06}. In other words, the Weyl symbol of an operator at a given phase-space location does not depend on how this location is parameterized, as long as different parameterizations (here, $\Vect{z}$ and $\Vect{Z}$) are connected via symplectic transformations.

Since symplectic transformations preserve the Poisson bracket, they also preserve the Moyal star product (\App{app:Weyl}). Hence, the GO limit of \Eq{eq:waveMET} can be obtained using the procedure outlined in \App{app:GO}, but replacing $\Symb{D}(\Vect{z})$ with $\Symb{D}(\Mat{S}^{-1} \Vect{Z})$ and $\Vect{z}$ by $\Vect{Z}$. This yields
\begin{align}
    &\hspace{-3mm}\left\{
        \Symb{D}\left[ \Mat{S}^{-1} \Stroke{\Vect{Z}}(\VectOp{Q}) \right]
        + \Vect{V}(\VectOp{Q})^\intercal \VectOp{P}
        - \frac{i}{2} \pd_\Vect{Q} \cdot \Vect{V}(\VectOp{Q})
    \right. \nonumber\\
    &\left.\hspace{38mm}
        - \frac{1}{2} \deltaQO(\Stroke{\VectOp{Z}})
    \right\} \ket{\Phi} 
    = \ket{0} \, , 
    \label{eq:approxENVmet}
\end{align}

\noindent where, given recent interest~\cite{Dodin19,Yanagihara19a,Yanagihara19b}, we have included the quasioptical (QO) term which governs diffraction as
\begin{align}
    \deltaQO(\Stroke{\VectOp{Z}}) &\doteq 
    - \Mat{M}(\VectOp{Q}) \dubdot \VectOp{P}\VectOp{P} 
    + i \left[ \pd_\Vect{Q} \cdot \Mat{M}(\VectOp{Q}) \right] \VectOp{P} 
    \, .
    \label{eq:deltaQO}
\end{align}

\noindent We have also defined the following quantities:
\begin{subequations}
    \begin{align}
        \Vect{P}(\Vect{Q}) 
        &\doteq \pd_{\Vect{Q}} \Theta(\Vect{Q}) \, , \\
        \Vect{V}(\Vect{Q}) 
        &\doteq \left.\pd_\Vect{P} \Symb{D}\left(\Mat{S}^{-1} \Stroke{\Vect{Z}} \right) \right|_{ \Vect{P} = \Vect{P}(\Vect{Q}) } \, , \\
        \Mat{M}(\Vect{Q}) &\doteq \left.
        \pd^2_{\Vect{P}\Vect{P}} \Symb{D}\left( \Mat{S}^{-1} \Stroke{\Vect{Z}} \right)
        \right|_{\Vect{P} = \Vect{P}(\Vect{Q})} \, .
    \end{align}
\end{subequations}

Dynamical equations that govern $\Phi(\Vect{Q}) \doteq \braket{\Vect{Q}}{\Phi}$ are obtained by projecting \Eq{eq:approxENVmet} onto $\{ \ket{\Vect{Q}} \}$. Neglecting diffraction, the GO equations on this projective plane are
\begin{subequations}
    \begin{gather}
        \label{eq:GOrayMET}
        \Symb{D}\left[ \Mat{S}^{-1} \Stroke{\Vect{Z}}(\Vect{Q}) \right] = 0 \, , \\
        \label{eq:GOenvMET}
        \Vect{V}(\Vect{Q})^\intercal \pd_\Vect{Q} \Phi(\Vect{Q}) 
        + \frac{1}{2} \left[ \pd_\Vect{Q} \cdot \Vect{V}(\Vect{Q}) \right] \Phi(\Vect{Q}) = 0 \, .
    \end{gather}
\end{subequations}

\noindent As before, \Eq{eq:GOrayMET} is solved via ray tracing, while \Eq{eq:GOenvMET} can be formally solved as
\begin{equation}
    \Phi(\Vect{\tau}) = \phi_0(\Vect{\tau}_\perp) 
    \sqrt{\frac{ \det \pd_\Vect{\tau } \Vect{Q}(0, \Vect{\tau}_\perp) }
    { \det \pd_\Vect{\tau} \Vect{Q}(\Vect{\tau}) }} \, .
    \label{eq:ENVjacobMET}
\end{equation}

However, let us make an observation regarding the rays generated by \Eq{eq:GOrayMET}. These rays satisfy
\begin{equation}
    \pd_{\tau_1} \Stroke{\Vect{Z}} 
    = \Mat{J} \, \pd_{\Stroke{\Vect{Z}}} \Symb{D}\left( \Mat{S}^{-1} \Stroke{\Vect{Z}} \right) \, .
\end{equation}

\noindent Since $\Mat{S}$ is constant, the chain rule yields
\begin{equation}
    \pd_{\tau_1} \Stroke{\Vect{Z}} 
    = \Mat{J} \, 
    \left( \Mat{S}^{-1} \right)^\intercal 
    \left. 
        \pd_{\Vect{z}} \Symb{D}\left( \Vect{z} \right) 
    \right|_{\Vect{z} = \Mat{S}^{-1}\Stroke{\Vect{Z}}} \, .
\end{equation}

\noindent Moreover, since $\Mat{S}$ is symplectic, multiplication by $\Mat{S}^{-1}$ from the left yields
\begin{equation}
    \pd_{\tau_1} \left( \Mat{S}^{-1} \Stroke{\Vect{Z}} \right)
    = \Mat{J} \, 
    \left. 
        \pd_{\Vect{z}} \Symb{D}\left( \Vect{z} \right) 
    \right|_{\Vect{z} = \Mat{S}^{-1}\Stroke{\Vect{Z}}} \, .
\end{equation}

\noindent Finally, a comparison with \Eq{eq:rayHAM} yields the relationship between the original and the transformed rays:
\begin{equation}
    \Stroke{\Vect{Z}}(\Vect{\tau}) 
    = \Mat{S} \, \Vect{z}(\Vect{\tau}) \, .
    \label{eq:sRAYS}
\end{equation}

\noindent Thus, one does not need to re-trace rays every time the projection plane is changed, but rather, one can simply perform the same symplectic transformation to the rays that one applies to the ambient phase space.

The wavefield on the projective plane is constructed as
\begin{equation}
    \Psi(\Vect{Q}) \doteq \Phi(\Vect{Q}) 
    e^{ i \Theta(\Vect{Q}) } \, ,
\end{equation}

\noindent summed over all branches of $\Theta(\Vect{Q})$ if $\Theta(\Vect{Q})$ is multivalued. The wavefield on $\Vect{q}$-space can then be obtained by taking the inverse MT of $\Psi(\Vect{Q})$ using \Eq{eq:invMET}.

\subsection{Sequenced geometrical optics in a piecewise-linear tangent space}

We are now equipped to discuss the new method of caustic removal outlined in \Fig{fig:metMASLOV}. In this subsection we shall discuss steps (i) and (ii) of \Fig{fig:metMASLOV}, that is, performing GO in the various tangent planes of the dispersion manifold and linking the obtained GO solutions using the NIMT, while in the following subsection we shall discuss step (iii). We shall assume that the dispersion manifold $\Vect{z}(\Vect{\tau})$ has already been obtained via \Eq{eq:rayHAM}. This is not a restrictive assumption, though, because the ray trajectories themselves are unaffected by caustics. 

Let us consider some point $\Vect{q}$ in configuration space and attempt to construct $\psi(\Vect{q})$. To do so, we shall map $\Vect{q}$ to the dispersion manifold using the ray map $\Vect{\tau}(\Vect{q})$, solve for $\Psi(\Vect{Q})$ in the optimal projection plane, that is, the tangent plane of the dispersion manifold at $\Vect{\tau}$, and then map $\Psi(\Vect{Q})$ to $\psi(\Vect{q})$ using an inverse MT. In general, however, $\Vect{\tau}(\Vect{q})$ will be multi-valued, and for the aforementioned scheme to work, the contributions to $\psi(\Vect{q})$ from each branch must be considered separately. 

Let $\Vect{t} \in \Vect{\tau}(\Vect{q})$ be a branch of $\Vect{\tau}(\Vect{q})$, and let us construct the GO wavefield in the tangent plane at $\Vect{t}$. Equations \eq{eq:uMAT} and \eq{eq:tangMAT} yield the $\Mat{S}_\Vect{t}$ that transforms $\Vect{q}$-space to the tangent plane. The rays are transformed into the new coordinates using \Eq{eq:sRAYS} as
\begin{equation}
    \Stroke{\Vect{Z}}_\Vect{t}(\Vect{\tau}) = \Mat{S}_\Vect{t} \, \Vect{z}(\Vect{\tau})  \, .
\end{equation}

\noindent Using the rays, $\Vect{P}_\Vect{t}(\Vect{Q})$ can be constructed as 
\begin{equation}
    \Vect{P}_\Vect{t}(\Vect{Q}) \doteq \Vect{P}_\Vect{t}\left[ \Vect{\tau}_\Vect{t}(\Vect{Q}) \right] \, ,
    \label{eq:PFIELD}
\end{equation}

\noindent where $\Vect{\tau}_\Vect{t}(\Vect{Q})$ is the function inverse of $\Vect{Q}_\Vect{t}(\Vect{\tau})$. Then, the phase on the tangent plane is computed as
\begin{equation}
    \Theta_\Vect{t}(\Vect{Q}) = 
    \int_{ \Vect{Q}_\Vect{t}(\Vect{t}) }^\Vect{Q} 
    \left( \dd \Vect{Q}' \right)^\intercal
    \Vect{P}_\Vect{t}\left( \Vect{Q}' \right) \, ,
    \label{eq:phase}
\end{equation}

\noindent while the envelope is computed using \Eq{eq:GOenvMET} or its QO analogue. Importantly, when $\Vect{\tau}(\Vect{q})$ is multi-valued, then $\Vect{P}_\Vect{t}(\Vect{Q})$ should be restricted to the branch containing $\Vect{P}_\Vect{t}(\Vect{t})$. This is because we are only interested in the contribution near $\Vect{t}$ on the dispersion manifold.

Let $\Phi_\Vect{t}(\Vect{Q})$ be the solution to \Eq{eq:GOenvMET} (or its QO analogue) subject to the initial condition $\Phi_\Vect{t}\left[ \Vect{Q}_\Vect{t}(\Vect{t}) \right] = 1$. The wavefield on the tangent plane is
\begin{equation}
    \Psi_\Vect{t}(\Vect{Q}) = \alpha_\Vect{t} \Phi_\Vect{t}(\Vect{Q}) e^{i \Theta_\Vect{t}(\Vect{Q})} \, ,
    \label{eq:wavefieldMET}
\end{equation}

\noindent where 
\begin{equation}
    \alpha_\Vect{t} \doteq \Psi_\Vect{t}\left[ \Vect{Q}_\Vect{t}(\Vect{t}) \right]
\end{equation}

\noindent is a function of $\Vect{t}$ which is required to make $\psi(\Vect{q})$ continuous. It can be found from
\begin{equation}
    \alpha_{\Vect{t}+\Vect{h}} = 
    \alpha_\Vect{t}
    \left.\NIMT
    {
        \Mat{S}_{\Vect{t}+\Vect{h}} \Mat{S}^{-1}_\Vect{t}
    }\left[ 
        \Phi_\Vect{t}(\Vect{Q}) \,
        e^{i \Theta_\Vect{t}(\Vect{Q}) }
    \right]\right|_{\Vect{Q}_{\Vect{t}+\Vect{h}}( \Vect{t}+\Vect{h} )}
    \, ,
    \label{eq:nimtCONT}
\end{equation}

\noindent where $\NIMT{\Mat{S}}\left[ \psi(\Vect{q}) \right]$ denotes the NIMT of $\psi(\Vect{q})$ with respect to $\Mat{S}$ via \Eq{eq:NIMT}. In other words, the arbitrary constant in $\Psi_{\Vect{t}+\Vect{h}}(\Vect{Q})$ is obtained by projecting the neighboring $\Psi_\Vect{t}(\Vect{Q})$ onto the tangent plane at $\Vect{\tau} = \Vect{t}+\Vect{h}$.

In the continuous limit, \Eq{eq:nimtCONT} can also be written as a differential equation. Using \Eqs{eq:NIMT} and results presented in Appendix C of \Ref{Lopez19a}, one can show that
\begin{align}
    &\left.\NIMT
    {
        \Mat{S}_{\Vect{t}+\Vect{h}} \Mat{S}^{-1}_\Vect{t}
    }\left[ 
        \Phi_\Vect{t}(\Vect{Q}) \,
        e^{i \Theta_\Vect{t}(\Vect{Q}) }
    \right]\right|_{\Vect{Q}_{\Vect{t}+\Vect{h}}( \Vect{t}+\Vect{h} )}
    \approx
    1 
    + h \, \eta_\Vect{t}
    \, ,
\end{align}

\noindent where we have defined $\eta_\Vect{t}$ as
\begin{align}
    \eta_\Vect{t}
    &\doteq
    \left[\pd_h \Vect{Q}_\Vect{t}(\Vect{t}) 
        - \Mat{V}_\Vect{t}^\intercal \Vect{Q}_\Vect{t}(\Vect{t}) 
        \nullFrac
    \right]^\intercal \left\{
        \pd_\Vect{Q} \Phi_\Vect{t}\left[
            \Vect{Q}_\Vect{t}(\Vect{t})
        \right]
        + i \Vect{P}_\Vect{t}(\Vect{t}) 
        \nullFrac
    \right\}
    \nonumber\\
    &\hspace{3mm} - \left[
        \Vect{P}_\Vect{t}(\Vect{t}) 
    \right]^\intercal 
    \Mat{W}_\Vect{t} \left\{ 
        \pd_\Vect{Q} \Phi_\Vect{t}\left[
            \Vect{Q}_\Vect{t}(\Vect{t})
        \right]
        + \frac{i}{2} \Vect{P}_\Vect{t}(\Vect{t})
    \right\}
    \nonumber\\
    &\hspace{3mm} - \frac{1}{2}
    \Tr\left(\Mat{V}_\Vect{t} \right) 
    - \frac{i}{2} \left[
        \Vect{Q}_\Vect{t}(\Vect{t}) 
    \right]^\intercal 
    \Mat{U}_\Vect{t} \Vect{Q}_\Vect{t}(\Vect{t})
    \, .
    \label{eq:etaDEF}
\end{align}

\noindent The $N\times N$ matrices $\Mat{U}$, $\Mat{V}$, and $\Mat{W}$ are obtained through the matrix decomposition
\begin{equation}
    \left(\pd_h \Mat{S}_\Vect{t}\right)\Mat{S}_\Vect{t}^{-1}
    \doteq
    \begin{pmatrix}
        \Mat{V}_\Vect{t}^\intercal & \Mat{W}_\Vect{t} \\
        - \Mat{U}_\Vect{t} & - \Mat{V}_\Vect{t}
    \end{pmatrix}
    \, ,
    \label{eq:UVWdef}
\end{equation}

\noindent which is possible because $\left(\pd_h \Mat{S}_\Vect{t}\right)\Mat{S}_\Vect{t}^{-1}$ is a Hamiltonian matrix~\cite{Lopez19a}. Consequently, $\Mat{U}_\Vect{t}$ and $\Mat{W}_\Vect{t}$ are both symmetric. We have also defined the directional derivative as
\begin{equation}
    h \, \pd_h 
    \doteq 
    \Vect{h}^\intercal \pd_\Vect{t} \, .
\end{equation}

\noindent Note that $\pd_h$ is a total (directional) derivative in $\Vect{t}$, so it acts on both arguments of $\Vect{Q}_\Vect{t}(\Vect{t})$, including the subscript. Then, \Eq{eq:nimtCONT} yields
\begin{equation}
    \pd_h \log \alpha_\Vect{t} 
    = 
    \eta_\Vect{t} \, .
    \label{eq:constDE}
\end{equation}

\noindent When $\alpha_\Vect{t}$ is evolved along a ray, then $\partial_h = \partial_{t_1}$, and \Eq{eq:constDE} is trivially integrated as
\begin{equation}
    \alpha_\Vect{t} = \alpha_{\left(0,\Vect{t}_\perp \right)} \exp\left[
        \int_0^{t_1} \dd h \, \eta_{\left(h, \Vect{t}_\perp \right)}
    \right] \, .
    \label{eq:alphaINT}
\end{equation}

\noindent The remaining constant function $\alpha_{\left(0,\Vect{t}_\perp \right)}$ is determined by initial conditions.

\subsection{Projecting tangent-space GO solution to configuration space}

Equations \eq{eq:wavefieldMET} and \eq{eq:alphaINT}, along with initial conditions, completely specify the wavefield in the tangent space of the dispersion manifold as $\Psi(\Vect{\tau}) \doteq \Psi_\Vect{\tau}\left[ Q_\Vect{\tau}(\Vect{\tau}) \right]$. This wavefield is free from caustics, and may be sufficient for certain applications. When $\psi(\Vect{q})$ is required, then $\Psi_\Vect{t}(\Vect{Q})$ must be projected back onto $\Vect{q}$-space using \Eq{eq:invMET} as%
\footnote{In using \Eq{eq:invMET}, we assume that $\det{\Mat{B}_t} \neq 0$. In particular, the following analysis is unsuitable for wave propagation in homogeneous media, because $\pd_\Vect{q} \Symb{D} = \Vect{0}$ implies that $\det{\Mat{B}_t} = 0$. }
\begin{equation}
    \psi_\Vect{t}(\Vect{q}) = 
    \int \frac{
        \dd \Vect{Q} \, \sigma_\Vect{t} \, \alpha_\Vect{t} \, \Phi_\Vect{t}(\Vect{Q})
    }
    {
        (-2\pi i)^{N/2} \sqrt{ \det{\Mat{B}_\Vect{t}} }
    } \,  
    e^{i \Theta_\Vect{t}(\Vect{Q}) - i G_1(\Vect{q},\Vect{Q} ) } \, ,
    \label{eq:tINVmet}
\end{equation}

\noindent and the contribution to $\psi_\Vect{t}(\Vect{q})$ from the field near $\Vect{t}$ must be isolated by using the SPA. Indeed, we calculate
\begin{align}
    &\pd_\Vect{Q} \Theta_\Vect{t}(\Vect{Q}) 
    - \pd_\Vect{Q} G_1(\Vect{q}, \Vect{Q}) \nonumber\\
    &\hspace{2cm}=
    \Vect{P}_\Vect{t}(\Vect{Q}) 
    + \left(\Mat{B}_\Vect{t}^{-1} \right)^\intercal \Vect{q} - \Mat{D}_\Vect{t} \Mat{B}_\Vect{t}^{-1}\Vect{Q} \, .
    \label{eq:phaseDERIV}
\end{align}

\noindent The roots to \Eq{eq:phaseDERIV} are $\Vect{Q} = \Vect{Q}_\Vect{t}\left[\Vect{\tau}(\Vect{q}) \right]$. Hence, restricting the integration domain near $\Vect{Q}_\Vect{t}( \Vect{t} )$ will isolate the contribution from $\Vect{t}$. To this end, let us define
\begin{equation}
    \Vect{\epsilon} \doteq \Vect{Q} - \Vect{Q}_\Vect{t}(\Vect{t}) \, .
\end{equation}

\noindent After performing a change in variables, \Eq{eq:tINVmet} becomes
\begin{equation}
    \psi_\Vect{t}(\Vect{q}) = 
    \frac{
        \sigma_\Vect{t} \, \alpha_\Vect{t} \exp\left( -i \beta_\Vect{t} \right)
    }
    {
        (-2\pi i)^{N/2} \sqrt{ \det{\Mat{B}_\Vect{t}}}
    } \,
    \Upsilon_\Vect{t}
    \, ,
    \label{eq:invMTmet}
\end{equation}

\noindent where we have defined
\begin{subequations}
    \begin{align}
    \label{eq:betaMET}
        \beta_\Vect{t} 
        &\doteq 
            \frac{1}{2} \Vect{q}^\intercal \Mat{B}_\Vect{t}^{-1} \Mat{A}_\Vect{t} \Vect{q} 
            - 
            \Vect{q}^\intercal \Mat{B}_\Vect{t}^{-1} \Vect{Q}_\Vect{t}(\Vect{t}) \nonumber \\
            &\hspace{23mm} +
            \frac{1}{2}
            \left[\Vect{Q}_\Vect{t}(\Vect{t})\right]^\intercal \Mat{D}_\Vect{t} \Mat{B}_\Vect{t}^{-1} \Vect{Q}_\Vect{t}(\Vect{t})
        \, , \\
        \Upsilon_\Vect{t} 
        &\doteq 
            \int \dd \Vect{\epsilon} \,
            \Phi_\Vect{t}\left[ 
                \Vect{\epsilon} + \Vect{Q}_\Vect{t}(\Vect{t})
            \right] \nonumber \\
            &\hspace{5mm}\times \exp\left\{ 
                i \Theta_\Vect{t}\left[    
                    \Vect{\epsilon} + \Vect{Q}_\Vect{t}(\Vect{t}) 
                \right] 
                - i \gamma_\Vect{t}(\Vect{\epsilon})
                \nullFrac
            \right\} 
        \, , \\
        \label{eq:gammaMET}
        \gamma_\Vect{t}(\Vect{\epsilon}) 
        &\doteq 
            \frac{1}{2} \Vect{\epsilon}^\intercal \Mat{D}_\Vect{t} \Mat{B}_\Vect{t}^{-1} \Vect{\epsilon}
            +
            \left[   
                \Mat{D}_\Vect{t}^\intercal \Vect{Q}_\Vect{t}(\Vect{t})
                -
                \Vect{q}
            \right]^\intercal \Mat{B}_\Vect{t}^{-1}\Vect{\epsilon}
        \, .
    \end{align}
\end{subequations}

\noindent The overall sign $\sigma_\Vect{t}$ is constant unless $\det{\Mat{B}_\Vect{t}}$ crosses the branch cut for the MT. Then, $\sigma_\Vect{t}$ changes sign to ensure that $\psi_\Vect{t}$ evolves smoothly in $\Vect{t}$. (See Sec.~2 of \Ref{Lopez19a} for details or \Sec{sec:examplesQHO} for an example.)

Invoking the SPA, the integration domain is restricted to the neighborhood of $\Vect{\epsilon} = \Vect{0}$, denoted $\delta \Vect{\epsilon}$. This yields
\begin{align}
    \Upsilon_\Vect{t} &\approx 
    \int_{\delta \Vect{\epsilon}} \dd \Vect{\epsilon} \,
        \Phi_\Vect{t}\left[ 
            \Vect{\epsilon} + \Vect{Q}_\Vect{t}(\Vect{t})
        \right] \nonumber \\
        &\hspace{5mm}\times \exp\left\{ 
            i \Theta_\Vect{t}\left[    
                \Vect{\epsilon} + \Vect{Q}_\Vect{t}(\Vect{t}) 
            \right] 
            - i \gamma_\Vect{t}(\Vect{\epsilon})
            \nullFrac
        \right\} 
    \, .
    \label{eq:intRESTRICT}
\end{align}

\noindent No further approximations to $\Upsilon_\Vect{t}$ can be made without additional knowledge of the caustic structure. We shall discuss some of these additional approximations in \Sec{sec:examples}. Nevertheless, by properly choosing $\delta \Vect{\epsilon}$, it should be possible to numerically evaluate \Eq{eq:intRESTRICT} to sufficiently high accuracy in the general case. Having isolated the field contribution from a single branch of the dispersion manifold, we sum over all such contributions to obtain
\begin{equation}
    \psi(\Vect{q}) = \sum_{ \Vect{t} \in \Vect{\tau}(\Vect{q}) } \psi_\Vect{t}\left[ \Vect{q}(\Vect{t}) \right] \, .
    \label{eq:psiSUM}
\end{equation}

\subsection{Curvature-dependent adaptive discretization}

Since the accuracy of the NIMT depends on the rotation angle between neighboring tangent planes, the discretization of the dispersion manifold would ideally have smaller step sizes in regions with higher curvature. This can be achieved by replacing the ray Hamiltonian $\Symb{D}(\Vect{z})$ in \Eq{eq:rayHAM} with a new function $\widetilde{\Symb{D}}(\Vect{z})$ that generates the same ray trajectories in phase space but with a different `time' parameterization. In this manner, adaptive integration schemes can be developed which are self-supervising (do not need to be error-controlled in the traditional sense~\cite{Press07}), and amenable to symplectic methods of numerical integration~\cite{Richardson12,Hairer97}.

Let us consider a modified ray Hamiltonian
\begin{equation}
    \widetilde{\Symb{D}}(\Vect{z}) \doteq f(\Vect{z}) \Symb{D}(\Vect{z}) \, ,
\end{equation}

\noindent where $f(\Vect{z})$ is some smooth function. Let $\Vect{z}_{\Symb{D}}$ and $\Vect{z}_f$ denote the zero sets of $\Symb{D}(\Vect{z})$ and $f(\Vect{z})$ respectively, \ie
\begin{equation}
    \Vect{z}_{\Symb{D}} \doteq \left\{ \Vect{z} \, | \, \Symb{D}(\Vect{z}) = 0 \right\} \, ,
    \quad
    \Vect{z}_{f} \doteq \left\{ \Vect{z} \, | \, f(\Vect{z}) = 0 \right\} \, .
\end{equation}

\noindent Clearly, the zero set of $\widetilde{\Symb{D}}(\Vect{z})$ is 
\begin{equation}
    \Vect{z}_{\widetilde{\Symb{D}}} \doteq \left\{ \Vect{z} \, | \, f(\Vect{z})\Symb{D}(\Vect{z}) = 0 \right\}
    = \Vect{z}_{\Symb{D}} \bigcup \Vect{z}_f \, ,
\end{equation}

\noindent where $\bigcup$ denotes the set union. For $\widetilde{\Symb{D}}(\Vect{z})$ and $\Symb{D}(\Vect{z})$ to generate the same set of rays, $f(\Vect{z})$ and $\Symb{D}(\Vect{z})$ cannot be zero simultaneously. Hence, we require
\begin{equation}
    \Vect{z}_{\Symb{D}} \bigcap \Vect{z}_f = \emptyset \, ,
    \label{eq:zeroINTER}
\end{equation}

\noindent where $\bigcap$ denotes the set intersection and $\emptyset$ is the empty set. Equation \eq{eq:zeroINTER} is most easily satisfied by requiring $f(\Vect{z})$ be sign-definite, say, positive (so $\Vect{z}_f = \emptyset$).

Let us now compute the rays generated by $\widetilde{D}(\Vect{z})$. Analogous to \Eq{eq:rayHAM}, the new rays satisfy
\begin{equation}
    \pd_{\widetilde{\tau}_1} \Vect{z} = \Mat{J} \, \pd_\Vect{z} \widetilde{\Symb{D}}(\Vect{z})
    = f(\Vect{z}) \Mat{J} \, \pd_\Vect{z} \Symb{D}(\Vect{z})
    + \Symb{D}(\Vect{z}) \Mat{J} \, \pd_\Vect{z} f(\Vect{z})
    \label{eq:modHAM}
\end{equation}

\noindent for some new parameterization $\widetilde{\tau}_1$. From \Eq{eq:modHAM}, rays initialized within $\Vect{z}_{\Symb{D}}$ will always remain in $\Vect{z}_{\Symb{D}}$, at least in exact arithmetic. For such rays, the second term in \Eq{eq:modHAM} is identically zero, making \Eq{eq:modHAM} simply a reparameterization of \Eq{eq:rayHAM} with
\begin{equation}
    \tau_1 = f(\Vect{z}) \widetilde{\tau}_1 \, .
    \label{eq:reparam}
\end{equation}

\noindent For inexact arithmetic, however, the second term in \Eq{eq:modHAM} is not exactly zero, and therefore must be retained to preserve the Hamiltonian structure~\cite{Hairer97,Zare75}.

By \Eq{eq:reparam}, $\tau_1$ is non-uniformly discretized when $\widetilde{\tau}_1$ is uniformly discretized; hence, a curvature-dependent adaptive discretization can be achieved by properly designing $f(\Vect{z})$. First, we restrict $f(\Vect{z})$ to only depend on the local curvature of the dispersion manifold, denoted $\curv(\Vect{z})$%
\footnote{$\curv(\Vect{z})$ may be difficult to calculate numerically, since obtaining the dispersion manifold is often the \textit{result} of ray-tracing, not the prerequisite as we suggest here. An iterative approach might be possible; however, this is outside the scope of the present work.}.
Next, we impose that the uniform and the adaptive discretizations are equivalent when $\curv(\Vect{z}) = 0$. Hence,
\begin{equation}
    \lim_{\curv \to 0} f(\curv) = 1 \, .
    \label{eq:limit}
\end{equation}

\begin{figure}
    \centering
    \begin{overpic}[width=0.9\linewidth]{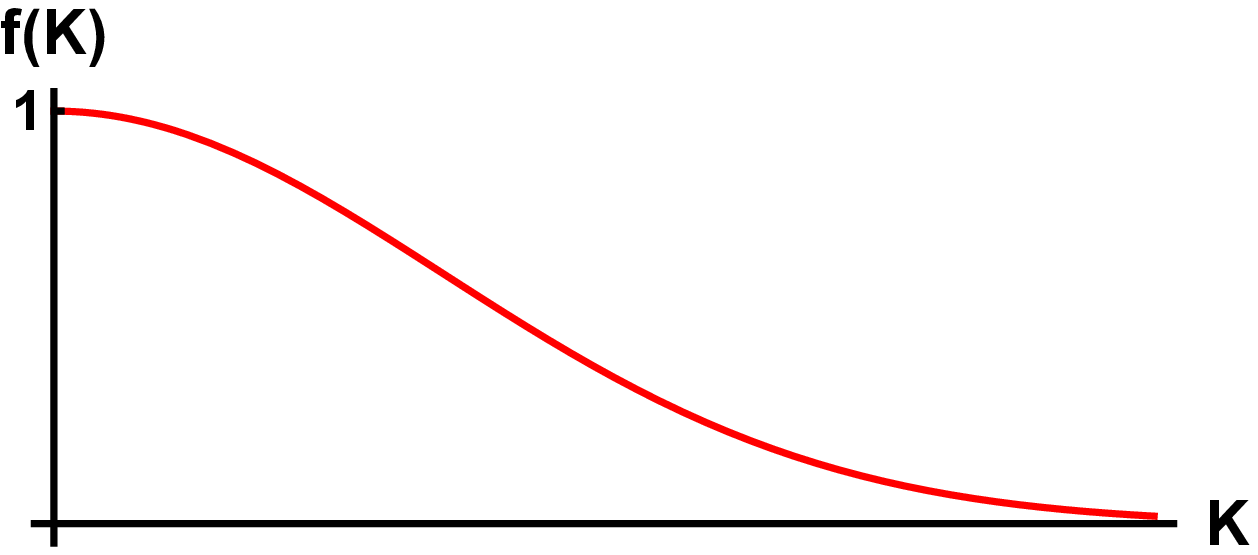}
		\put(90,9){\textbf{\large(a)}}
	\end{overpic}
    
    \begin{overpic}[width=0.9\linewidth]{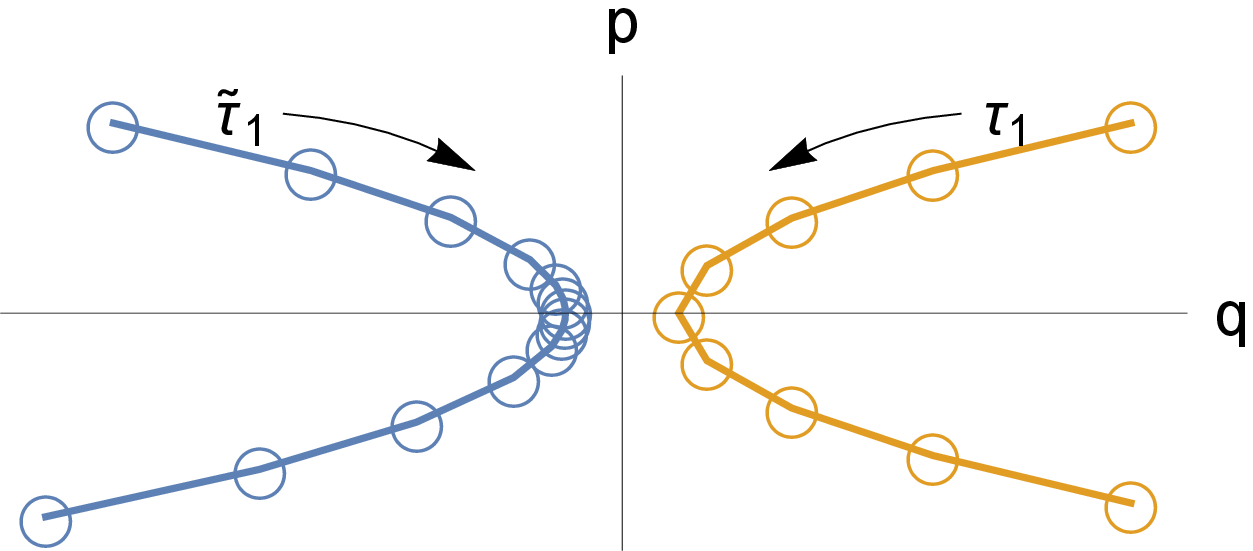}
        \put(90,9){\textbf{\large(b)}}
    \end{overpic}
    \caption{\textbf{(a)} An example $f(\curv)$, where $\curv$ is the local curvature of the dispersion manifold, which satisfies the requirements \eq{eq:zeroINTER}, \eq{eq:limit}, and \eq{eq:decrease}. \textbf{(b)} Comparison of the uniform discretization $\tau_1$ and the adaptive discretization $\widetilde{\tau}_1$ for the Airy dispersion manifold [\Eq{eq:airyDman}] using $f(\curv)$ provided by \Eq{eq:fADAPT} with $\mu = 1$. The values of $\tau_1$ and $\widetilde{\tau}_1$ are uniformly discretized on the intervals $[0,4]$ and $[0,6.5]$ respectively. $\curv(\Vect{z})$ is calculated in the usual manner for a $1$-D planar curve, \ie $\curv(\Vect{z}) = \|\pd_\Vect{z} \Symb{D}(\Vect{z}) \|^{-3} \, \left| \Mat{J} \,  \pd_\Vect{z} \Symb{D}(\Vect{z}) \dubdot \pd^2_{\Vect{z}\Vect{z}} \Symb{D}(\Vect{z}) \right| $~\cite{Goldman05}. For visualization purposes, the dispersion manifolds are displaced slightly from the origin, and the $\tau_1$ discretization is reflected about the $p$ axis.}
    \label{fig:reparam}
\end{figure}

\noindent Finally, for the adaptive discretization to congregate in regions of high curvature, we require $f(\curv)$ to be a strictly decreasing function of $\curv$, that is
\begin{equation}
    f'(\curv \neq 0) < 0 \, ,
    \quad
    f'(0) \le 0 \, .
    \label{eq:decrease}
\end{equation}

Any function which satisfies \Eqs{eq:zeroINTER}, \eq{eq:limit}, and \eq{eq:decrease} will be a suitable choice, for example,
\begin{equation}
    f\left[ \curv(\Vect{z}) \right] = \frac{1}{1 + \mu \left[\curv(\Vect{z})\right]^2} \, ,
    \label{eq:fADAPT}
\end{equation}

\noindent with $\mu \ge 0$ a free parameter. Figure \ref{fig:reparam} shows an example $f(\curv)$ which satisfies these three requirements, and shows the adaptive discretization generated by \Eq{eq:fADAPT}. As a final remark, when reparameterized rays are used to calculate $\Phi(\Vect{Q})$, additional terms related to $\pd_\Vect{z}f(\Vect{z})$ will arise. These can be interpreted as the `gravitational' forces associated with the `time dilation' $\tau_1 \to \widetilde{\tau}_1$~\cite{Dodin19,Dodin10c}.


\section{Examples}
\label{sec:examples}

We now illustrate our methodology with a pair of examples, performed in $1$-D for simplicity.

\subsection{Airy's equation in one dimension}

As a first example, let us consider a simple fold caustic in $1$-D, which occurs when a wave encounters an isolated cutoff. For slowly varying media, this situation is often modeled with Airy's equation~\cite{Kravtsov93},
\begin{equation}
    \pd^2_q \psi(q) - q \psi(q) = 0 \, .
    \label{eq:airy}
\end{equation}

\noindent Like with \Eq{eq:hilbertWAVE}, we can also write \Eq{eq:airy} as
\begin{equation}
    \oper{D}(\VectOp{z}) \ket{\psi} = \ket{0} \, ,
    \quad
    \oper{D}(\VectOp{z}) \doteq \oper{p}^2 + \oper{q} \, ,
\end{equation}

\noindent where we have first multiplied \Eq{eq:airy} by minus one for convenience. Using results presented in \App{app:Weyl}, the Weyl symbol of $\oper{D}(\VectOp{z})$ is calculated to be
\begin{equation}
    \Symb{D}(\Vect{z}) \doteq \WeylInv \left[ \oper{D}(\VectOp{z}) \right] = p^2 + q \, .
    \label{eq:airyDman}
\end{equation}

\noindent In this case, the dispersion manifold $\Symb{D}(\Vect{z}) = 0$ is a parabola which opens along the negative $q$-axis.

By \Eq{eq:rayHAM}, the rays are obtained by integrating
\begin{equation}
    \begin{pmatrix}
        \pd_\tau q \\
        \pd_\tau p
    \end{pmatrix}
    = 
    \begin{pmatrix}
        0 & 1 \\
        -1 & 0
    \end{pmatrix}
    \begin{pmatrix}
        \pd_q \Symb{D}(\Vect{z}) \\
        \pd_p \Symb{D}(\Vect{z})
    \end{pmatrix}
    =
    \begin{pmatrix} 
        2p \\
        -1
    \end{pmatrix}
    \, .
    \label{eq:airyRAYeq}
\end{equation}

\noindent The solution to \Eq{eq:airyRAYeq} is
\begin{equation}
    q(\tau) = -(p_0 - \tau)^2
    \, ,
    \quad
    p(\tau) = p_0 - \tau
    \, ,
    \label{eq:airyRAYS}
\end{equation}

\noindent where $p_0$ is a constant determined by initial conditions. From \Eqs{eq:tangVEC1} and \eq{eq:airyRAYS}, we compute the unit tangent vector to the dispersion manifold at some $\tau = t$ as
\begin{equation}
    \unit{\Vect{T}}(t) 
    = \frac{1}{ \vartheta(t) }
    \begin{pmatrix}
        2p(t) \\
        -1
    \end{pmatrix} \, ,
\end{equation}

\noindent where we have defined
\begin{equation}
    \vartheta(t) \doteq \sqrt{1 + 4 p^2(t) } \, .
\end{equation}

\noindent Correspondingly, the normal vector to the dispersion manifold is calculated via \Eq{eq:sympNORM} as
\begin{equation}
    \unit{\Vect{N}}(t) 
    = \frac{1}{ \vartheta(t) }
    \begin{pmatrix}
        1 \\
        2p(t)
    \end{pmatrix} \, .
\end{equation}

\noindent Using \Eqs{eq:uMAT} and \eq{eq:tangMAT}, we therefore construct
\begin{equation}
    \Mat{S}_t = \frac{1}{\vartheta(t)}
    \begin{pmatrix}
        2p(t) & -1 \\
        1 & 2p(t)
    \end{pmatrix}
\end{equation}

\noindent and identify the $1 \times 1$ block matrices
\begin{equation}
    \Mat{A}_t = \Mat{D}_t = \frac{2p(t)}{\vartheta(t)}
    \, ,
    \quad 
    \Mat{B}_t = - \Mat{C}_t = - \frac{1}{\vartheta(t)}
    \, .
\end{equation}

\noindent Since $\Mat{B}_t$ never changes sign, then the sign of the inverse MT never changes either, and we can choose
\begin{equation}
    \sigma_t = 1 \, .
\end{equation}

Using \Eq{eq:sRAYS}, the rays are transformed by $\Mat{S}_t$ as
\begin{subequations}
    \label{eq:airyMETrays}
    \begin{align}
        Q_t(\tau) &= -\frac{p(\tau) + 2p(t) p^2(\tau) }{ \vartheta(t) }
        \, , \\
        P_t(\tau) &= \frac{2p(t) p(\tau) - p^2(\tau)}{ \vartheta(t) }
        \, .
    \end{align}
\end{subequations}

\noindent Equations \eq{eq:airyMETrays} can be combined to obtain $P_t(Q)$ as
\begin{equation}
    P_t(Q) = 
    \frac
    {
        4p(t) Q + \vartheta(t) \left[-1 \pm \sqrt{1 - 8p(t) \vartheta(t) Q} \right]
    }
    {
        8p^2(t)
    }
    \, .
    \label{eq:airyPFIELD}
\end{equation}

\noindent Hence, $P_t(Q)$ is double-valued. As discussed following \Eq{eq:phase}, we must restrict $P_t(Q)$ to the branch where $P_t[Q_t(t)] = P_t(t)$. This is accomplished by choosing the $(+)$~sign in \Eq{eq:airyPFIELD}. The wavefield phase near $Q_t(t)$ is obtained by integrating \Eq{eq:phase}; this ultimately yields
\begin{align}
    \Theta_t\left[ \epsilon + Q_t(t) \right] &= 
    \frac{8p^4(t) - \vartheta^4(t)}{8p^2(t) \vartheta(t)} \epsilon
    + \frac{1}{4p(t)} \epsilon^2
    \nonumber\\
    &\hspace{3mm} + \frac{\vartheta^6(t) - \left[\vartheta^4(t) - 8p(t) \vartheta(t) \epsilon \right]^{3/2}}{96p^3(t)} \, ,
    \label{eq:airyTHETAmet}
\end{align}

\noindent where as a reminder, $\epsilon \doteq Q - Q_t(t)$.

\begin{figure*}
    \centering
    \includegraphics[width=0.24\linewidth]{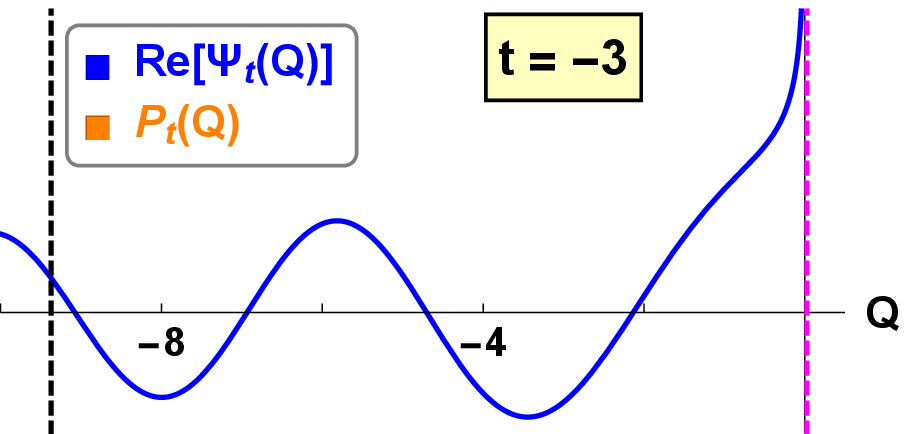}
    \includegraphics[width=0.24\linewidth]{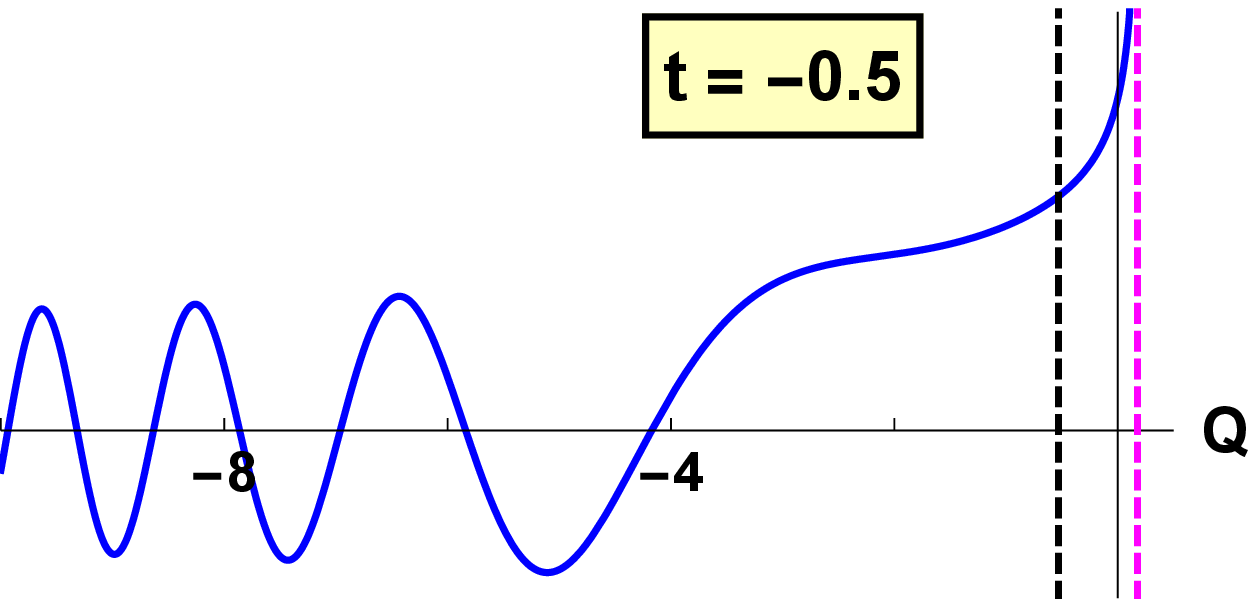}
    \includegraphics[width=0.24\linewidth]{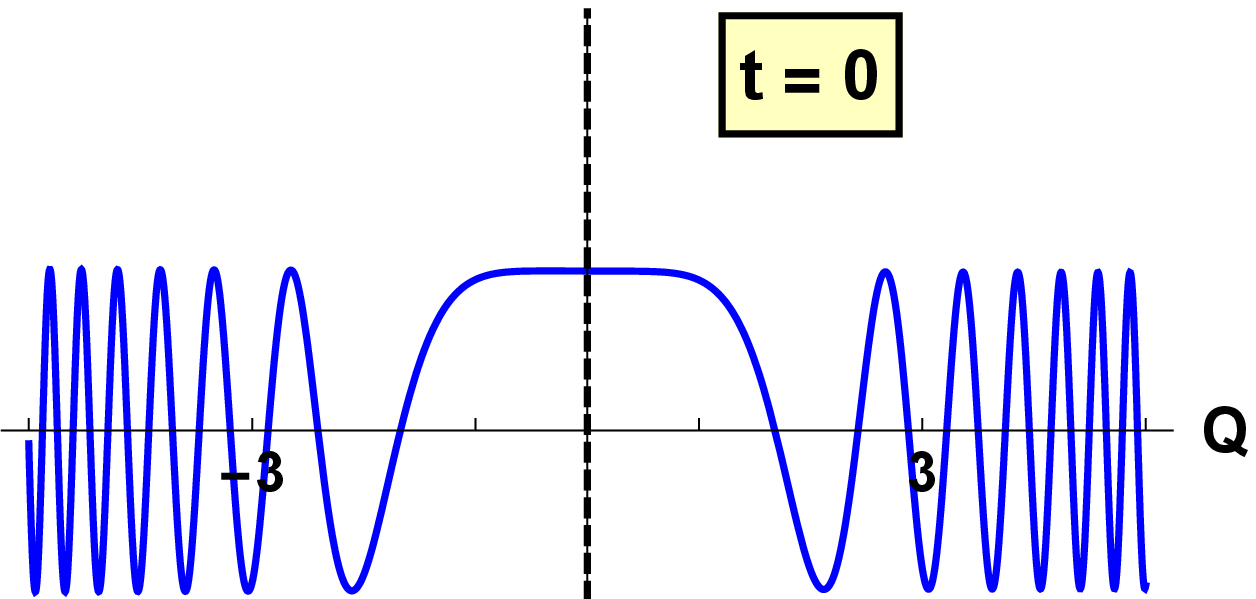}
    \includegraphics[width=0.24\linewidth]{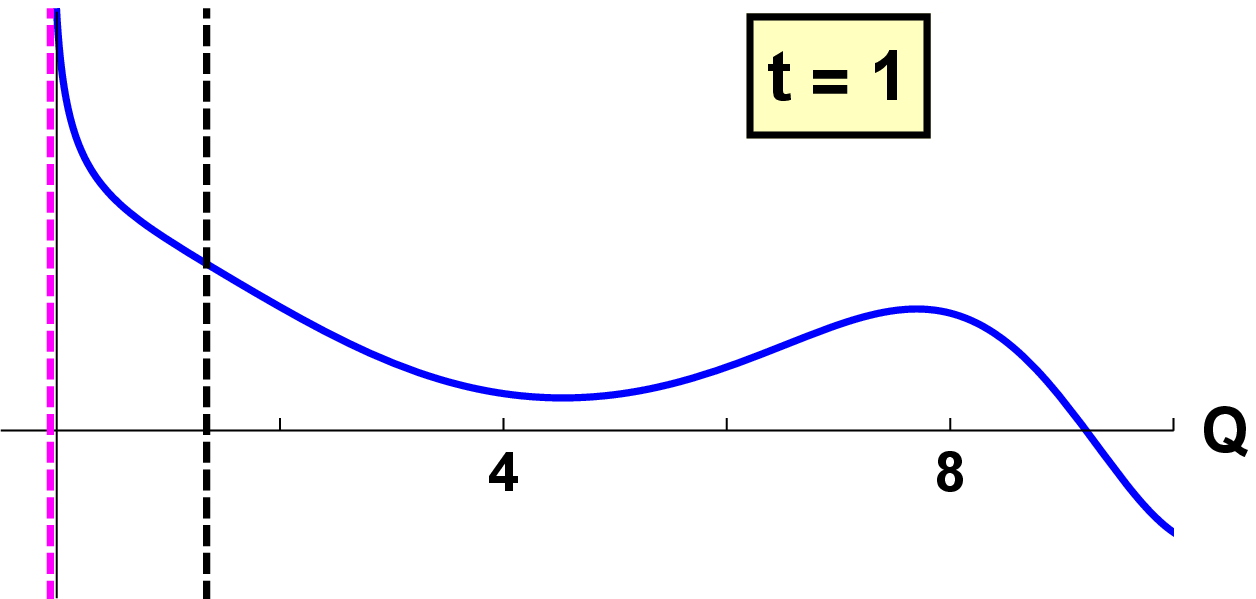}
    
    \includegraphics[width=0.24\linewidth]{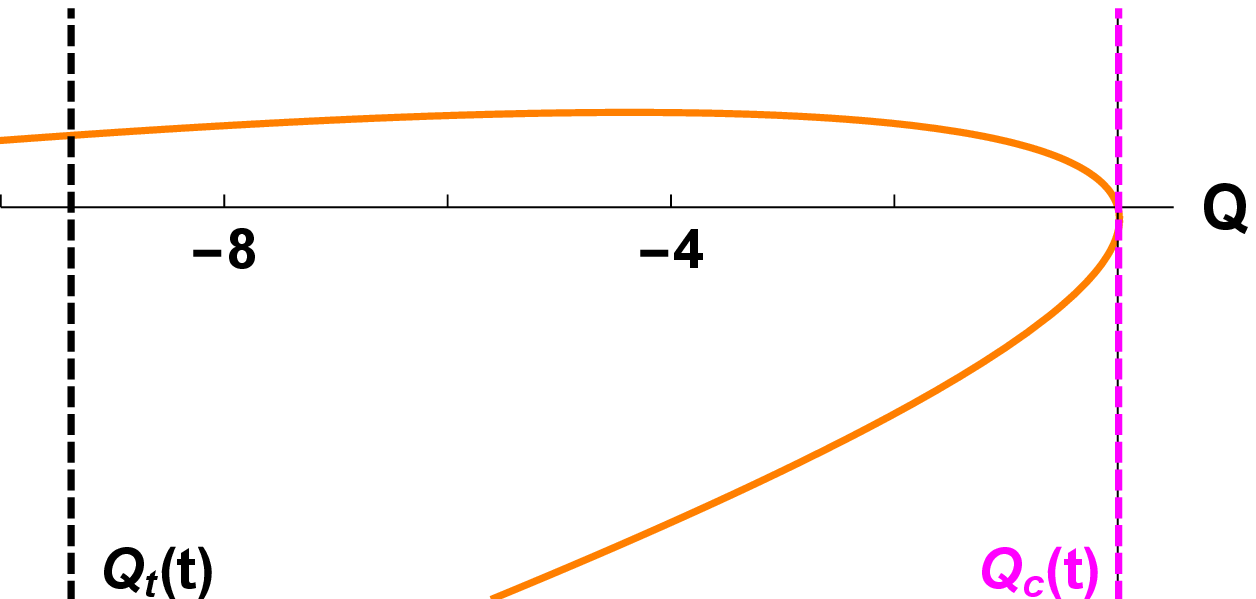}
    \includegraphics[width=0.24\linewidth]{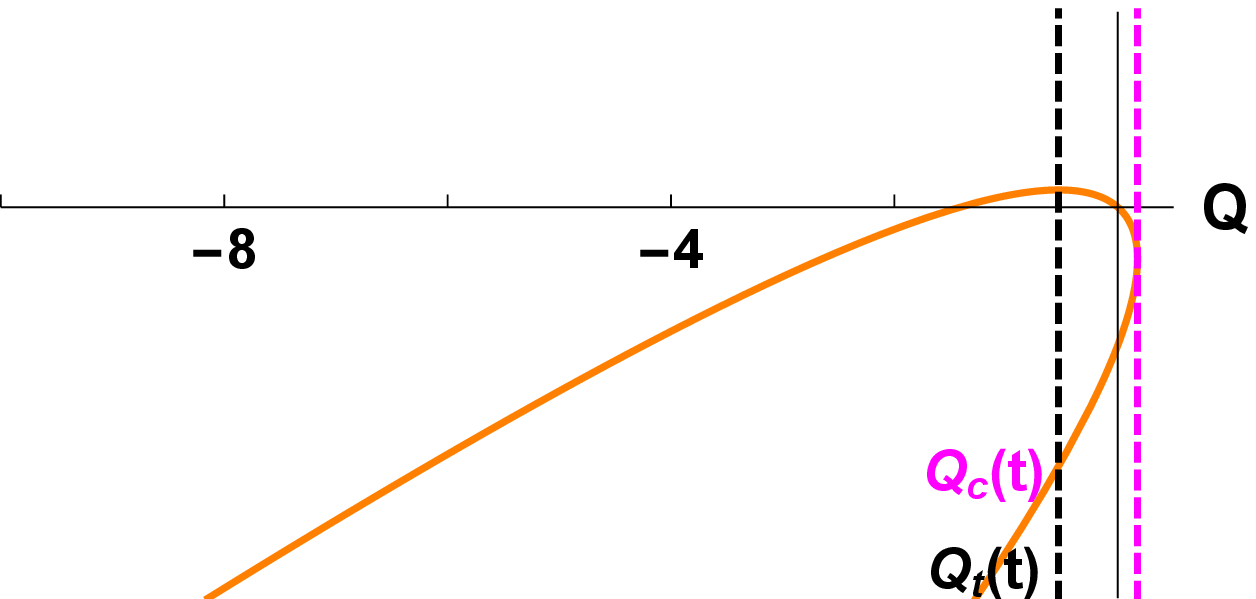}
    \includegraphics[width=0.24\linewidth]{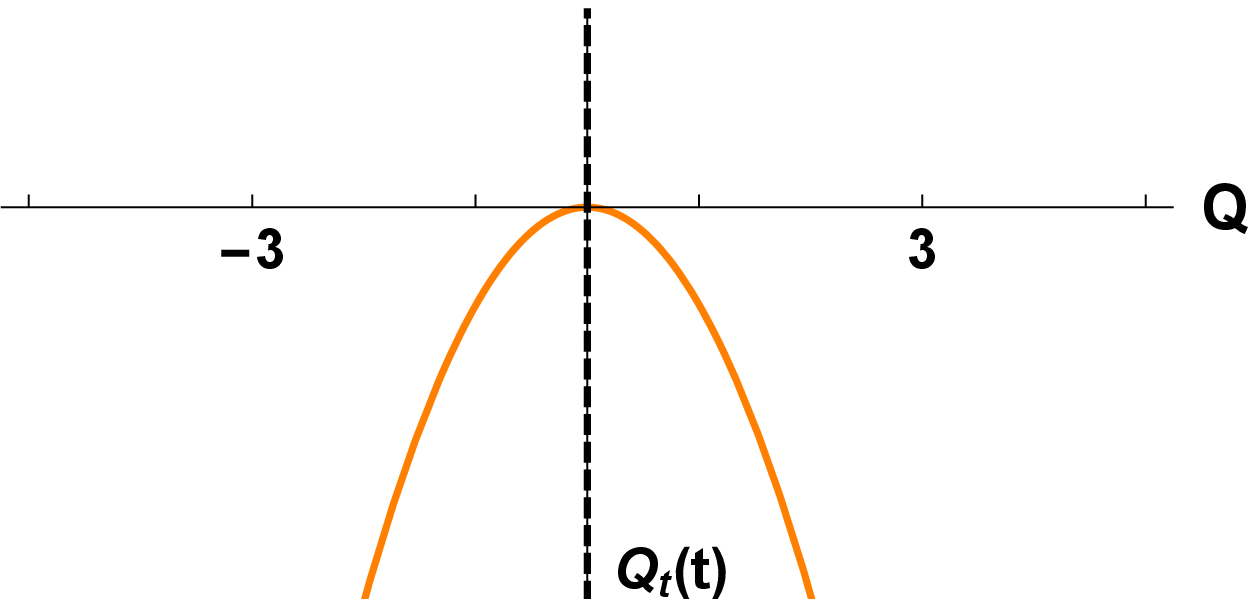}
    \includegraphics[width=0.24\linewidth]{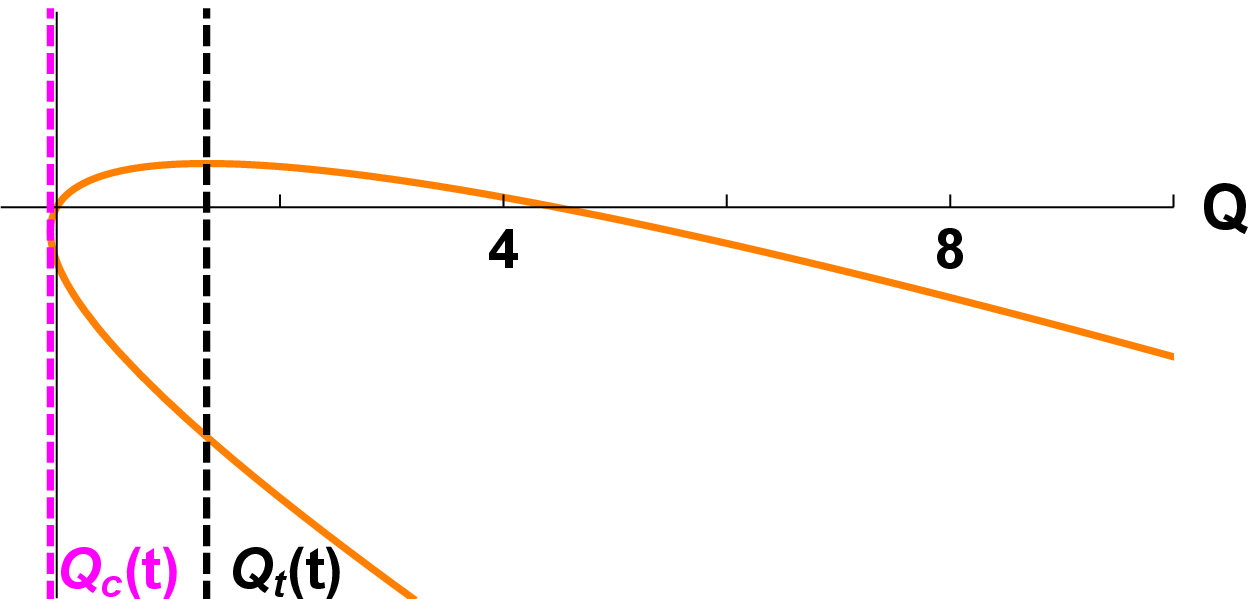}
    \caption{The MGO solution $\Psi_t(Q)$ and the dispersion manifold $P_t(Q)$ for Airy's equation, given respectively by \Eqs{eq:airyPSImet} and \eq{eq:airyPFIELD}, at $t = -3$, $t = -0.5$, $t = 0$, and $t = 1$. For each value of $t$, the projective plane $Q$ is the tangent plane at $Q_t(t)$, denoted by the black dashed line. The caustic, denoted by the red dashed line, is located at $Q_c(t)$ where the curve $P_t(Q)$ has a vertical tangent line. We therefore see that under this rotation scheme, the neighborhood of $Q_t(t)$ is always free from caustics.}
    \label{fig:airyTANG}
\end{figure*}

It is easier analytically to obtain $\Phi_t(Q)$ via \Eq{eq:GOenvMET} rather than \Eq{eq:ENVjacobMET}, since using \Eq{eq:ENVjacobMET} would require inverting $Q_t(\tau)$. (In numerical implementations, though, this may not be an issue). Therefore, we must compute
\begin{equation}
    V_t(Q) \doteq \left.
        \pd_P \Symb{D}\left( \Mat{S}_t^{-1} \Stroke{\Vect{Z}} \right)
    \right|_{P = P_t(Q)} \, .
\end{equation}

\noindent Since
\begin{equation}
    \Symb{D}\left( \Mat{S}^{-1}_t \Stroke{\Vect{Z}} \right)
    = \frac{\left[ 2p(t) P - Q \right]^2}{\vartheta^2(t)} 
    + \frac{P + 2p(t) Q}{\vartheta(t)} 
    \, ,
\end{equation}

\noindent we compute
\begin{equation}
    V_t(Q) 
    = \frac{ \sqrt{1 - 8p(t)\vartheta(t)Q} }{\vartheta(t)} \, .
\end{equation}

\noindent Depending on implementation, it may be more convenient to compute $V_t(Q)$ by using the chain rule to express
\begin{equation}
    \hspace{-1mm}
    \pd_P \Symb{D}\left( \Mat{S}_t \Stroke{\Vect{Z}} \right)
    = \left. 
        \left[
            - \Mat{B}_t \, \pd_q \Symb{D}\left( \Vect{z} \right)
            + \Mat{A}_t \, \pd_p \Symb{D}\left( \Vect{z} \right)
        \right]
    \right|_{\Vect{z} = \Mat{S}^{-1}_t\Stroke{\Vect{Z}}} \, .
    \label{eq:chainRULE}
\end{equation}

\noindent Using \Eq{eq:GOenvMET}, we therefore obtain
\begin{equation}
    \Phi_t\left[ \epsilon + Q_t(t) \right] 
    =
    \frac{ \vartheta(t) }{\left[ 
        \vartheta^4(t) - 8p(t) \vartheta(t) \epsilon 
    \right]^{1/4}} \, ,
    \label{eq:airyPHImet}
\end{equation}

\noindent where we have imposed that $\Phi_t\left[Q_t(t) \right] = 1$ in accordance with the discussion preceding \Eq{eq:wavefieldMET}.

As our final step in constructing $\Psi_t(Q)$, we must calculate $\alpha_t$. From \Eq{eq:UVWdef}, we compute
\begin{equation}
    \Mat{U}_t = \Mat{W}_t = \frac{2}{\vartheta^2(t)} \, \pd_t p(t)
    \, ,
    \quad
    \Mat{V}_t = 0
    \, .
\end{equation}

\noindent We also compute
\begin{subequations}
    \begin{align}
        \pd_t Q_t(t) &= \frac{2p^2(t) - \vartheta^4(t)}{\vartheta^3(t)} \, \pd_t p(t)
        \, , \\
        \pd_Q \Phi_t\left[ Q_t(t) \right] &= \frac{2 p(t)}{\vartheta^3(t)}
        \, .
    \end{align}
\end{subequations}

\noindent Hence, using \Eq{eq:etaDEF} we compute
\begin{align}
    \eta_t &= -\pd_t p(t)\left[
        \frac{2 p(t)}{ \vartheta^2(t) } 
        \right. \nonumber\\
        &\left.\hspace{18mm} 
        + i \frac{20p^6(t) + 11p^4(t) + 2p^2(t)}{ \vartheta^4(t) } 
    \right] \, .
\end{align}

\noindent Performing a change in variables allows the integral in \Eq{eq:alphaINT} to be evaluated as
\begin{align}
    \int_{0}^{t} \dd h \, \eta_h 
    &=
    - \int_{0}^{p(t)} \dd p \, \frac{2p}{1+4p^2}
    \nonumber\\
    &\hspace{4mm}- i \int_{0}^{p(t)} \dd p \, \frac{20p^6 + 11p^4 + 2p^2}{\left(1+4p^2 \right)^2}
    \nonumber\\
    &= - \log \sqrt{\vartheta(t)} - i \frac{2 p^3(t)}{3} + i \frac{p^5(t)}{\vartheta^2(t)}
    \, ,
    \label{eq:airyALPHAint}
\end{align}

\noindent where we have chosen $p(0) = 0$, \ie $p_0 = 0$. 

Thus, using \Eqs{eq:wavefieldMET}, \eq{eq:alphaINT}, and \eq{eq:airyALPHAint}, we obtain the wavefield on the tangent plane,
\begin{equation}
    \hspace{-2mm}\Psi_t(Q)
    = \frac{\alpha_0 \Phi_t(Q)}{\sqrt{\vartheta(t)}} 
    \exp\left[
        i \frac{p^5(t)}{\vartheta^2(t)} - i \frac{2 p^3(t)}{3} + i \Theta_t(Q)
    \right]
    \, ,
    \label{eq:airyPSImet}
\end{equation}

\noindent where $\Theta_t(Q)$ and $\Phi_t(Q)$ are provided in \Eqs{eq:airyTHETAmet} and \eq{eq:airyPHImet} respectively. For reference, $\Psi_t\left[Q_t(t) \right]$ is plotted for select values of $t$ in \Fig{fig:airyTANG}.

Next, we perform the inverse MT via \Eq{eq:invMTmet} to obtain $\psi_t\left[ q(t) \right]$. From \Eq{eq:betaMET}, we compute
\begin{equation}
    \beta_t = \frac{p^5(t)}{\vartheta^2(t)} \, .
\end{equation}

\noindent Hence, \Eq{eq:invMTmet} yields
\begin{equation}
    \psi_t\left[q(t)\right] 
    = 
    \frac{i \alpha_0 }{\sqrt{- 2\pi i}}
    \exp\left[ - i \frac{2}{3}p^3(t) \right]
    \Upsilon_t
    \, ,
    \label{eq:airyPSIt}
\end{equation}

\noindent where $\Upsilon_t$ is given in \Eq{eq:intRESTRICT}. We can now make further approximations to $\Upsilon_t$. First, since the integral of $\Upsilon_t$ is restricted to $\delta \epsilon$, let us approximate 
\begin{equation}
    \Phi_t\left[ \epsilon + Q_t(t) \right] 
    \approx 
    \Phi_t\left[ Q_t(t) \right]
    = 1 \, .
\end{equation}

\noindent Second, let us expand the exponential term about $\epsilon = 0$. From \Eq{eq:gammaMET}, we compute
\begin{equation}
    \gamma_t(\epsilon) = \frac{p^2(t)}{\vartheta(t)} \epsilon - p(t) \epsilon^2 \, .
\end{equation}

\noindent Therefore, 
\begin{equation}
    \Theta_t\left[ \epsilon + Q_t(t) \right] - \gamma_t(\epsilon) 
    \approx p(t) \epsilon^2 - \frac{\epsilon^3}{3 \vartheta^3(t) } \, ,
\end{equation}

\noindent and consequently,
\begin{equation}
    \Upsilon_t \approx \int_{\delta\epsilon} \dd \epsilon \,
    \exp\left[ i p(t) \epsilon^2 - i \frac{\epsilon^3}{3 \vartheta^3(t) } \right] \, .
    \label{eq:airyUPSILONsaddles}
\end{equation}

Since the coefficient of $\epsilon^2$ vanishes when $t = p_0$, while that of $\epsilon^3$ remains nonzero for all $t$, \Eq{eq:airyUPSILONsaddles} contains a pair of coalescing saddle points; isolating the contribution from the saddlepoint at $\epsilon = 0$ analytically is therefore nontrivial. For now, let us suppose there is some `saddlepoint filter' $\filter$ which can perform this operation. [We shall soon show that $\filter$ is related to choosing a steepest descent curve to evaluate \Eq{eq:airyUPSILONsaddles}.] Then, by definition of $\filter$, we can extend the integration bounds to infinity without incurring large errors. This yields
\begin{equation}
    \Upsilon_t \approx \filter \left\{ 
        \int_{-\infty}^\infty \dd \epsilon \,
        \exp\left[ i p(t) \epsilon^2 - i \frac{\epsilon^3}{3 \vartheta^3(t) } \right] 
    \right\} \, .
    \label{eq:airyFILTER}
\end{equation}

By making the substitution $\varepsilon \doteq i \epsilon/ \vartheta(t) - i p(t) \vartheta^2(t)$ and using Cauchy's integral theorem, the integral of \Eq{eq:airyFILTER} is placed into standard form
\begin{align}
    &\int_{-\infty}^\infty \dd \epsilon \,
    \exp\left[ i p(t) \epsilon^2 - i \frac{\epsilon^3}{3 \vartheta^3(t) } \right] = - i \vartheta(t) \nonumber\\
    &\times
    \int_{\cont{0}} \dd \varepsilon \,
    \exp\left[
        i \frac{2}{3} p^3(t) \vartheta^6(t) 
        + p^2(t)\vartheta^4(t) \varepsilon 
        + \frac{\varepsilon^3}{3 } 
    \right] \, ,
    \label{eq:airySTANDARD}
\end{align}

\begin{figure}
    \centering
    \includegraphics[width=0.95\linewidth,trim={0mm 0mm 0mm 0mm},clip]{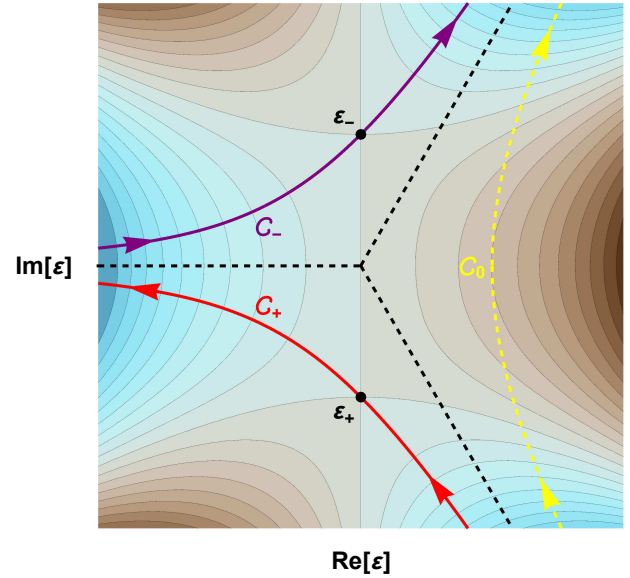}
    \caption{The function $f(\varepsilon; x) = x \varepsilon + \frac{\varepsilon^3}{3}$ in the complex $\varepsilon$ plane with $x \ge 0$, which serves as the phase in Airy's integral \eq{eq:airySTANDARD}. The background color depicts $\text{Re}\left[ f(\varepsilon; x) \right]$, with blue denoting negative values. The points $\varepsilon_\pm = \mp i \sqrt{|x|}$ are the saddlepoints where $\pd_\varepsilon f(\varepsilon; x) = 0$. Correspondingly, the standard contour for Airy's integral, $\cont{0}$, can be decomposed into the two steepest descent contours $\cont{\pm}$.}
    \label{fig:airyCONT}
\end{figure}

\noindent where $\cont{0}$ is a contour from $r e^{-i \pi/3}$ to $r e^{i\pi/3}$, where $r \to \infty$ (see \Fig{fig:airyCONT}). There are two saddlepoints in \Eq{eq:airySTANDARD}, located respectively at $\varepsilon = \varepsilon_{\pm} \doteq \mp i | p(t) \vartheta^2(t)|$. When \Eq{eq:airySTANDARD} is evaluated using the method of steepest descent~\cite{Bender78}, the integration contour must be split into the two pieces denoted $\cont{+}$ and $\cont{-}$, each of which encounters only one of the two saddle points (\Fig{fig:airyCONT}).

It is now clear how the `saddlepoint filter' $\filter$ can be designed: when $\filter$ acts on \Eq{eq:airySTANDARD}, rather than evaluating the integral over \textit{both} $\cont{+}$ \textit{and} $\cont{-}$, the integral is evaluated over \textit{either} $\cont{+}$ \textit{or} $\cont{-}$. To determine which, note that when $p > 0$, $\epsilon = 0$ corresponds to $\varepsilon = -ip\vartheta^2 = -i|p\vartheta^2| = \epsilon_+$, while when $p < 0$, $\epsilon = 0$ corresponds to $\varepsilon = -ip\vartheta^2 = +i|p\vartheta^2| = \epsilon_-$. Hence,
\begin{align}
    &\hspace{-1mm}\filter\left\{
        \int_{-\infty}^\infty \dd \epsilon \,
        \exp\left[ i p(t) \epsilon^2 - i \frac{\epsilon^3}{3 \vartheta^3(t) } \right]
    \right\} \doteq -i \vartheta(t) \nonumber\\ 
    &\hspace{-1mm}\times 
    \int_{\cont{s(t)}} \hspace{-2mm}\dd \varepsilon \,
    \exp\left[
        i \frac{2}{3} p^3(t) \vartheta^6(t) 
        + p^2(t)\vartheta^4(t) \varepsilon 
        + \frac{\varepsilon^3}{3 } 
    \right] \, ,
    \label{eq:airyFILTERfinal}
\end{align}

\noindent with
\begin{equation}
    s(t) \doteq \text{sign}\left[ p(t) \right] = \text{sign}\left( p_0 - t \right) \, .
\end{equation}

As can readily be shown,
\begin{equation}
    \int_{\cont{\pm}} \hspace{-2mm}\dd \varepsilon \,
    \exp\left(
        \frac{\varepsilon^3}{3 }
        - x \varepsilon 
    \right)
    = i \pi \left[\textrm{Ai}(x) \pm i \textrm{Bi}(x) \nullFrac\right] \, ,
\end{equation}

\noindent where $\textrm{Ai}(x)$ and $\textrm{Bi}(x)$ are the Airy functions of the first and second kind, respectively~\cite{Olver10}. Thus, using \Eqs{eq:airyPSIt}, \eq{eq:airyFILTER}, and \eq{eq:airyFILTERfinal}, we obtain
\begin{align}
    &\psi_t\left[q(t) \right] = i \alpha_0 \frac{\sqrt{\pi}}{\sqrt{- 2i}} \vartheta(t)
    \exp\left\{
        i \frac{2}{3} p^3(t) \left[ \vartheta^6(t) - 1 \right] 
    \right\} \nonumber\\
    &\times\left\{
        \textrm{Ai}\left[-p^2(t)\vartheta^4(t) \right] 
        +i \, s(t) \textrm{Bi}\left[-p^2(t)\vartheta^4(t) \right]
        \nullFrac
    \right\}\, .
\end{align}

\noindent We next sum over both branches of $\tau(q)$, per \Eq{eq:psiSUM}. Upon setting $\alpha_0 = - \sqrt{i/2\pi}$, we obtain
\begin{align}
    \psi(q) &=
    \sqrt{1 - 4 q} \, \textrm{Ai}\left[- \varrho^2(q) \right]
    \cos[ \varpi(q)]
    \nonumber\\
    &\hspace{5mm}- \sqrt{1 - 4 q} \, \textrm{Bi}\left[ - \varrho^2(q) \right]
    \sin[ \varpi(q) ]
     \, ,
     \label{eq:airyMGO}
\end{align}

\noindent where we have defined
\begin{subequations}
    \begin{align}
        \varrho(q) &\doteq
        (1 - 4q) \sqrt{-q}
        \, , \\
        \varpi(q) &\doteq 
        \frac{2}{3} \varrho^3(q) - \frac{2}{3}(-q)^{3/2}
        \, .
    \end{align}
\end{subequations}

\begin{figure}
    \centering
    \includegraphics[width=0.95\linewidth]{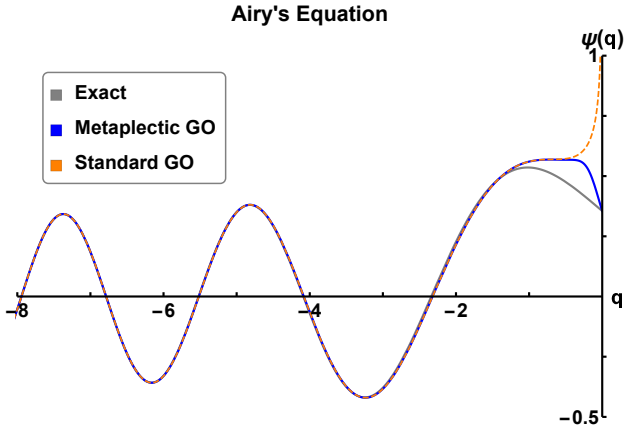}
    \caption{Comparison between the Airy function $\textrm{Ai}(q)$, its standard GO approximation [\Eq{eq:airyWKB}], and its MGO approximation [\Eq{eq:airyMGO}]. The caustic is located at $q = 0$. The error in the MGO solution is largely due to the approximations made when performing the inverse MT (which may not be needed for certain applications).}
    \label{fig:airySOL}
\end{figure}

Figure~\ref{fig:airySOL} compares \Eq{eq:airyMGO} with the exact solution,
\begin{equation}
    \psi_\text{exact}(q) = \textrm{Ai}(q) \, ,
\end{equation}

\noindent and with the standard GO approximation~\cite{Olver10},
\begin{equation}
    \psi_\text{GO} = \pi^{-1/2} (-q)^{-1/4} 
    \sin\left[\frac{2}{3} (-q)^{3/2} + \frac{\pi}{4} \right] \, .
    \label{eq:airyWKB}
\end{equation}

\noindent As can be seen, the MGO solution is almost indistinguishable from the standard GO approximation far from the caustic at $q = 0$. However, in contrast to the standard GO solution, our solution remains finite for all $q \le 0$, like the exact solution of \Eq{eq:airy}.

\subsection{Weber's equation in one dimension}
\label{sec:examplesQHO}

Next, let us consider a bounded wave in a $1$-D harmonic potential, which exhibits two adjacent fold caustics. This situation is described mathematically by Weber's equation,
\begin{equation}
    \pd^2_q \psi(q) + \left(2\nu + 1 - q^2 \right)\psi(q) = 0 \, ,
    \label{eq:parabEQ}
\end{equation}

\noindent which is also the Schr\"odinger equation for a quantum harmonic oscillator~\cite{Shankar94}. Equation \eq{eq:parabEQ} can be written as
\begin{equation}
    \oper{D}(\VectOp{z})\ket{\psi} = \ket{0} \, ,
    \quad
    \oper{D}(\VectOp{z}) \doteq \oper{p}^2 + \oper{q}^2 - 2E \, \IdentOp \, ,
\end{equation}

\noindent where $E \doteq \nu + 1/2$. As before, we have multiplied \Eq{eq:parabEQ} by minus one for convenience. The Weyl symbol is readily computed to be
\begin{equation}
    \Symb{D}(\Vect{z}) = p^2 + q^2 - 2E \, .
\end{equation}

\noindent In this case, the dispersion manifold $\Symb{D}(\Vect{z}) = 0$ is a circle of radius $R \doteq \sqrt{2E}$.

From \Eq{eq:rayHAM}, the ray equations are
\begin{equation}
    \pd_\tau q = 2 p \, ,
    \quad
    \pd_\tau p = -2 q \, .
\end{equation}

\noindent Their solutions have the form
\begin{equation}
    \hspace{-1mm}
    q(\tau) = R \cos \left( 2 \tau \right) \, ,
    \quad
    p(\tau) = - R \sin \left( 2 \tau \right) \, ,
\end{equation}

\noindent where we have assumed the initial condition $q(0) = R$. The unit tangent and normal vectors at $\tau = t$ are calculated from \Eqs{eq:tangVEC1} and \eq{eq:sympNORM} as
\begin{equation}
    \unit{\Vect{T}}(t) = 
    \begin{pmatrix}
        - \sin \left( 2 t \right) \\
        - \cos \left( 2 t \right)
    \end{pmatrix} \, ,
    \quad
    \unit{\Vect{N}}(t) = 
    \begin{pmatrix}
        \cos \left( 2 t \right) \\
        - \sin \left( 2 t \right)
    \end{pmatrix} \, .
\end{equation}

\noindent Using \Eqs{eq:uMAT} and \eq{eq:tangMAT}, we therefore construct
\begin{equation}
    \Mat{S}_t = 
    \begin{pmatrix}
        - \sin \left( 2 t \right) & - \cos \left( 2 t \right) \\
        \cos \left( 2 t \right) & - \sin \left( 2 t \right)
    \end{pmatrix}
\end{equation}

\noindent and identify the $1 \times 1$ block matrices
\begin{equation}
    \Mat{A}_t = \Mat{D}_t = - \sin \left( 2 t \right)
    \, ,
    \quad 
    \Mat{B}_t = - \Mat{C}_t = - \cos \left( 2 t \right)
    \, .
\end{equation}

Unlike the previous example, $\Mat{B}_t$ can now change sign, and consequently, $\sigma_t$ will change sign as well. Let us choose to have $\Mat{B}_t$ cross the branch cut whenever $\Mat{B}_t$ changes from positive to negative. This is encapsulated by the phase convention
\begin{equation}
    \Mat{B}_t 
    = |\Mat{B}_t| \exp
    \left( 
        i
        \left\lfloor
            \frac{4t - \pi}{2\pi}
        \right\rfloor
        \pi
    \right) \, ,
\end{equation}

\noindent where $\lfloor \, \rfloor$ denotes the floor operation. Hence, choosing
\begin{equation}
    \sigma_t = 
    \exp
    \left( 
        -i
        \left\lfloor
            \frac{4t + \pi}{4\pi} 
        \right\rfloor
        \pi
    \right)
\end{equation}

\noindent will ensure continuity in $t$ across the branch cut.

Using \Eq{eq:sRAYS}, the rays are transformed by $\Mat{S}_t$ as
\begin{subequations}
    \label{eq:parabMETrays}
    \begin{align}
        Q_t(\tau) &= R \sin \left( 2\tau - 2 t \right)
        \, , \\
        P_t(\tau) &= R \cos \left( 2\tau - 2 t \right)
        \, .
    \end{align}
\end{subequations}

\noindent Equations \eq{eq:parabMETrays} can be combined to obtain $P_t(Q)$ as
\begin{equation}
    P_t(Q) = 
    \pm \sqrt{R^2 - Q^2} \, .
    \label{eq:parabPFIELD}
\end{equation}

\noindent Hence, $P_t(Q)$ is double-valued. As before, we restrict $P_t(Q)$ to the (+) branch. After integrating \Eq{eq:phase}, we obtain the wavefield phase
\begin{align}
    \Theta_t\left( \epsilon \right) &= 
    \frac{\epsilon}{2}\sqrt{R^2 - \epsilon^2} 
    + \frac{R^2}{2} \tan^{-1}\left( \frac{\epsilon}{\sqrt{R^2 - \epsilon^2}} \right) \, ,
    \label{eq:parabTHETAmet}
\end{align}

\noindent where $\epsilon \doteq Q - Q_t(t) = Q$, since per \Eqs{eq:parabMETrays}, $Q_t(t) = 0$. Next, since
\begin{equation}
    \Symb{D}\left( \Mat{S}^{-1}_t \Stroke{\Vect{Z}} \right)
    = P^2 + Q^2 - R^2
    \, ,
\end{equation}

\noindent we compute
\begin{equation}
    V_t(Q) 
    = 2 \sqrt{R^2 - Q^2} \, .
\end{equation}

\noindent Thus, using \Eq{eq:GOenvMET} we obtain the wavefield envelope
\begin{equation}
    \Phi_t\left( \epsilon \right) 
    =
    \left[1 - \left( \epsilon / R \right)^2 \right]^{-1/4} \, ,
    \label{eq:parabPHImet}
\end{equation}

\noindent where as before, we set $\Phi_t\left[Q_t(t) \right] = \Phi(0) = 1$.

\begin{figure*}
    \centering
    \includegraphics[width=0.24\linewidth]{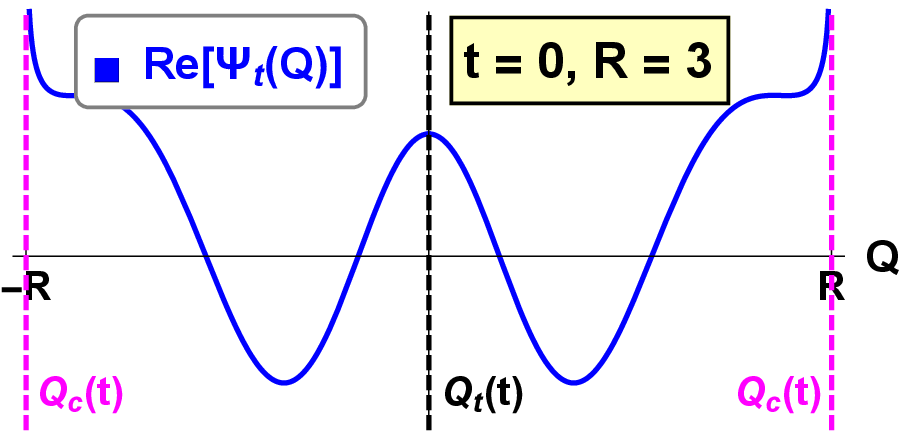}
    \includegraphics[width=0.24\linewidth]{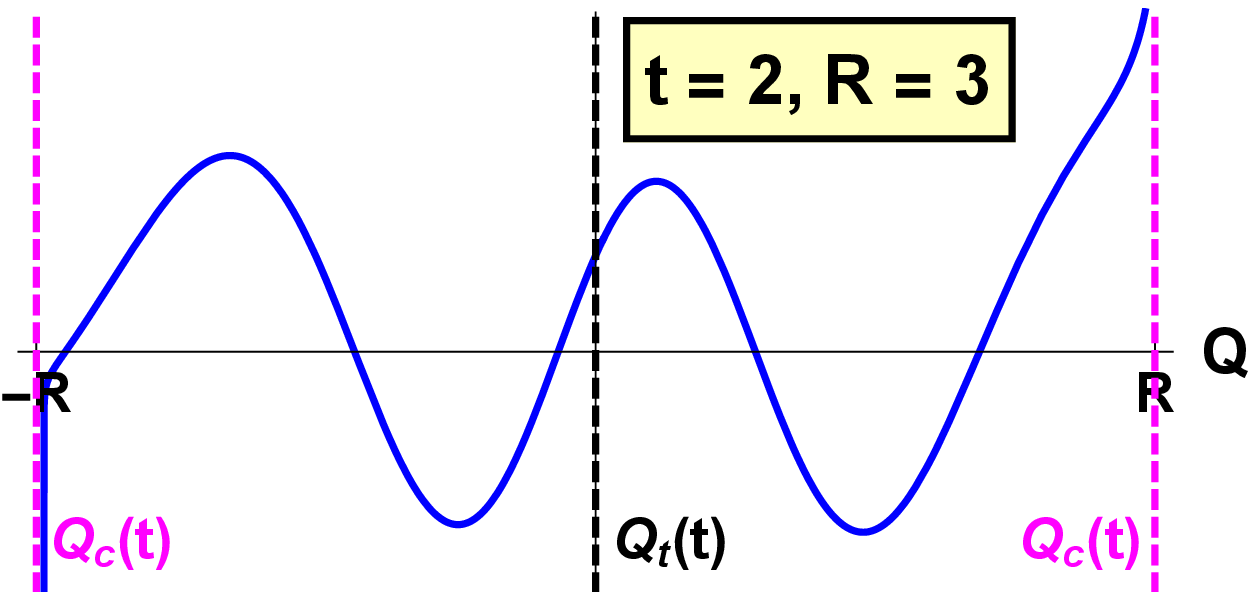}
    \includegraphics[width=0.24\linewidth]{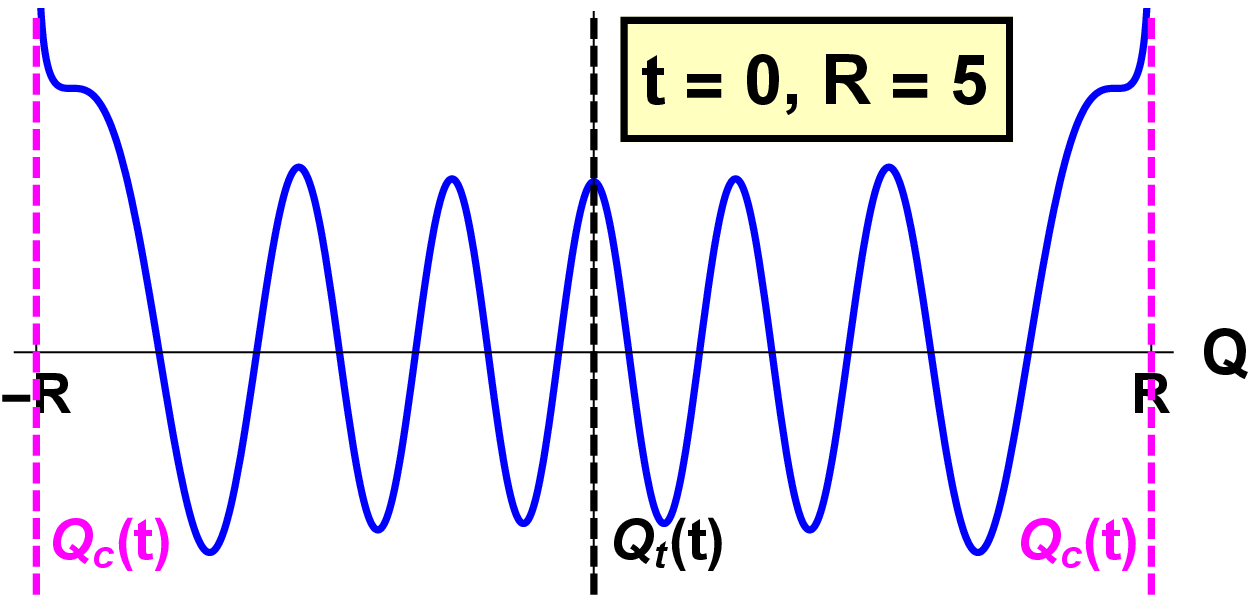}
    \includegraphics[width=0.24\linewidth]{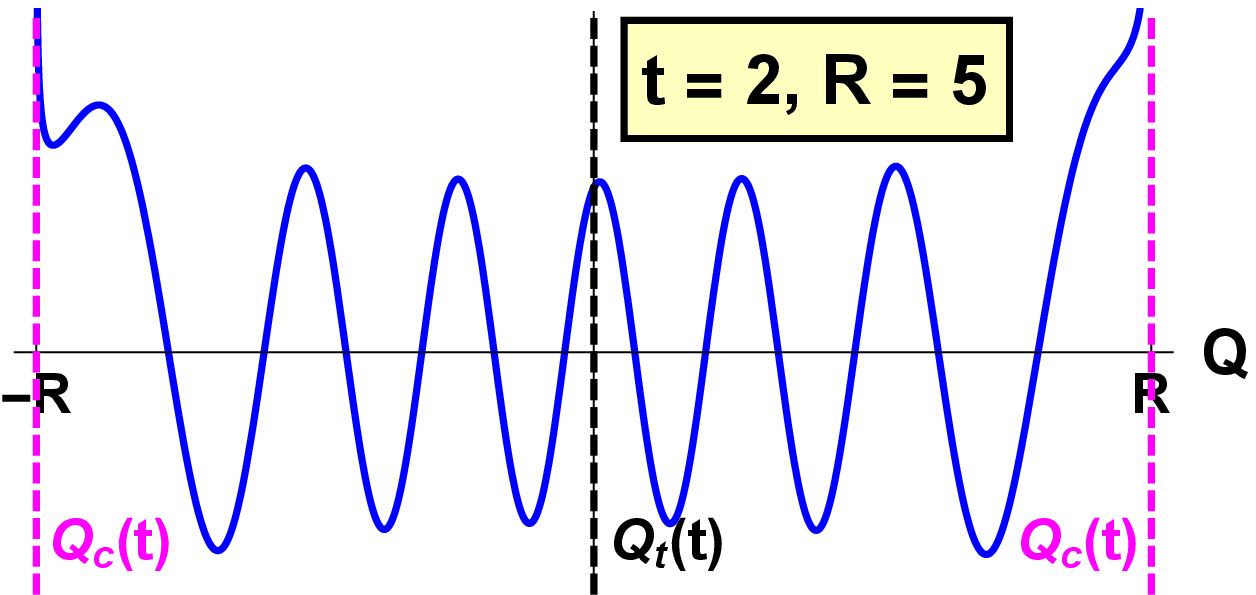}
    \caption{Same as \Fig{fig:airyTANG} but for Weber's equation \eq{eq:parabEQ} for two different values of $t$ and $R$. In all cases, $P_t(Q)$ is a circle of radius $R$ centered at the origin.}
    \label{fig:parabTANG}
\end{figure*}

We now calculate $\alpha_t$. From \Eq{eq:UVWdef}, we compute
\begin{equation}
    \Mat{U}_t = \Mat{W}_t = -2
    \, ,
    \quad
    \Mat{V}_t = 0
    \, .
\end{equation}

\noindent We also compute
\begin{equation}
    \pd_t Q_t(t) = 0
    \, ,
    \quad
    \pd_Q \Phi_t\left[ Q_t(t) \right] = 0
    \, .
\end{equation}

\noindent Hence, using \Eq{eq:etaDEF} we compute
\begin{equation}
    \eta_t = i R^2 \, .
\end{equation}

\noindent Thus, we evaluate
\begin{equation}
    \int_0^t \dd h \, \eta_h = i R^2 t \, .
\end{equation}

\noindent The wavefield on the tangent plane is therefore
\begin{equation}
    \Psi_t(Q) = \alpha_0 \Phi_t(Q) \, \exp\left[iR^2 t + i \Theta_t(Q) \right] \, .
\end{equation}

\noindent Figure~\ref{fig:parabTANG} shows $\Psi_t (Q)$ at select values of $t$ and $R$.

Next, we perform the inverse MT via \Eq{eq:invMTmet} to obtain $\psi_t\left[ q(t) \right]$. From \Eq{eq:betaMET}, we compute
\begin{equation}
    \beta_t = \frac{1}{2} \, \tan\left( 2t \right) q^2 \, .
\end{equation}

\noindent Hence, \Eq{eq:invMTmet} yields
\begin{equation}
    \psi_t\left[q(t)\right] 
    = 
    \frac{\alpha_0 \, \sigma_t \, \exp\left[
        i R^2 t
        - i \frac{R^2}{4} \, \sin\left( 4t \right)
    \right]}{\sqrt{-2\pi i} \, \sqrt{- \cos \left(2 t \right) }} \,  \Upsilon_t
    \, ,
    \label{eq:parabPSIt}
\end{equation}

\noindent where $\Upsilon_t$ is given in \Eq{eq:intRESTRICT}. Before proceeding, note that $\Upsilon_t$ only involves functions which are either constant ($\Phi_t$, $\Theta_t$) or $\pi$-periodic in time ($\Mat{S}_t$, $q$). Thus,
\begin{equation}
    \psi_{t + \pi} \left[q(t + \pi) \right]
    = \psi_t\left[q(t) \right] \exp\left[i \pi (R^2 - 1) \right] \, ,
\end{equation}

\noindent where we have used $\sigma_{t+\pi} = \sigma_t \exp\left( - i \pi \right)$. For $\psi_t$ to be single-valued over the dispersion manifold, $R^2 - 1$ must be an even integer, which in turn requires $\nu$ to be an integer. Since $E \ge 0$ is also needed for $R$ to be real, the integer must be nonnegative. All together, this leads to the Bohr--Sommerfeld quantization of Weber's equation, more commonly known as~\cite{Shankar94}
\begin{equation}
    E = \nu + 1/2 
    \, ,
    \quad
    \nu = 0, 1, 2, \ldots
    \, .
\end{equation}

To evaluate $\Upsilon_t$, note that \Eq{eq:gammaMET} leads to
\begin{equation}
    \gamma_t(\epsilon) = \epsilon R + \frac{\tan \left(2 t \right)}{2} \epsilon^2 \, .
\end{equation}

\noindent We then approximate $\Upsilon_t$ in the same manner as in the previous example; namely, we approximate 
\begin{equation}
    \Phi_t\left( \epsilon \right) 
    \approx 
    \Phi_t (0)
    = 1 \, ,
\end{equation}

\noindent and we approximate
\begin{equation}
    \Theta_t\left( \epsilon \right) - \gamma_t(\epsilon) 
    \approx - \frac{\tan \left( 2t \right)}{2} \epsilon^2 - \frac{\epsilon^3}{6 R} \, .
\end{equation}

\noindent Consequently,
\begin{equation}
    \Upsilon_t \approx \int_{\delta\epsilon} \dd \epsilon \,
    \exp\left[- i \frac{\tan \left(2 t \right)}{2} \epsilon^2 - i \frac{\epsilon^3}{6 R} \right] \, .
    \label{eq:parabUPSILONsaddles}
\end{equation}

Equation \eq{eq:parabUPSILONsaddles} is of the same basic form as \Eq{eq:airyUPSILONsaddles}. Therefore, we immediately conclude that
\begin{align}
    \Upsilon_t &\approx 
    \pi (2R)^{1/3} \exp\left[ -i \frac{R^2}{3} \tan^3(2t) \right]
    \nonumber\\
    &\hspace{4mm}\times\left\{ 
        \textrm{Ai}\left[ - \frac{\tan^2(2t)}{4} (2R)^{4/3} \right]
        \right.
    \nonumber\\
    &\hspace{10mm}\left.
        + i \, s(t) \textrm{Bi}\left[ - \frac{\tan^2(2t)}{4} (2R)^{4/3} \right]
    \right\} \, ,
\end{align}

\noindent where we have defined
\begin{equation}
    s(t) \doteq - \text{sign}\left[ \tan(2t) \right] = \text{sign}\left[ p(t) \right]
    \,
    \text{sign}\left[ q(t) \right] \, .
\end{equation}

\noindent Hence,
\begin{align}
    \psi_t\left[ q(t) \right] &= \alpha_0
    \frac{ \sigma_t \, \sqrt{i \pi}}{ \sqrt{-\cos (2t)}} 
    \left(\frac{R}{\sqrt{2}} \right)^{1/3}
    \nonumber\\
    &\hspace{4mm}\times\exp\left\{
        i R^2 \left[
            t 
            - \frac{\sin(4t)}{4} 
            - \frac{\tan^3(2t)}{3} 
        \right] 
    \right\}
    \nonumber\\
    &\hspace{4mm}\times\left\{ 
        \textrm{Ai}\left[ - \frac{\tan^2(2t)}{4} (2R)^{4/3} \right]
        \right.
    \nonumber\\
    &\hspace{10mm}\left.
        + i \, s(t) \textrm{Bi}\left[ - \frac{\tan^2(2t)}{4} (2R)^{4/3} \right]
    \right\}
    \, .
    \label{eq:parabPSIpresum}
\end{align}

\noindent Upon setting $\alpha_0 = -i (i \pi)^{-1/2} \, (2R)^{-5/6}$, we sum over both branches of $\tau(q)$ to obtain the MGO solution
\begin{align}
    \psi(q) &= \frac{ \textrm{Ai}\left[ -\varrho^2(q) \right]\cos \varpi(q)}{\sqrt{|q|}}
    \nonumber\\
    &\hspace{4mm}- \text{sign}(q) \frac{ \textrm{Bi}\left[ -\varrho^2(q) \right] \sin \varpi(q)}{\sqrt{|q|}}
    \, ,
    \label{eq:parabMGO}
\end{align}

\begin{figure*}
    \centering
    \begin{overpic}[width=0.46\linewidth]{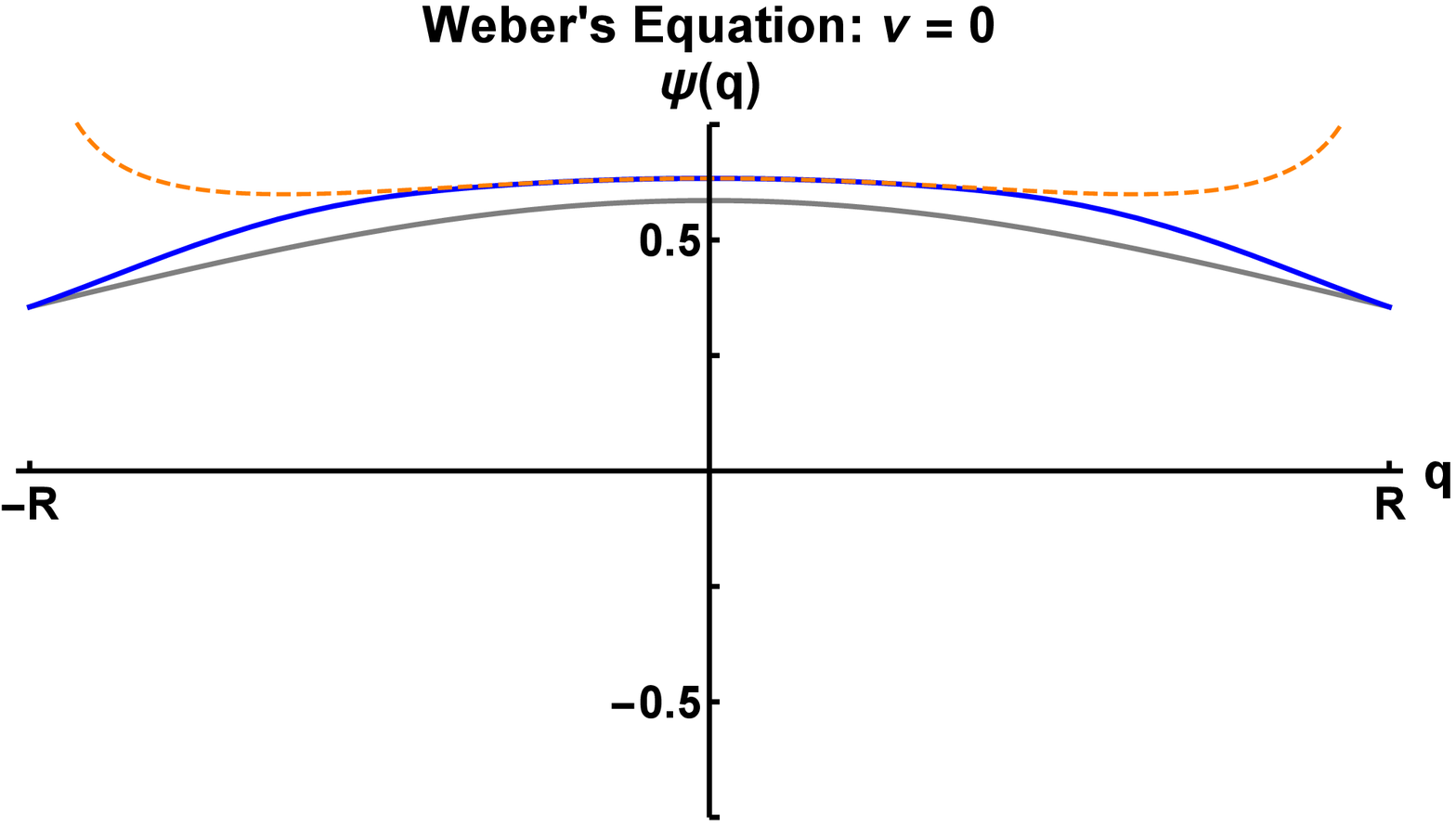}
        \put(5,5){\textbf{\large(a)}}
    \end{overpic}
    \hspace{1mm}
    \begin{overpic}[width=0.46\linewidth]{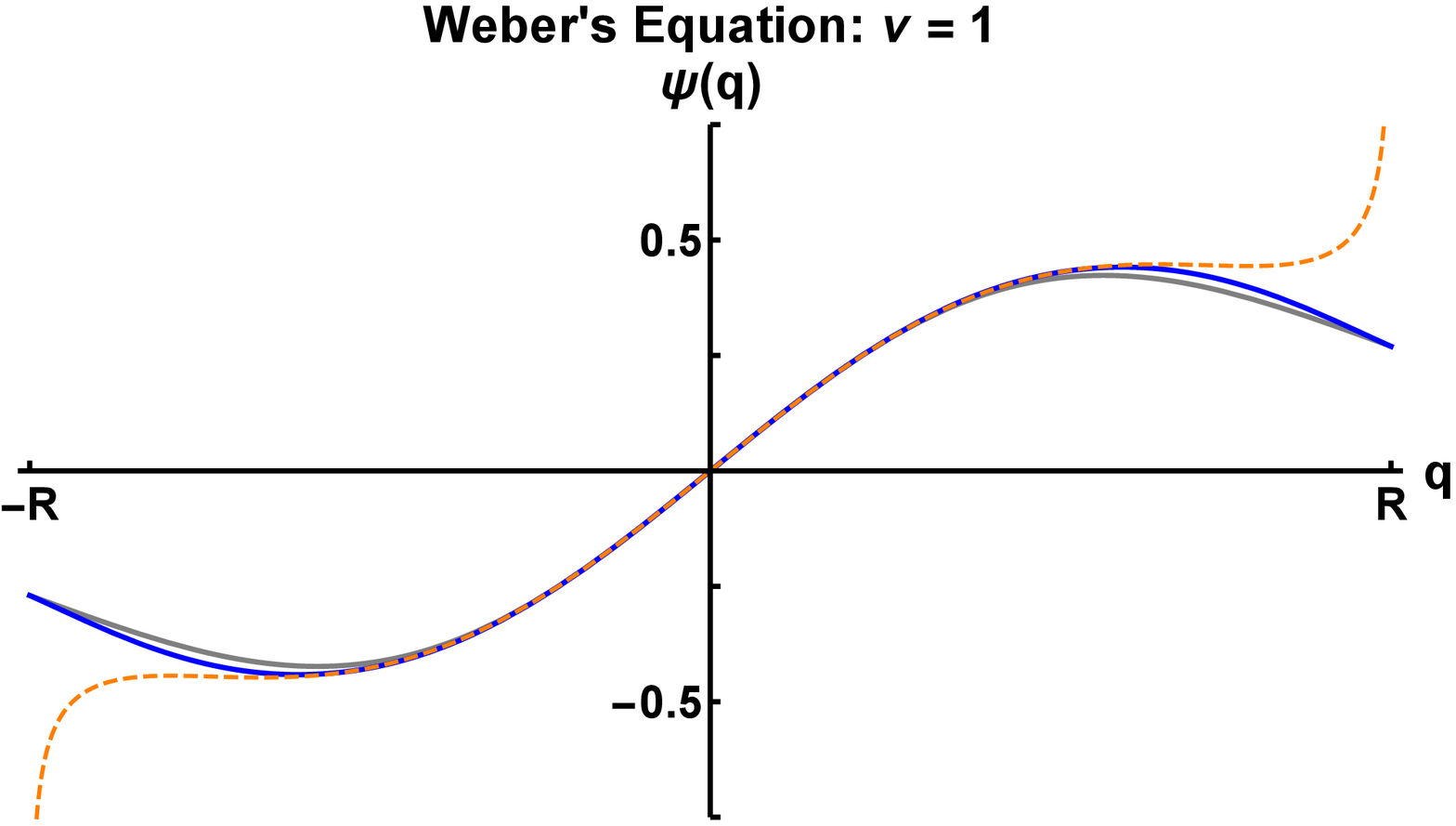}
        \put(5,5){\textbf{\large(b)}}
    \end{overpic}
    
    \vspace{3mm}
    \begin{overpic}[width=0.46\linewidth]{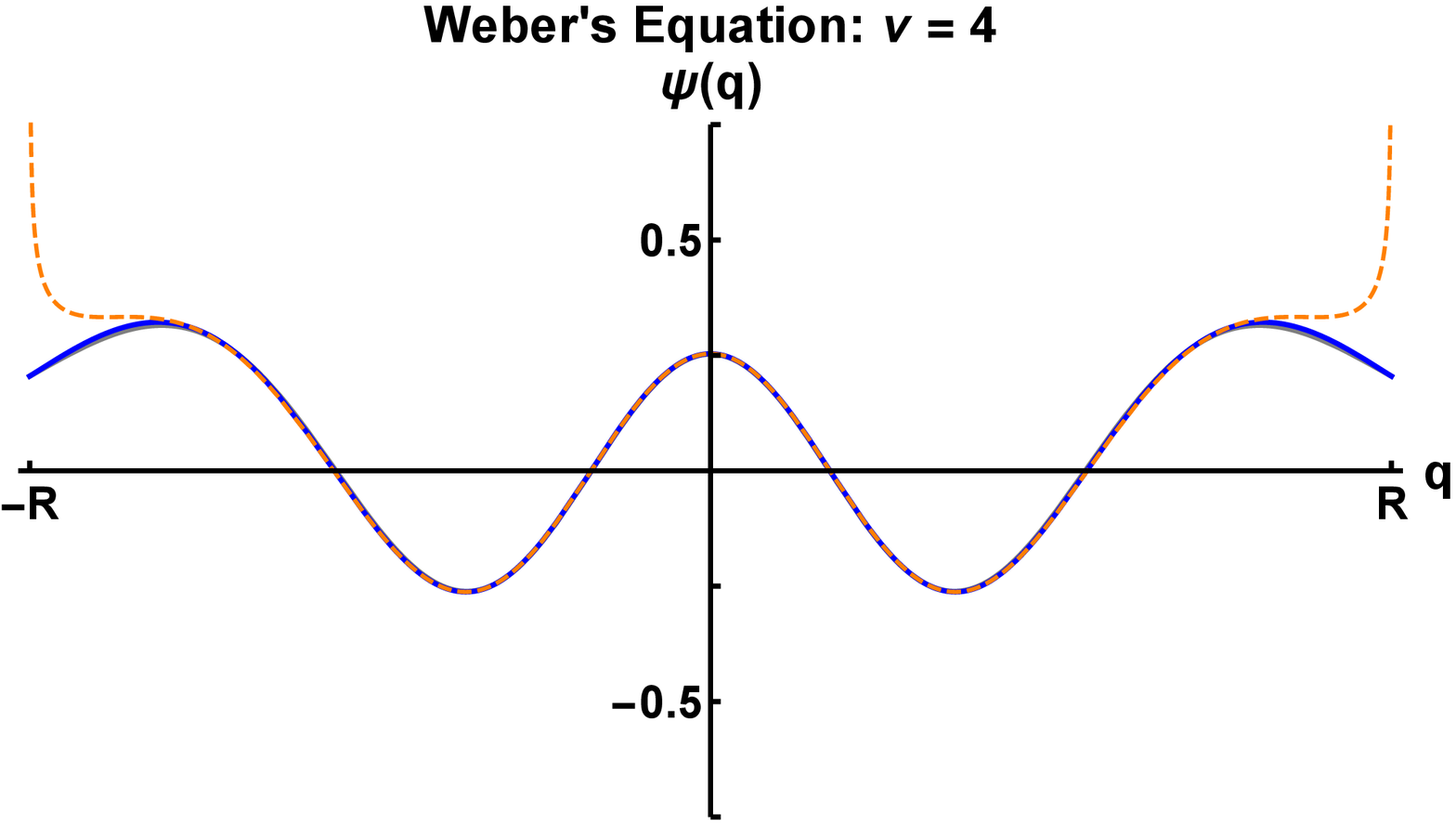}
        \put(5,5){\textbf{\large(c)}}
    \end{overpic}
    \hspace{1mm}
    \begin{overpic}[width=0.46\linewidth]{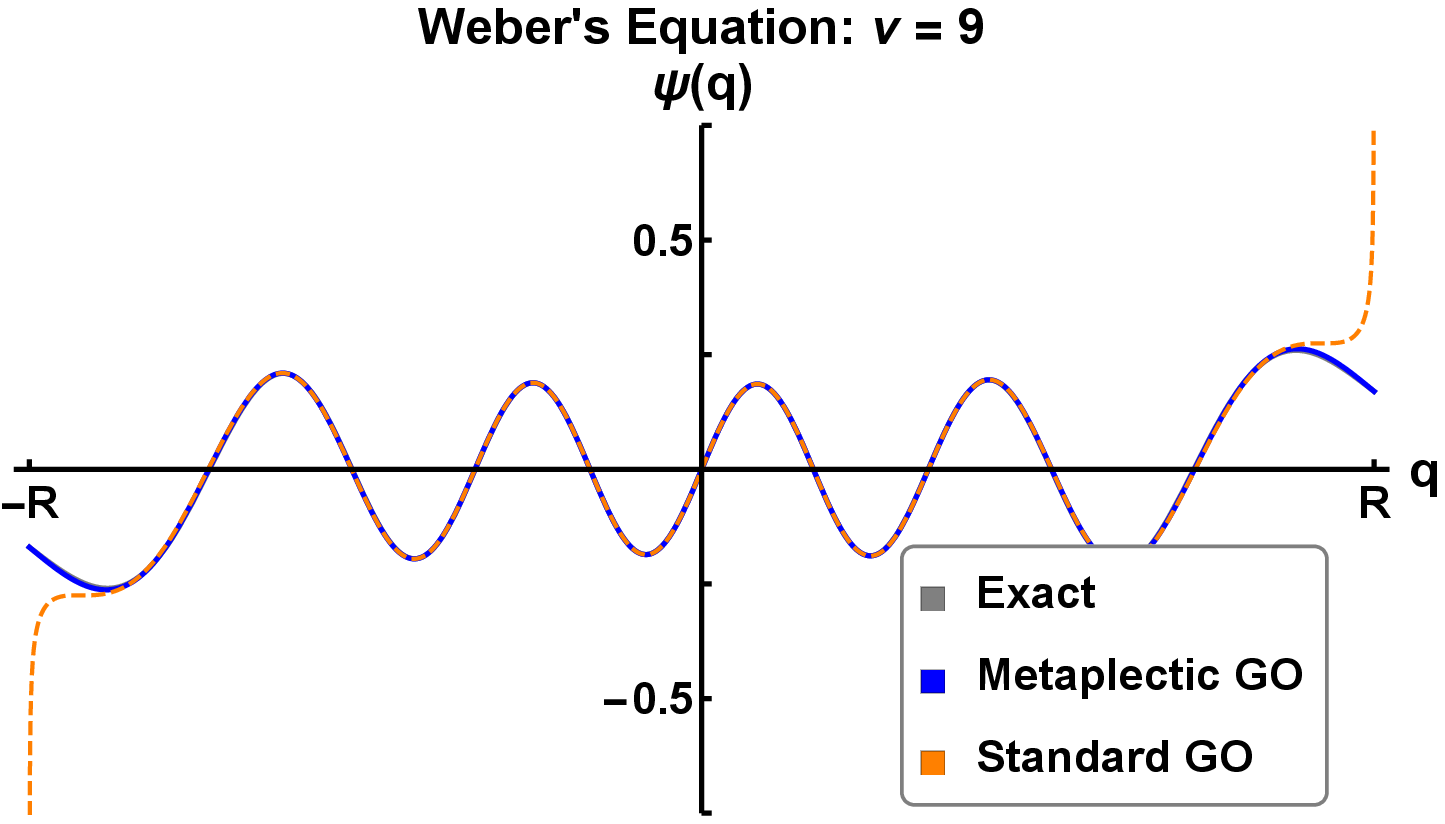}
        \put(5,5){\textbf{\large(d)}}
    \end{overpic}
    \caption{Comparison between the parabolic cylinder function [\Eq{eq:parabEXACT}], its standard GO approximation [\Eq{eq:parabWKB}], and its MGO approximation [\Eq{eq:parabMGO}] for $\nu = 0$, $\nu = 1$, $\nu = 4$, and $\nu = 9$. Two caustics are located at $\pm R$, where $R \doteq \sqrt{2 \nu + 1}$.}
    \label{fig:parabSOL}
\end{figure*}

\noindent where we have defined
\begin{subequations}
    \begin{align}
        \varrho(q) &\doteq \frac{ R^{2/3} \sqrt{R^2 - q^2} }{2^{1/3}q}
        \, , \\
        \varpi(q) &\doteq \frac{q\sqrt{R^2 - q^2}}{2} 
        - \frac{R^2 \cos^{-1} \left(\frac{q}{R} \right)}{2}
        \nonumber\\
        &\hspace{4mm}+ \frac{2}{3}\varrho^3(q) 
        + \frac{\pi}{4} \left[1 - \text{sign}(q) \right]
        \, .
    \end{align}
\end{subequations}

Figure~\ref{fig:parabSOL} compares \Eq{eq:parabMGO} with the exact solution for the boundary condition $\psi(R) = \textrm{Ai}(0)/\sqrt{R}$,
\begin{equation}
    \psi_\text{exact}(q) = \frac{\textrm{Ai}(0)}{\sqrt{R}} \, 
    \frac{
        \textrm{D}_{\nu} (\sqrt{2} \, q)
    }{
        \textrm{D}_\nu(\sqrt{2}\,R) 
    } \, ,
    \label{eq:parabEXACT}
\end{equation}

\noindent where $\textrm{D}_\nu(x)$ is Whittaker's parabolic cylinder function~\cite{Olver10}, and with the standard GO approximation~\cite{Grunwald71},
\begin{equation}
    \psi_\text{GO} = \frac{2^{1/6} \cos\left[ \frac{q}{2}\sqrt{R^2 - q^2} - \frac{R^2 }{2}\cos^{-1}\left(\frac{q}{R} \right)  + \frac{\pi}{4} \right]}{\sqrt{\pi} \, R^{1/3} (R^2 - q^2)^{1/4}} 
    \, .
    \label{eq:parabWKB}
\end{equation}

\noindent Already for $\nu = 0$, our MGO solution generally captures the exact solution behavior, albeit with some small error. Beginning from $\nu = 1$, the agreement with the exact solution becomes even more remarkable, for either parity. Just like the previous example, our solution remains finite everywhere, whereas the standard GO solution becomes singular near the cutoffs at $q = \pm R$. 


\section{Conclusion}
\label{sec:conclusion}

In this work, we develop a reformulation of GO, called `metaplectic GO' or MGO, that is well-behaved near caustics and can be applied to any linear wave equation. MGO uses sequenced MTs to rotate the phase space continually along a ray such that caustics are never encountered. For each point on the dispersion manifold, (i) the phase space is rotated to align configuration space with the local tangent plane, (ii) GO is applied in the rotated phase space, (iii) the GO solution $\Psi(\Vect{Q})$ is linked to previous and subsequent GO calculations via NIMTs to ensure continuity, and (iv) $\Psi(\Vect{Q})$ is transformed back to the original phase space using an MT, summing over distinct branches of the dispersion manifold if applicable. This procedure should also be suitable for quasioptical modeling when generalized to non-Euclidean ray-based coordinates.

Our auxiliary results include: (i) an explicit construction of the rotation matrix for the tangent plane based on Gram--Schmidt orthogonalization with symplectic modifications; (ii) the GO equations for an arbitrarily rotated phase space (and more generally, a phase space obtained by an arbitrary linear symplectic transformation); and (iii) a simplified version of the MT to obtain $\psi(\Vect{q})$ from $\Psi(\Vect{Q})$ using the stationary phase approximation. This final result is restricted to $\det{\Mat{B}} \neq 0$, which prohibits its use to analyze wave propagation in homogeneous media; however, a generalization that would allow for $\det B = 0$ is straightforward and will be discussed in a future publication. We present two examples of 1-D linear wave problems and show analytically how MGO successfully approximates the exact wave dynamics. Based on these promising analytical results, future work will explore a computational implementation of MGO.


\section*{Acknowledgements}

The work was supported by the U.S. DOE through Contract No.~DE-AC02-09CH11466.


\appendix


\section{Weyl symbol calculus}
\label{app:Weyl}

Here, we briefly review some aspects of the Weyl symbol calculus. The Weyl transform along with its inverse, the Wigner transform, provide a mapping between operators on a Hilbert space and functions on the corresponding phase space. Equations \eq{eq:Weyl} and \eq{eq:Wigner} present this mapping explicitly, which we shall repeat here for convenience. We shall only consider the mapping between scalar functions and scalar operators; matrix-valued functions and matrix-valued operators can be transformed elementwise using the scalar formulae. Also, we assume all functions are square integrable, and all operators are Hilbert--Schmidt normalizable.

Let $\Symb{A}(\Vect{z})$ be a phase-space function. The Weyl transform maps $\Symb{A}(\Vect{z})$ to the Hilbert-space operator
\begin{align}
    \Weyl \left[ \Symb{A}(\Vect{z}) \right]
    \doteq
    \int \frac{\dd \Vect{z}' \, \dd \Vect{\zeta}}{(2\pi)^{2N}} \, 
    \Symb{A}(\Vect{z}') 
    e^{-i\Vect{\zeta}^\intercal \Mat{J} \Vect{z}'} 
    e^{i\Vect{\zeta}^\intercal \Mat{J} \VectOp{z}} \, ,
\end{align}

\noindent where both integrals are taken over phase space. Similarly, let $\oper{A}$ be a Hilbert-space operator. The Wigner transform maps $\oper{A}$ to the phase-space function
\begin{equation}
    \WeylInv \left[\oper{A} \right] 
    \doteq \int \frac{\dd \Vect{z}'}{(2\pi)^N} \, e^{i\left(\Vect{z}'\right)^\intercal \Mat{J} \Vect{z}} \, \Tr \left[
    e^{-i\left(\Vect{z}'\right)^\intercal \Mat{J} \VectOp{z}}
    \oper{A}\right] \, ,
    \label{eq:wignerTRACE}
\end{equation}

\noindent where $\Tr$ is the matrix trace, and the integral is taken over phase space. If the $\Vect{q}$-space matrix elements of $\oper{A}$ are known, then the Wigner transform is also given by
\begin{equation}
    \WeylInv \left[\oper{A} \right] 
    \doteq \int \dd \Vect{s} \, e^{i\Vect{p}^\intercal \Vect{s}} 
    \bra{\Vect{q}-\Vect{s}/2} \oper{A} \ket{\Vect{q} + \Vect{s}/2} \, ,
    \label{eq:wignerKERN}
\end{equation}

\noindent where the integral is now taken over $\Vect{q}$-space. Note that \Eq{eq:wignerKERN} is derived from \Eq{eq:wignerTRACE} using the $\Vect{q}$-space matrix elements~\cite{Littlejohn86a}
\begin{equation}
    \bra{\Vect{q}}e^{-i\left(\Vect{z}'\right)^\intercal \Mat{J} \VectOp{z}} \ket{\Vect{q}''} 
    = e^{\frac{i}{2}(\Vect{p}')^\intercal (\Vect{q}+\Vect{q}'')}
    \delta(\Vect{q} - \Vect{q}' - \Vect{q}'' ) \, .
\end{equation}

Here, we summarize some important properties of the Weyl transform:%
\footnote{For proofs, see \Ref{Pool66}. For some extensions to non-Euclidean coordinates, see also the Supplementary Material in \Ref{Dodin19}.}:%
\begin{subequations}
    \label{eq:weylPROPS}
    \begin{align}
        \Weyl\left[\Symb{A}^*(\Vect{z}) \right] &= \Weyl\left[\Symb{A}(\Vect{z}) \right]^\dagger \, , \\
        \Weyl \left[ \alpha \Symb{A}(\Vect{z}) + \beta \Symb{B}(\Vect{z}) \right] &= \alpha \Weyl \left[\Symb{A}(\Vect{z}) \right] + \beta \Weyl \left[\Symb{B}(\Vect{z}) \right]\, , \\
        \|\Weyl\left[\Symb{A}(\Vect{z}) \right] \|_\text{HS} &= \left(2\pi\right)^{-N} \| \Symb{A}(\Vect{z}) \|_{L_2} \, ,
    \end{align}
\end{subequations}

\noindent where $\| \cdot \|_\text{HS}$ is the Hilbert--Schmidt norm on the space of operators, defined as
\begin{equation}
    \|\Weyl\left[\Symb{A}(\Vect{z}) \right] \|_\text{HS} \doteq \Tr\left\{ \Weyl\left[\Symb{A}^*(\Vect{z}) \right] \Weyl\left[\Symb{A}(\Vect{z}) \right] \right\} \, ,
\end{equation}

\noindent and $\| \cdot \|_{L_2}$ is the $L_2$ norm on the space of functions. Also, 
\begin{equation}
    \WeylInv[\oper{A}\oper{B}] = \Symb{A}(\Vect{z}) \star \Symb{B}(\Vect{z}) \, ,
    \label{eq:weylMOYAL}
\end{equation}

\noindent where $\star$ denotes the Moyal star product. The Moyal star product is an associative non-commutative product rule for phase-space functions given explicitly as
\begin{align}
    \Symb{A}(\Vect{z}) \star \Symb{B}(\Vect{z}) 
    &= \left. 
        \sum_{s = 0}^\infty 
        \frac{\left( \frac{i}{2} \pd_\Vect{z}^\intercal \, \Mat{J} \, \pd_\Vect{\zeta} \right)^{s}}
        {s!}  
        \Symb{A}(\Vect{z}) \Symb{B}(\Vect{\zeta}) 
    \right|_{\Vect{\zeta} = \Vect{z}} \nonumber\\
    &= \Symb{A}(\Vect{z}) \Symb{B}(\Vect{z}) + \frac{i}{2} \left\{ \Symb{A}(\Vect{z}), \Symb{B}(\Vect{z}) \right\} + \ldots \, ,
    \label{eq:moyal}
\end{align}

\noindent where we have introduced the Poisson bracket
\begin{equation}
     \left\{ \Symb{A}(\Vect{z}), \Symb{B}(\Vect{z}) \right\} = 
     \left[ \pd_\Vect{z} \Symb{A}(\Vect{z}) \right]^\intercal \Mat{J} \, \pd_\Vect{z} \Symb{B}(\Vect{z}) \, ,
\end{equation}

\noindent with $\Mat{J}$ provided by \Eq{eq:jMAT}.

Importantly, \Eqs{eq:weylPROPS} imply that the Weyl transform preserves both hermiticity and locality: the Weyl symbol of a hermitian operator is a function that is purely real, and the Weyl transforms of two functions which are `close' to each other in the function space will also be `close' in the operator space. Both of these properties make the Weyl symbol calculus an attractive means to approximate wave equations.

The following are some relevant Weyl transforms:
\begin{subequations}
    \label{eq:weylEX}
    \begin{align}
        \label{eq:weylQ}
        \Weyl\left[ f(\Vect{q}) \right] &= f(\VectOp{q}) \, , \\
        \label{eq:weylP}
        \Weyl\left[ f(\Vect{p}) \right] &= f(\VectOp{p}) \, , \\
        \Weyl\left[ \Vect{v}(\Vect{q})^\intercal \Vect{p} \right] &= \Vect{v}(\VectOp{q})^\intercal \VectOp{p} - \frac{i}{2} \pd_\Vect{q} \cdot \Vect{v}(\VectOp{q}) \, , \\
        \Weyl\left[ \Vect{p}^\intercal \Mat{M}(\Vect{q}) \Vect{p} \right] &= \Mat{M}(\VectOp{q}) \dubdot \VectOp{p} \VectOp{p} - i \left[ \pd_\Vect{q} \cdot \Mat{M}(\VectOp{q}) \right]^\intercal \VectOp{p} \nonumber\\
        &\hspace{18mm} - \frac{1}{4} \pd^2_{\Vect{q}\Vect{q}} \dubdot \Mat{M}(\VectOp{q}) \, ,
    \end{align}
\end{subequations}

\noindent where $\dubdot$ denotes the double contraction.


\section{Approximating the envelope equation in the GO limit via the Weyl symbol calculus}
\label{app:GO}

Here, we derive \Eq{eq:approxENV} from \Eq{eq:hilbertENV} using the Weyl symbol calculus. (See also \Refs{Dodin19,McDonald88a}.) To obtain the GO envelope operator, we shall (i) calculate the Weyl symbol of the envelope operator, (ii) approximate the Weyl symbol in the GO limit, and (iii) take the Weyl transform of the GO Weyl symbol.

Using \Eqs{eq:weylMOYAL} and \eq{eq:weylQ}, one obtains
\begin{equation}
    \WeylInv\left[ e^{-i \theta(\VectOp{q})}
    \oper{D}(\VectOp{z})
    e^{i \theta(\VectOp{q})} \right] = e^{-i \theta(\Vect{q})} \star \Symb{D}(\Vect{z}) \star e^{i \theta(\Vect{q})} \, ,
\end{equation}

\noindent where $\Symb{D}(\Vect{z}) \doteq \WeylInv\left[\oper{D}(\VectOp{z}) \right]$. Since $e^{i \theta(\Vect{q})}$ is independent of $\Vect{p}$, one readily finds using \Eq{eq:moyal} that
\begin{equation}
    \Symb{D}(\Vect{z}) \star e^{ i\theta(\Vect{q})} = 
    \left.
    \sum_{s = 0}^\infty 
    \frac{1}{s!} \left(
    \frac{\pd_{\Vect{p}}^\intercal \pd_{\Vect{\varsigma}} }
    {2 i \kappa} 
    \right)^s 
    \Symb{D}(\Vect{z}) e^{i \kappa \theta(\Vect{\varsigma})} 
    \right|_{\Vect{\varsigma} = \Vect{q}} \, .
\end{equation}

\noindent Here, $\kappa$ is a dimensionless wavenumber that has been formally introduced to elucidate the GO ordering, where $\kappa \to \infty$. We shall keep terms up to $O(\kappa^{-2})$, rather than $O(\kappa^{-1})$ as is traditionally done, to demonstrate the ease with which `full-wave' effects such as diffraction can be included into reduced wave models.

Let us consider $1$-D for simplicity. (The $N$-D case is analogous.) Using Faa di Bruno's formula~\cite{Comtet74},
\begin{equation}
    \pd^s_\varsigma e^{i \kappa \theta(\varsigma)} = e^{i \kappa \theta(\varsigma)} \sum_{j = 1}^s (i \kappa)^j B_{s,j}\left[\pd_\varsigma \theta(\varsigma), \ldots, \pd^{s - j + 1}_\varsigma \theta(\varsigma) \right] \, ,
\end{equation}

\noindent where $B_{s,j}$ are the incomplete, or partial, Bell polynomials. Some important Bell polynomials are
\begin{subequations}
    \begin{align}
        B_{s,s}(f_1) &= (f_1)^s \, , \\
        B_{s, s - 1}(f_1, f_2) &= \binom{s}{2} (f_1)^{s-2}f_2 \, , \\
        B_{s, s - 2}(f_1, f_2, f_3) &= \binom{s}{3}(f_1)^{s-3} f_3 \nonumber\\
        &\hspace{5mm} + 3\binom{s}{4} (f_1)^{s-4} (f_2)^2 \, ,
    \end{align}
\end{subequations}

\noindent where $\binom{n}{m} \doteq \frac{n!}{m!(n-m)!}$. Hence, to $O(\kappa^{-2})$,
\begin{align}
     \Symb{D}(\Vect{z}) \star e^{i \theta(q)} 
     &\approx e^{i \kappa \theta(q)} \left\{ \sum_{s = 0}^\infty \frac{ \left[\theta'(q) /2\right]^s}{s!} \pd^s_p \Symb{D}(\Vect{z}) \right.\nonumber\\
     & - \frac{i \theta''(q)}{8\kappa} \sum_{s = 2}^\infty \frac{ \left[ \theta'(q) /2\right]^{s-2}}{(s-2)!} 
     \pd^s_p \Symb{D}(\Vect{z}) \nonumber\\
     & - \frac{\theta'''(q)}{48 \kappa^2} \sum_{s = 3}^\infty \frac{ \left[ \theta'(q) /2\right]^{s-3}}{(s-3)!} 
     \pd^s_p \Symb{D}(\Vect{z}) \nonumber\\
     &\left. - \frac{\left[\theta''(q) \right]^2}{128 \kappa^2} \sum_{s = 4}^\infty \frac{ \left[ \theta'(q) /2\right]^{s-4}}{(s-4)!} 
     \pd^s_p \Symb{D}(\Vect{z}) \right\} \, ,
\end{align}

\noindent where $\theta'(q) \doteq \pd_q \theta(q)$, since $\theta(q)$ is univariate. 

Upon shifting the indices back to $s = 0$, we obtain
\begin{align}
     \Symb{D}(\Vect{z}) \star e^{i\theta(q)} 
     &\approx e^{i \kappa \theta(q)} \left\{ \sum_{s = 0}^\infty \frac{ \left[\theta'(q) /2\right]^s}{s!} \pd^s_p \Symb{D}(\Vect{z}) \right.\nonumber\\
     & - \frac{i \theta''(q)}{8\kappa} \sum_{s = 0}^\infty \frac{ \left[ \theta'(q) /2\right]^s}{s!} 
     \pd^{s+2}_p \Symb{D}(\Vect{z}) \nonumber\\
     & - \frac{\theta'''(q)}{48 \kappa^2} \sum_{s = 0}^\infty \frac{ \left[ \theta'(q) /2\right]^s}{s!} 
     \pd^{s+3}_p \Symb{D}(\Vect{z}) \nonumber\\
     &\left. - \frac{\left[\theta''(q) \right]^2}{128 \kappa^2} \sum_{s = 0}^\infty \frac{ \left[ \theta'(q) /2\right]^s}{s!} 
     \pd^{s+4}_p \Symb{D}(\Vect{z}) \right\} \, .
\end{align}

\noindent We recognize these summations as the Taylor expansions of $\Symb{D}$, $\pd^2_p \Symb{D}$, $\pd^3_p \Symb{D}$, and $\pd^4_p \Symb{D}$ about $p + \theta'/2$. Therefore,
\begin{align}
    \Symb{D}(\Vect{z}) \star e^{i \theta(q)} &\approx
    e^{i \kappa \theta(q)} \left\{ \Symb{D}\left[q, p + \frac{\theta'(q)}{2} \right] \right. \nonumber\\
    &- \frac{i \theta''(q)}{8 \kappa} \pd^2_p \Symb{D}\left[q, p + \frac{\theta'(q)}{2} \right] \nonumber\\
    &- \frac{\theta'''(q)}{48 \kappa^2} \pd^3_p \Symb{D}\left[q, p + \frac{\theta'(q)}{2} \right] \nonumber\\
    &\left. - \frac{\left[\theta''(q) \right]^2}{128 \kappa^2} \pd^4_p \Symb{D}\left[q, p + \frac{\theta'(q)}{2} \right] 
    \right\} \, .
\end{align}

\noindent A similar calculation will show that
\begin{align}
    e^{-i \theta(q)} \star \Symb{D}(\Vect{z}) \star e^{i \theta(q)}
    &= e^{-i \theta(q)} \star \left[ \Symb{D}(\Vect{z}) \star e^{i \theta(q)} \right] \nonumber\\
    &\approx \Symb{D}\left[q, p + \theta'(q) \right] \nonumber\\
    &- \frac{\theta'''(q)}{24 \kappa^2} \pd^3_p \Symb{D}\left[q, p + \theta'(q) \right] \, .
    \label{eq:symbTRANS1D}
\end{align}

\noindent In multiple dimensions, \Eq{eq:symbTRANS1D} readily generalizes to
\begin{equation}
    e^{-i \theta(\Vect{q})} \star \Symb{D}(\Vect{z}) \star e^{i \theta(\Vect{q})}
    \approx \QoSymb\left[ \Vect{q}, \Vect{p} + \pd_\Vect{q} \theta(\Vect{q}) \right] \, ,
\end{equation}

\noindent where
\begin{equation}
    \QoSymb(\Vect{z}) \doteq
    \Symb{D}(\Vect{z}) - \frac{1}{24 \kappa^2} \pd^3_{\Vect{q}\Vect{q}\Vect{q} } \theta(\Vect{q}) \, \tripdot \, \pd^3_{\Vect{p}\Vect{p}\Vect{p} } \Symb{D}( \Vect{z} )
\end{equation}

\noindent is the correction to $\Symb{D}(\Vect{z})$ found in \Ref{Dodin19}. Here, $\tripdot$ denotes the triple contraction.

Since $\Weyl\left[\Vect{p} \right] = \VectOp{p} \sim \kappa^{-1} \pd_\Vect{q}$ on $\Vect{q}$-space, a power series expansion of $\QoSymb\left[\Vect{q}, \Vect{p} + \pd_\Vect{q} \theta(\Vect{q}) \right]$ in $\kappa^{-1}$ is equivalent to a power series expansion in $\Vect{p}$. Hence,
\begin{align}
    \hspace{-1mm} \QoSymb\left[\Vect{q}, \Vect{p} + \pd_\Vect{q} \theta(\Vect{q}) \right] 
    &\approx \QoSymb\left[\Vect{q}, \pd_\Vect{q} \theta (\Vect{q}) \right] \nonumber\\
    &+ \frac{1}{\kappa} \Vect{v}(\Vect{q})^\intercal \Vect{p}
    + \frac{1}{2 \kappa^2} \Vect{p}^\intercal \Mat{m}(\Vect{q}) \Vect{p} \, ,
    \label{eq:approxSYMB}
\end{align}

\noindent where we have defined
\begin{subequations}
    \begin{align}
        \Vect{v}(\Vect{q}) &\doteq \left. 
            \pd_\Vect{p} \QoSymb( \Vect{z} )
        \right|_{\Vect{p} = \pd_\Vect{q} \theta (\Vect{q})} \, , \\
        \Mat{m}(\Vect{q}) &\doteq \left.
            \pd^2_{\Vect{p} \Vect{p}} \QoSymb( \Vect{z} )
        \right|_{\Vect{p} = \pd_\Vect{q} \theta (\Vect{q})} \, .
    \end{align}
\end{subequations}

\noindent Finally, taking the Weyl transform of \Eq{eq:approxSYMB} using \Eqs{eq:weylEX} yields the reduced envelope operator
\begin{align}
    & \Weyl\left[e^{-i\theta(\Vect{q})} \star \Symb{D}(\Vect{z}) \star e^{i\theta(\Vect{q})} \right] 
    \approx \nonumber\\
    &\QoSymb\left[ \VectOp{q}, \pd_\Vect{q} \theta(\VectOp{q}) \right] 
    + \frac{\Vect{v}(\VectOp{q})^\intercal \VectOp{p} - \frac{i}{2} \pd_\Vect{q} \cdot \Vect{v}(\VectOp{q}) }
    {\kappa} \nonumber\\
    &+ \frac{\Mat{m}(\VectOp{q}) \dubdot \VectOp{p} \VectOp{p} 
        - i \left[ \pd_\Vect{q} \cdot \Mat{m}(\VectOp{q}) \right]^\intercal \VectOp{p}
        - \frac{1}{4} \pd^2_{\Vect{q}\Vect{q}} \dubdot \Mat{m}(\VectOp{q})}
    {2\kappa^2} \, .
    \label{eq:approxOP}
\end{align}

Section \ref{sec:raysCAUSTICS} only considers the lowest order GO approximation. Hence, \Eq{eq:approxENV} is obtained from \Eq{eq:approxOP} by dropping all $O(\kappa^{-2})$ terms. Also, $\pd^3_{\Vect{p}\Vect{p}\Vect{p}} \Symb{D}(\Vect{z})$ and $\pd^2_{\Vect{q}\Vect{q}} \dubdot \Mat{m}(\Vect{q})$ are often negligibly small. (Both are identically zero for the Helmholtz equation, for example.) We therefore drop these two terms in obtaining \Eq{eq:deltaQO}.


\section{Symplectic covariance of the Weyl symbol}
\label{app:metWEYL}

Here we demonstrate the symplectic covariance of the Weyl symbol. Consider some operator $\oper{D}$ with symbol
\begin{equation}
    \Symb{D}(\Vect{z}) = 
    \int \frac{\dd \Vect{z}'}{(2\pi)^N} \, e^{i\left(\Vect{z}'\right)^\intercal \Mat{J} \Vect{z}} \, \Tr \left[
    e^{-i\left(\Vect{z}'\right)^\intercal \Mat{J} \VectOp{z}}
    \oper{D}\right] \, .
\end{equation}

\noindent Correspondingly, the symbol of $\oper{M}^\dagger \oper{D} \oper{M}$ is 
\begin{align}
    \WeylInv\left[ 
        \oper{M}^\dagger 
        \oper{D}
        \oper{M} 
    \right]
    &= \int \frac{\dd \Vect{z}'}{(2\pi)^N} \, e^{i\left(\Vect{z}'\right)^\intercal \Mat{J} \Vect{z}} \, \nonumber\\
    &\hspace{3mm}\times \Tr \left[
        e^{-i\left(\Vect{z}'\right)^\intercal \Mat{J} \VectOp{z}}
        \oper{M}^\dagger \oper{D} \oper{M}
    \right] \, .
\end{align}

\noindent Since $\oper{M}$ is unitary,
\begin{align}
    \Tr \left[
        e^{-i\left(\Vect{z}'\right)^\intercal \Mat{J} \VectOp{z}}
        \oper{M}^\dagger \oper{D} \oper{M} 
    \right]
    &= \Tr \left[
        \oper{M}
        e^{-i\left(\Vect{z}'\right)^\intercal \Mat{J} \VectOp{z}}
        \oper{M}^\dagger \oper{D} 
    \right] \nonumber \\
    &= \Tr \left[
        e^{-i\left(\Vect{z}'\right)^\intercal \Mat{J} \oper{M} \VectOp{z} \oper{M}^\dagger}
        \oper{D} 
    \right] \nonumber \\
    &= \Tr \left[
        e^{-i\left(\Vect{z}'\right)^\intercal \Mat{J} \Mat{S}^{-1} \VectOp{z}}
        \oper{D} 
    \right] \, .
\end{align}

\noindent Since $\Mat{S}$ is symplectic,
\begin{equation}
    \left(\Vect{z}'\right)^\intercal \Mat{J} \Mat{S}^{-1} \VectOp{z}
    = 
    \left(\Vect{z}'\right)^\intercal \Mat{S}^\intercal \left( \Mat{S}^{-1} \right)^\intercal \Mat{J} \Mat{S}^{-1} \VectOp{z}
    = \left(\Mat{S} \Vect{z}'\right)^\intercal \Mat{J} \, \VectOp{z} \, .
\end{equation}

\noindent Hence, after making the variable substitution $\Vect{\zeta} \doteq \Mat{S}\Vect{z}'$,
\begin{align}
    \hspace{-1mm}\WeylInv\left[ 
        \oper{M}^\dagger 
        \oper{D}
        \oper{M} 
    \right]
    &= \int \frac{\dd \Vect{\zeta}}{(2\pi)^N} \, e^{i\left(\Mat{S}^{-1} \Vect{\zeta}\right)^\intercal \Mat{J} \Vect{z}} \, 
    \Tr \left(
        e^{-i\Vect{\zeta}^\intercal \Mat{J} \VectOp{z}}
        \oper{D}
    \right) \nonumber \\
    &= \int \frac{\dd \Vect{\zeta}}{(2\pi)^N} \, e^{i\Vect{\zeta}^\intercal 
    \Mat{J} \,  
    \Mat{S} \, \Vect{z}} \,  
    \Tr \left(
        e^{-i\Vect{\zeta}^\intercal \Mat{J} \VectOp{z}}
        \oper{D}
    \right) \, ,
\end{align}

\noindent where we have used \Eq{eq:symplecS} and also that $\det \Mat{S} = 1$ for any symplectic matrix. We therefore obtain
\begin{equation}
    \WeylInv\left[ 
        \oper{M}^\dagger \oper{D} \oper{M} 
    \right]
    = \Symb{D}\left( \Mat{S} \Vect{z} \right)
    = \Symb{D}\left( \Stroke{\Vect{Z}} \right) \, ,
\end{equation}

\noindent using \Eq{eq:symplecZ}. Similarly, 
\begin{align}
    &\WeylInv\left[ 
        \oper{M}
        \oper{M}^\dagger 
        \oper{D} 
        \oper{M} 
        \oper{M}^\dagger
    \right]
    \nonumber\\
    &= \int \frac{\dd \Vect{\zeta}}{(2\pi)^N} \, e^{ i\Vect{\zeta}^\intercal \Mat{J} \, \Stroke{\Vect{Z}} } \,
    \Tr \left[
        e^{-i\Vect{\zeta}^\intercal \Mat{J} \VectOp{z}}
        \oper{M}
        \oper{D}
        \oper{M}^\dagger
    \right] \nonumber \\
    &= \int \frac{\dd \Vect{z}'}{(2\pi)^N} \, e^{ i\left( \Vect{z}'\right)^\intercal \Mat{J} \Mat{S}^{-1} \, \Stroke{\Vect{Z}} } \,
    \Tr \left[
        e^{-i\left(\Vect{z}' \right)^\intercal \Mat{J} 
        \VectOp{z}}
        \oper{D}
    \right] \, .
\end{align}

\noindent Therefore,
\begin{equation}
    \WeylInv\left[ 
        \oper{M}
        \oper{M}^\dagger 
        \oper{D} 
        \oper{M} 
        \oper{M}^\dagger
    \right]
    = \Symb{D}\left(\Mat{S}^{-1} \Stroke{\Vect{Z}} \right) = \Symb{D}(\Vect{z}) \, .
\end{equation}

\bibliography{Biblio.bib}

\begin{thebibliography}{43}%
\makeatletter
\providecommand \@ifxundefined [1]{%
 \@ifx{#1\undefined}
}%
\providecommand \@ifnum [1]{%
 \ifnum #1\expandafter \@firstoftwo
 \else \expandafter \@secondoftwo
 \fi
}%
\providecommand \@ifx [1]{%
 \ifx #1\expandafter \@firstoftwo
 \else \expandafter \@secondoftwo
 \fi
}%
\providecommand \natexlab [1]{#1}%
\providecommand \enquote  [1]{``#1''}%
\providecommand \bibnamefont  [1]{#1}%
\providecommand \bibfnamefont [1]{#1}%
\providecommand \citenamefont [1]{#1}%
\providecommand \href@noop [0]{\@secondoftwo}%
\providecommand \href [0]{\begingroup \@sanitize@url \@href}%
\providecommand \@href[1]{\@@startlink{#1}\@@href}%
\providecommand \@@href[1]{\endgroup#1\@@endlink}%
\providecommand \@sanitize@url [0]{\catcode `\\12\catcode `\$12\catcode
  `\&12\catcode `\#12\catcode `\^12\catcode `\_12\catcode `\%12\relax}%
\providecommand \@@startlink[1]{}%
\providecommand \@@endlink[0]{}%
\providecommand \url  [0]{\begingroup\@sanitize@url \@url }%
\providecommand \@url [1]{\endgroup\@href {#1}{\urlprefix }}%
\providecommand \urlprefix  [0]{URL }%
\providecommand \Eprint [0]{\href }%
\providecommand \doibase [0]{http://dx.doi.org/}%
\providecommand \selectlanguage [0]{\@gobble}%
\providecommand \bibinfo  [0]{\@secondoftwo}%
\providecommand \bibfield  [0]{\@secondoftwo}%
\providecommand \translation [1]{[#1]}%
\providecommand \BibitemOpen [0]{}%
\providecommand \bibitemStop [0]{}%
\providecommand \bibitemNoStop [0]{.\EOS\space}%
\providecommand \EOS [0]{\spacefactor3000\relax}%
\providecommand \BibitemShut  [1]{\csname bibitem#1\endcsname}%
\let\auto@bib@innerbib\@empty
\bibitem [{\citenamefont {Landau}\ and\ \citenamefont
  {Lifshitz}(1981)}]{Landau81}%
  \BibitemOpen
  \bibfield  {author} {\bibinfo {author} {\bibfnamefont {L.~D.}\ \bibnamefont
  {Landau}}\ and\ \bibinfo {author} {\bibfnamefont {E.~M.}\ \bibnamefont
  {Lifshitz}},\ }\href@noop {} {\emph {\bibinfo {title} {Quantum Mechanics:
  Non-relativistic Theory}}},\ \bibinfo {edition} {3rd}\ ed.\ (\bibinfo
  {publisher} {London: Pergamon},\ \bibinfo {year} {1981})\BibitemShut
  {NoStop}%
\bibitem [{\citenamefont {Kravtsov}\ and\ \citenamefont
  {Orlov}(1990)}]{Kravtsov90}%
  \BibitemOpen
  \bibfield  {author} {\bibinfo {author} {\bibfnamefont {Y.~A.}\ \bibnamefont
  {Kravtsov}}\ and\ \bibinfo {author} {\bibfnamefont {Y.~I.}\ \bibnamefont
  {Orlov}},\ }\href@noop {} {\emph {\bibinfo {title} {Geometrical Optics of
  Inhomogeneous Media}}}\ (\bibinfo  {publisher} {Berlin: Springer},\ \bibinfo
  {year} {1990})\BibitemShut {NoStop}%
\bibitem [{\citenamefont {Shankar}(1994)}]{Shankar94}%
  \BibitemOpen
  \bibfield  {author} {\bibinfo {author} {\bibfnamefont {R.}~\bibnamefont
  {Shankar}},\ }\href {\doibase 10.1007/978-1-4757-0576-8} {\emph {\bibinfo
  {title} {Principles of Quantum Mechanics}}},\ \bibinfo {edition} {2nd}\ ed.\
  (\bibinfo  {publisher} {New York: Plenum},\ \bibinfo {year}
  {1994})\BibitemShut {NoStop}%
\bibitem [{\citenamefont {Tracy}\ \emph {et~al.}(2014)\citenamefont {Tracy},
  \citenamefont {Brizard}, \citenamefont {Richardson},\ and\ \citenamefont
  {Kaufman}}]{Tracy14}%
  \BibitemOpen
  \bibfield  {author} {\bibinfo {author} {\bibfnamefont {E.~R.}\ \bibnamefont
  {Tracy}}, \bibinfo {author} {\bibfnamefont {A.~J.}\ \bibnamefont {Brizard}},
  \bibinfo {author} {\bibfnamefont {A.~S.}\ \bibnamefont {Richardson}}, \ and\
  \bibinfo {author} {\bibfnamefont {A.~N.}\ \bibnamefont {Kaufman}},\ }\href
  {\doibase 10.1017/CBO9780511667565} {\emph {\bibinfo {title} {Ray Tracing and
  Beyond: Phase Space Methods in Plasma Wave Theory}}}\ (\bibinfo  {publisher}
  {Cambridge: Cambridge University Press},\ \bibinfo {year} {2014})\BibitemShut
  {NoStop}%
\bibitem [{\citenamefont {Ford}\ and\ \citenamefont {Wheeler}(1959)}]{Ford59}%
  \BibitemOpen
  \bibfield  {author} {\bibinfo {author} {\bibfnamefont {K.~W.}\ \bibnamefont
  {Ford}}\ and\ \bibinfo {author} {\bibfnamefont {J.~A.}\ \bibnamefont
  {Wheeler}},\ }\href {\doibase 10.1016/0003-4916(59)90026-0} {\bibfield
  {journal} {\bibinfo  {journal} {Ann. Phys.}\ }\textbf {\bibinfo {volume}
  {7}},\ \bibinfo {pages} {259} (\bibinfo {year} {1959})}\BibitemShut {NoStop}%
\bibitem [{\citenamefont {Adam}(2002)}]{Adam02}%
  \BibitemOpen
  \bibfield  {author} {\bibinfo {author} {\bibfnamefont {J.~A.}\ \bibnamefont
  {Adam}},\ }\href {\doibase 10.1016/S0370-1573(01)00076-X} {\bibfield
  {journal} {\bibinfo  {journal} {Phys. Rep.}\ }\textbf {\bibinfo {volume}
  {356}},\ \bibinfo {pages} {229} (\bibinfo {year} {2002})}\BibitemShut
  {NoStop}%
\bibitem [{\citenamefont {Jaun}\ \emph {et~al.}(2007)\citenamefont {Jaun},
  \citenamefont {Tracy},\ and\ \citenamefont {Kaufman}}]{Jaun07}%
  \BibitemOpen
  \bibfield  {author} {\bibinfo {author} {\bibfnamefont {A.}~\bibnamefont
  {Jaun}}, \bibinfo {author} {\bibfnamefont {E.~R.}\ \bibnamefont {Tracy}}, \
  and\ \bibinfo {author} {\bibfnamefont {A.~N.}\ \bibnamefont {Kaufman}},\
  }\href {\doibase 10.1088/0741-3335/49/1/004} {\bibfield  {journal} {\bibinfo
  {journal} {Plasma Phys. Control. Fusion}\ }\textbf {\bibinfo {volume} {49}},\
  \bibinfo {pages} {43} (\bibinfo {year} {2007})}\BibitemShut {NoStop}%
\bibitem [{\citenamefont {Richardson}\ \emph {et~al.}(2010)\citenamefont
  {Richardson}, \citenamefont {Bonoli},\ and\ \citenamefont
  {Wright}}]{Richardson10}%
  \BibitemOpen
  \bibfield  {author} {\bibinfo {author} {\bibfnamefont {A.~S.}\ \bibnamefont
  {Richardson}}, \bibinfo {author} {\bibfnamefont {P.~T.}\ \bibnamefont
  {Bonoli}}, \ and\ \bibinfo {author} {\bibfnamefont {J.~C.}\ \bibnamefont
  {Wright}},\ }\href {\doibase 10.1063/1.3400217} {\bibfield  {journal}
  {\bibinfo  {journal} {Phys. Plasmas}\ }\textbf {\bibinfo {volume} {17}},\
  \bibinfo {pages} {052107} (\bibinfo {year} {2010})}\BibitemShut {NoStop}%
\bibitem [{\citenamefont {Myatt}\ \emph {et~al.}(2017)\citenamefont {Myatt},
  \citenamefont {Follett}, \citenamefont {Shaw}, \citenamefont {Edgell},
  \citenamefont {Froula}, \citenamefont {Igumenshchev},\ and\ \citenamefont
  {Goncharov}}]{Myatt17}%
  \BibitemOpen
  \bibfield  {author} {\bibinfo {author} {\bibfnamefont {J.~F.}\ \bibnamefont
  {Myatt}}, \bibinfo {author} {\bibfnamefont {R.~K.}\ \bibnamefont {Follett}},
  \bibinfo {author} {\bibfnamefont {J.~G.}\ \bibnamefont {Shaw}}, \bibinfo
  {author} {\bibfnamefont {D.~H.}\ \bibnamefont {Edgell}}, \bibinfo {author}
  {\bibfnamefont {D.~G.}\ \bibnamefont {Froula}}, \bibinfo {author}
  {\bibfnamefont {I.~V.}\ \bibnamefont {Igumenshchev}}, \ and\ \bibinfo
  {author} {\bibfnamefont {V.~N.}\ \bibnamefont {Goncharov}},\ }\href {\doibase
  10.1063/1.4982059} {\bibfield  {journal} {\bibinfo  {journal} {Phys.
  Plasmas}\ }\textbf {\bibinfo {volume} {24}},\ \bibinfo {pages} {056308}
  (\bibinfo {year} {2017})}\BibitemShut {NoStop}%
\bibitem [{\citenamefont {Lopez}\ and\ \citenamefont {Poli}(2018)}]{Lopez18a}%
  \BibitemOpen
  \bibfield  {author} {\bibinfo {author} {\bibfnamefont {N.~A.}\ \bibnamefont
  {Lopez}}\ and\ \bibinfo {author} {\bibfnamefont {F.~M.}\ \bibnamefont
  {Poli}},\ }\href {\doibase 10.1088/1361-6587/aabaa8} {\bibfield  {journal}
  {\bibinfo  {journal} {Plasma Phys. Control. Fusion}\ }\textbf {\bibinfo
  {volume} {60}},\ \bibinfo {pages} {065007} (\bibinfo {year}
  {2018})}\BibitemShut {NoStop}%
\bibitem [{\citenamefont {Kravtsov}\ and\ \citenamefont
  {Orlov}(1993)}]{Kravtsov93}%
  \BibitemOpen
  \bibfield  {author} {\bibinfo {author} {\bibfnamefont {Y.~A.}\ \bibnamefont
  {Kravtsov}}\ and\ \bibinfo {author} {\bibfnamefont {Y.~I.}\ \bibnamefont
  {Orlov}},\ }\href {\doibase 10.1007/978-3-642-59887-6} {\emph {\bibinfo
  {title} {Caustics, Catastrophes and Wave Fields}}}\ (\bibinfo  {publisher}
  {Berlin: Springer},\ \bibinfo {year} {1993})\BibitemShut {NoStop}%
\bibitem [{\citenamefont {Ludwig}(1966)}]{Ludwig66}%
  \BibitemOpen
  \bibfield  {author} {\bibinfo {author} {\bibfnamefont {D.}~\bibnamefont
  {Ludwig}},\ }\href {\doibase 10.1002/cpa.3160190207} {\bibfield  {journal}
  {\bibinfo  {journal} {Commun. Pure Appl. Math.}\ }\textbf {\bibinfo {volume}
  {19}},\ \bibinfo {pages} {215} (\bibinfo {year} {1966})}\BibitemShut
  {NoStop}%
\bibitem [{\citenamefont {Berry}\ and\ \citenamefont {Mount}(1972)}]{Berry72}%
  \BibitemOpen
  \bibfield  {author} {\bibinfo {author} {\bibfnamefont {M.~V.}\ \bibnamefont
  {Berry}}\ and\ \bibinfo {author} {\bibfnamefont {K.~E.}\ \bibnamefont
  {Mount}},\ }\href {\doibase 10.1088/0034-4885/35/1/306} {\bibfield  {journal}
  {\bibinfo  {journal} {Rep. Prog. Phys.}\ }\textbf {\bibinfo {volume} {35}},\
  \bibinfo {pages} {315} (\bibinfo {year} {1972})}\BibitemShut {NoStop}%
\bibitem [{\citenamefont {Jackson}(1975)}]{Jackson75}%
  \BibitemOpen
  \bibfield  {author} {\bibinfo {author} {\bibfnamefont {J.~D.}\ \bibnamefont
  {Jackson}},\ }\href@noop {} {\emph {\bibinfo {title} {Classical
  Electrodynamics}}},\ \bibinfo {edition} {2nd}\ ed.\ (\bibinfo  {publisher}
  {New York: Wiley},\ \bibinfo {year} {1975})\BibitemShut {NoStop}%
\bibitem [{\citenamefont {Born}\ and\ \citenamefont {Wolf}(1999)}]{Born99}%
  \BibitemOpen
  \bibfield  {author} {\bibinfo {author} {\bibfnamefont {M.}~\bibnamefont
  {Born}}\ and\ \bibinfo {author} {\bibfnamefont {E.}~\bibnamefont {Wolf}},\
  }\href {\doibase 10.1017/CBO9781139644181} {\emph {\bibinfo {title}
  {Principles of Optics}}},\ \bibinfo {edition} {7th}\ ed.\ (\bibinfo
  {publisher} {Cambridge: Cambridge University Press},\ \bibinfo {year}
  {1999})\BibitemShut {NoStop}%
\bibitem [{\citenamefont {Lopez}\ and\ \citenamefont {Ram}(2018)}]{Lopez18b}%
  \BibitemOpen
  \bibfield  {author} {\bibinfo {author} {\bibfnamefont {N.~A.}\ \bibnamefont
  {Lopez}}\ and\ \bibinfo {author} {\bibfnamefont {A.~K.}\ \bibnamefont
  {Ram}},\ }\href {\doibase 10.1088/1361-6587/aae95e} {\bibfield  {journal}
  {\bibinfo  {journal} {Plasma Phys. Control. Fusion}\ }\textbf {\bibinfo
  {volume} {60}},\ \bibinfo {pages} {125012} (\bibinfo {year}
  {2018})}\BibitemShut {NoStop}%
\bibitem [{\citenamefont {Maslov}\ and\ \citenamefont
  {Fedoriuk}(1981)}]{Maslov81}%
  \BibitemOpen
  \bibfield  {author} {\bibinfo {author} {\bibfnamefont {V.~P.}\ \bibnamefont
  {Maslov}}\ and\ \bibinfo {author} {\bibfnamefont {M.~V.}\ \bibnamefont
  {Fedoriuk}},\ }\href@noop {} {\emph {\bibinfo {title} {Semiclassical
  Approximation in Quantum Mechanics}}}\ (\bibinfo  {publisher} {Dordrecht,
  Netherlands: Reidel},\ \bibinfo {year} {1981})\BibitemShut {NoStop}%
\bibitem [{\citenamefont {Alonso}\ and\ \citenamefont
  {Forbes}(1997)}]{Alonso97b}%
  \BibitemOpen
  \bibfield  {author} {\bibinfo {author} {\bibfnamefont {M.~A.}\ \bibnamefont
  {Alonso}}\ and\ \bibinfo {author} {\bibfnamefont {G.~W.}\ \bibnamefont
  {Forbes}},\ }\href {\doibase 10.1364/JOSAA.14.001279} {\bibfield  {journal}
  {\bibinfo  {journal} {J. Opt. Soc. Am. A}\ }\textbf {\bibinfo {volume}
  {14}},\ \bibinfo {pages} {1279} (\bibinfo {year} {1997})}\BibitemShut
  {NoStop}%
\bibitem [{\citenamefont {Littlejohn}(1986)}]{Littlejohn86a}%
  \BibitemOpen
  \bibfield  {author} {\bibinfo {author} {\bibfnamefont {R.~G.}\ \bibnamefont
  {Littlejohn}},\ }\href {\doibase 10.1016/0370-1573(86)90103-1} {\bibfield
  {journal} {\bibinfo  {journal} {Phys. Rep.}\ }\textbf {\bibinfo {volume}
  {138}},\ \bibinfo {pages} {193} (\bibinfo {year} {1986})}\BibitemShut
  {NoStop}%
\bibitem [{\citenamefont {Huber}\ \emph {et~al.}(1988)\citenamefont {Huber},
  \citenamefont {Heller},\ and\ \citenamefont {Littlejohn}}]{Huber88a}%
  \BibitemOpen
  \bibfield  {author} {\bibinfo {author} {\bibfnamefont {D.}~\bibnamefont
  {Huber}}, \bibinfo {author} {\bibfnamefont {E.~J.}\ \bibnamefont {Heller}}, \
  and\ \bibinfo {author} {\bibfnamefont {R.~G.}\ \bibnamefont {Littlejohn}},\
  }\href {\doibase 10.1063/1.455714} {\bibfield  {journal} {\bibinfo  {journal}
  {J. Chem. Phys.}\ }\textbf {\bibinfo {volume} {89}},\ \bibinfo {pages} {2003}
  (\bibinfo {year} {1988})}\BibitemShut {NoStop}%
\bibitem [{\citenamefont {{de Gosson}}(2006)}]{deGosson06}%
  \BibitemOpen
  \bibfield  {author} {\bibinfo {author} {\bibfnamefont {M.}~\bibnamefont {{de
  Gosson}}},\ }\href {\doibase 10.1007/3-7643-7575-2} {\emph {\bibinfo {title}
  {Symplectic Geometry and Quantum Mechanics}}}\ (\bibinfo  {publisher} {Basel:
  Birkh{\"a}user},\ \bibinfo {year} {2006})\BibitemShut {NoStop}%
\bibitem [{\citenamefont {Lopez}\ and\ \citenamefont {Dodin}(2019)}]{Lopez19a}%
  \BibitemOpen
  \bibfield  {author} {\bibinfo {author} {\bibfnamefont {N.~A.}\ \bibnamefont
  {Lopez}}\ and\ \bibinfo {author} {\bibfnamefont {I.~Y.}\ \bibnamefont
  {Dodin}},\ }\href {\doibase 10.1364/JOSAA.36.001846} {\bibfield  {journal}
  {\bibinfo  {journal} {J. Opt. Soc. Am. A}\ }\textbf {\bibinfo {volume}
  {36}},\ \bibinfo {pages} {1846} (\bibinfo {year} {2019})}\BibitemShut
  {NoStop}%
\bibitem [{\citenamefont {Dodin}\ \emph {et~al.}(2019)\citenamefont {Dodin},
  \citenamefont {Ruiz}, \citenamefont {Yanagihara}, \citenamefont {Zhou},\ and\
  \citenamefont {Kubo}}]{Dodin19}%
  \BibitemOpen
  \bibfield  {author} {\bibinfo {author} {\bibfnamefont {I.~Y.}\ \bibnamefont
  {Dodin}}, \bibinfo {author} {\bibfnamefont {D.~E.}\ \bibnamefont {Ruiz}},
  \bibinfo {author} {\bibfnamefont {K.}~\bibnamefont {Yanagihara}}, \bibinfo
  {author} {\bibfnamefont {Y.}~\bibnamefont {Zhou}}, \ and\ \bibinfo {author}
  {\bibfnamefont {S.}~\bibnamefont {Kubo}},\ }\href {\doibase
  10.1063/1.5095076} {\bibfield  {journal} {\bibinfo  {journal} {Phys.
  Plasmas}\ }\textbf {\bibinfo {volume} {26}},\ \bibinfo {pages} {072110}
  (\bibinfo {year} {2019})}\BibitemShut {NoStop}%
\bibitem [{\citenamefont {Stoler}(1981)}]{Stoler81}%
  \BibitemOpen
  \bibfield  {author} {\bibinfo {author} {\bibfnamefont {D.}~\bibnamefont
  {Stoler}},\ }\href {\doibase 10.1364/JOSA.71.000334} {\bibfield  {journal}
  {\bibinfo  {journal} {J. Opt. Soc. Am.}\ }\textbf {\bibinfo {volume} {71}},\
  \bibinfo {pages} {334} (\bibinfo {year} {1981})}\BibitemShut {NoStop}%
\bibitem [{\citenamefont {McDonald}(1988)}]{McDonald88a}%
  \BibitemOpen
  \bibfield  {author} {\bibinfo {author} {\bibfnamefont {S.~W.}\ \bibnamefont
  {McDonald}},\ }\href {\doibase 10.1016/0370-1573(88)90012-9} {\bibfield
  {journal} {\bibinfo  {journal} {Phys. Rep.}\ }\textbf {\bibinfo {volume}
  {158}},\ \bibinfo {pages} {337} (\bibinfo {year} {1988})}\BibitemShut
  {NoStop}%
\bibitem [{\citenamefont {Oancea}\ \emph {et~al.}(2020)\citenamefont {Oancea},
  \citenamefont {Joudioux}, \citenamefont {Dodin}, \citenamefont {Ruiz},
  \citenamefont {Paganini},\ and\ \citenamefont {Andersson}}]{Oancea20}%
  \BibitemOpen
  \bibfield  {author} {\bibinfo {author} {\bibfnamefont {M.~A.}\ \bibnamefont
  {Oancea}}, \bibinfo {author} {\bibfnamefont {J.}~\bibnamefont {Joudioux}},
  \bibinfo {author} {\bibfnamefont {I.~Y.}\ \bibnamefont {Dodin}}, \bibinfo
  {author} {\bibfnamefont {D.~E.}\ \bibnamefont {Ruiz}}, \bibinfo {author}
  {\bibfnamefont {C.~F.}\ \bibnamefont {Paganini}}, \ and\ \bibinfo {author}
  {\bibfnamefont {L.}~\bibnamefont {Andersson}},\ }\href
  {https://arxiv.org/abs/2003.04553} {\bibfield  {journal} {\bibinfo  {journal}
  {arXiv:2003.04553}\ } (\bibinfo {year} {2020})}\BibitemShut {NoStop}%
\bibitem [{\citenamefont {Ziolkowski}\ and\ \citenamefont
  {Deschamps}(1984)}]{Ziolkowski84}%
  \BibitemOpen
  \bibfield  {author} {\bibinfo {author} {\bibfnamefont {R.~W.}\ \bibnamefont
  {Ziolkowski}}\ and\ \bibinfo {author} {\bibfnamefont {G.~A.}\ \bibnamefont
  {Deschamps}},\ }\href {\doibase 10.1029/RS019i004p01001} {\bibfield
  {journal} {\bibinfo  {journal} {Radio Sci.}\ }\textbf {\bibinfo {volume}
  {19}},\ \bibinfo {pages} {1001} (\bibinfo {year} {1984})}\BibitemShut
  {NoStop}%
\bibitem [{\citenamefont {Olver}\ \emph {et~al.}(2010)\citenamefont {Olver},
  \citenamefont {Lozier}, \citenamefont {Boisvert},\ and\ \citenamefont
  {Clark}}]{Olver10}%
  \BibitemOpen
  \bibfield  {author} {\bibinfo {author} {\bibfnamefont {F.~W.~J.}\
  \bibnamefont {Olver}}, \bibinfo {author} {\bibfnamefont {D.~W.}\ \bibnamefont
  {Lozier}}, \bibinfo {author} {\bibfnamefont {R.~F.}\ \bibnamefont
  {Boisvert}}, \ and\ \bibinfo {author} {\bibfnamefont {C.~W.}\ \bibnamefont
  {Clark}},\ }\href@noop {} {\emph {\bibinfo {title} {NIST Handbook of
  Mathematical Functions}}}\ (\bibinfo  {publisher} {Cambridge: Cambridge
  University Press},\ \bibinfo {year} {2010})\BibitemShut {NoStop}%
\bibitem [{\citenamefont {Goldstein}\ \emph {et~al.}(2002)\citenamefont
  {Goldstein}, \citenamefont {Poole},\ and\ \citenamefont
  {Safko}}]{Goldstein02}%
  \BibitemOpen
  \bibfield  {author} {\bibinfo {author} {\bibfnamefont {H.}~\bibnamefont
  {Goldstein}}, \bibinfo {author} {\bibfnamefont {C.~P.}\ \bibnamefont
  {Poole}}, \ and\ \bibinfo {author} {\bibfnamefont {J.~L.}\ \bibnamefont
  {Safko}},\ }\href@noop {} {\emph {\bibinfo {title} {Classical Mechanics}}},\
  \bibinfo {edition} {3rd}\ ed.\ (\bibinfo  {publisher} {New York:
  Addison-Wesley},\ \bibinfo {year} {2002})\BibitemShut {NoStop}%
\bibitem [{\citenamefont {Trefethen}\ and\ \citenamefont {{Bau,
  III}}(1997)}]{Trefethen97}%
  \BibitemOpen
  \bibfield  {author} {\bibinfo {author} {\bibfnamefont {L.~N.}\ \bibnamefont
  {Trefethen}}\ and\ \bibinfo {author} {\bibfnamefont {D.}~\bibnamefont {{Bau,
  III}}},\ }\href@noop {} {\emph {\bibinfo {title} {Numerical Linear
  Algebra}}}\ (\bibinfo  {publisher} {Philadelphia: SIAM},\ \bibinfo {year}
  {1997})\BibitemShut {NoStop}%
\bibitem [{\citenamefont {Arnold}(1989)}]{Arnold89}%
  \BibitemOpen
  \bibfield  {author} {\bibinfo {author} {\bibfnamefont {V.~I.}\ \bibnamefont
  {Arnold}},\ }\href {\doibase 10.1007/978-1-4757-2063-1} {\emph {\bibinfo
  {title} {Mathematical Methods of Classical Mechanics}}}\ (\bibinfo
  {publisher} {New York: Springer},\ \bibinfo {year} {1989})\BibitemShut
  {NoStop}%
\bibitem [{\citenamefont {Yanagihara}\ \emph
  {et~al.}(2019{\natexlab{a}})\citenamefont {Yanagihara}, \citenamefont
  {Dodin},\ and\ \citenamefont {Kubo}}]{Yanagihara19a}%
  \BibitemOpen
  \bibfield  {author} {\bibinfo {author} {\bibfnamefont {K.}~\bibnamefont
  {Yanagihara}}, \bibinfo {author} {\bibfnamefont {I.~Y.}\ \bibnamefont
  {Dodin}}, \ and\ \bibinfo {author} {\bibfnamefont {S.}~\bibnamefont {Kubo}},\
  }\href {\doibase 10.1063/1.5095173} {\bibfield  {journal} {\bibinfo
  {journal} {Phys. Plasmas}\ }\textbf {\bibinfo {volume} {26}},\ \bibinfo
  {pages} {072111} (\bibinfo {year} {2019}{\natexlab{a}})}\BibitemShut
  {NoStop}%
\bibitem [{\citenamefont {Yanagihara}\ \emph
  {et~al.}(2019{\natexlab{b}})\citenamefont {Yanagihara}, \citenamefont
  {Dodin},\ and\ \citenamefont {Kubo}}]{Yanagihara19b}%
  \BibitemOpen
  \bibfield  {author} {\bibinfo {author} {\bibfnamefont {K.}~\bibnamefont
  {Yanagihara}}, \bibinfo {author} {\bibfnamefont {I.~Y.}\ \bibnamefont
  {Dodin}}, \ and\ \bibinfo {author} {\bibfnamefont {S.}~\bibnamefont {Kubo}},\
  }\href {\doibase 10.1063/1.5095174} {\bibfield  {journal} {\bibinfo
  {journal} {Phys. Plasmas}\ }\textbf {\bibinfo {volume} {26}},\ \bibinfo
  {pages} {072112} (\bibinfo {year} {2019}{\natexlab{b}})}\BibitemShut
  {NoStop}%
\bibitem [{\citenamefont {Press}\ \emph {et~al.}(2007)\citenamefont {Press},
  \citenamefont {Teukolsky}, \citenamefont {Vetterling},\ and\ \citenamefont
  {Flannery}}]{Press07}%
  \BibitemOpen
  \bibfield  {author} {\bibinfo {author} {\bibfnamefont {W.~H.}\ \bibnamefont
  {Press}}, \bibinfo {author} {\bibfnamefont {S.~A.}\ \bibnamefont
  {Teukolsky}}, \bibinfo {author} {\bibfnamefont {W.~T.}\ \bibnamefont
  {Vetterling}}, \ and\ \bibinfo {author} {\bibfnamefont {B.~P.}\ \bibnamefont
  {Flannery}},\ }\href@noop {} {\emph {\bibinfo {title} {Numerical Recipes}}},\
  \bibinfo {edition} {3rd}\ ed.\ (\bibinfo  {publisher} {Cambridge: Cambridge
  University Press},\ \bibinfo {year} {2007})\BibitemShut {NoStop}%
\bibitem [{\citenamefont {Richardson}\ and\ \citenamefont
  {Finn}(2012)}]{Richardson12}%
  \BibitemOpen
  \bibfield  {author} {\bibinfo {author} {\bibfnamefont {A.~S.}\ \bibnamefont
  {Richardson}}\ and\ \bibinfo {author} {\bibfnamefont {J.~M.}\ \bibnamefont
  {Finn}},\ }\href {\doibase 10.1088/0741-3335/54/1/014004} {\bibfield
  {journal} {\bibinfo  {journal} {Plasma Phys. Control. Fusion}\ }\textbf
  {\bibinfo {volume} {54}},\ \bibinfo {pages} {014004} (\bibinfo {year}
  {2012})}\BibitemShut {NoStop}%
\bibitem [{\citenamefont {Hairer}(1997)}]{Hairer97}%
  \BibitemOpen
  \bibfield  {author} {\bibinfo {author} {\bibfnamefont {E.}~\bibnamefont
  {Hairer}},\ }\href {\doibase 10.1016/S0168-9274(97)00061-5} {\bibfield
  {journal} {\bibinfo  {journal} {Appl. Numer. Math}\ }\textbf {\bibinfo
  {volume} {25}},\ \bibinfo {pages} {219} (\bibinfo {year} {1997})}\BibitemShut
  {NoStop}%
\bibitem [{\citenamefont {Zare}\ and\ \citenamefont
  {Szebehely}(1975)}]{Zare75}%
  \BibitemOpen
  \bibfield  {author} {\bibinfo {author} {\bibfnamefont {K.}~\bibnamefont
  {Zare}}\ and\ \bibinfo {author} {\bibfnamefont {V.}~\bibnamefont
  {Szebehely}},\ }\href {\doibase 10.1007/BF01650285} {\bibfield  {journal}
  {\bibinfo  {journal} {Celest. Mech.}\ }\textbf {\bibinfo {volume} {11}},\
  \bibinfo {pages} {469} (\bibinfo {year} {1975})}\BibitemShut {NoStop}%
\bibitem [{\citenamefont {Goldman}(2005)}]{Goldman05}%
  \BibitemOpen
  \bibfield  {author} {\bibinfo {author} {\bibfnamefont {R.}~\bibnamefont
  {Goldman}},\ }\href {\doibase 10.1016/j.cagd.2005.06.005} {\bibfield
  {journal} {\bibinfo  {journal} {Comput. Aided Geom. Des.}\ }\textbf {\bibinfo
  {volume} {22}},\ \bibinfo {pages} {632} (\bibinfo {year} {2005})}\BibitemShut
  {NoStop}%
\bibitem [{\citenamefont {Dodin}\ and\ \citenamefont {Fisch}(2010)}]{Dodin10c}%
  \BibitemOpen
  \bibfield  {author} {\bibinfo {author} {\bibfnamefont {I.~Y.}\ \bibnamefont
  {Dodin}}\ and\ \bibinfo {author} {\bibfnamefont {N.~J.}\ \bibnamefont
  {Fisch}},\ }\href {\doibase 10.1063/1.3497005} {\bibfield  {journal}
  {\bibinfo  {journal} {Phys. Plasmas}\ }\textbf {\bibinfo {volume} {17}},\
  \bibinfo {pages} {112118} (\bibinfo {year} {2010})}\BibitemShut {NoStop}%
\bibitem [{\citenamefont {Bender}\ and\ \citenamefont
  {Orszag}(1978)}]{Bender78}%
  \BibitemOpen
  \bibfield  {author} {\bibinfo {author} {\bibfnamefont {C.~M.}\ \bibnamefont
  {Bender}}\ and\ \bibinfo {author} {\bibfnamefont {S.~A.}\ \bibnamefont
  {Orszag}},\ }\href {\doibase 10.1007/978-1-4757-3069-2} {\emph {\bibinfo
  {title} {Advanced Mathematical Methods for Scientists and Engineers}}}\
  (\bibinfo  {publisher} {New York: McGraw-Hill},\ \bibinfo {year}
  {1978})\BibitemShut {NoStop}%
\bibitem [{\citenamefont {Grunwald}\ and\ \citenamefont
  {Milano}(1971)}]{Grunwald71}%
  \BibitemOpen
  \bibfield  {author} {\bibinfo {author} {\bibfnamefont {H.~P.}\ \bibnamefont
  {Grunwald}}\ and\ \bibinfo {author} {\bibfnamefont {F.}~\bibnamefont
  {Milano}},\ }\href {\doibase 10.1119/1.1976668} {\bibfield  {journal}
  {\bibinfo  {journal} {Am. J. Phys.}\ }\textbf {\bibinfo {volume} {39}},\
  \bibinfo {pages} {1394} (\bibinfo {year} {1971})}\BibitemShut {NoStop}%
\bibitem [{\citenamefont {Pool}(1966)}]{Pool66}%
  \BibitemOpen
  \bibfield  {author} {\bibinfo {author} {\bibfnamefont {J.~C.~T.}\
  \bibnamefont {Pool}},\ }\href {\doibase 10.1063/1.1704817} {\bibfield
  {journal} {\bibinfo  {journal} {J. Math. Phys.}\ }\textbf {\bibinfo {volume}
  {7}},\ \bibinfo {pages} {1966} (\bibinfo {year} {1966})}\BibitemShut
  {NoStop}%
\bibitem [{\citenamefont {Comtet}(1974)}]{Comtet74}%
  \BibitemOpen
  \bibfield  {author} {\bibinfo {author} {\bibfnamefont {L.}~\bibnamefont
  {Comtet}},\ }\href {\doibase 10.1007/978-94-010-2196-8} {\emph {\bibinfo
  {title} {Advanced Combinatorics}}}\ (\bibinfo  {publisher} {Dordrecht,
  Netherlands: Reidel},\ \bibinfo {year} {1974})\BibitemShut {NoStop}%
\end{thebibliography}%
\bibliographystyle{apsrev4-1}
\end{document}